\documentclass[fleqn,usenatbib]{mnras}

\usepackage{newtxtext,newtxmath}

\usepackage[T1]{fontenc}

\DeclareRobustCommand{\VAN}[3]{#2}
\let\VANthebibliography\thebibliography
\def\thebibliography{\DeclareRobustCommand{\VAN}[3]{##3}\VANthebibliography}

\usepackage{graphicx}	
\usepackage{amsmath}	
\usepackage{CJKutf8}    

\usepackage{tikz,xcolor,hyperref}

\newcommand{\hl}{}

\newcommand{\cii}{C\,\textsc{ii}]}
\newcommand{\ciii}{C\,\textsc{iii}]}
\newcommand{\mgii}{Mg\,\textsc{ii}}
\newcommand{\civ}{C\,\textsc{iv}}
\newcommand{\feii}{Fe\,\textsc{ii}}
\newcommand{\nv}{N\,\textsc{v}}
\newcommand{\niv}{N\,\textsc{iv}]}
\newcommand{\siii}{Si\,\textsc{ii}}
\newcommand{\siiii}{Si\,\textsc{iii}}
\newcommand{\siiv}{Si\,\textsc{iv}}
\newcommand{\oiv}{O\,\textsc{iv}}
\newcommand{\heii}{He\,\textsc{ii}}
\newcommand{\oi}{O\,\textsc{i}]}
\newcommand{\oii}{O\,\textsc{ii}]}
\newcommand{\oiii}{O\,\textsc{iii}]}
\newcommand{\alii}{Al\,\textsc{ii}}
\newcommand{\aliii}{Al\,\textsc{iii}}
\newcommand{\lya}{Ly\,\textsc{$\alpha$}}

\newcommand{\ciiil}{C \textsc{iii}] $\lambda$1909}
\newcommand{\mgiil}{Mg \textsc{ii} $\lambda$2799}
\newcommand{\civl}{C \textsc{iv} $\lambda$1549}
\newcommand{\nvl}{N \textsc{v} $\lambda$1240}
\newcommand{\nivl}{N \textsc{iv}] $\lambda$1486}
\newcommand{\siiil}{Si \textsc{ii} $\lambda$1263}
\newcommand{\siiiil}{Si \textsc{iii} $\lambda$1887}
\newcommand{\siivl}{Si \textsc{iv} $\lambda$1398}
\newcommand{\oivl}{O \textsc{iv} $\lambda$1402}
\newcommand{\heiil}{He \textsc{ii} $\lambda$1640}
\newcommand{\oiiil}{O \textsc{iii}] $\lambda$1663}
\newcommand{\aliil}{Al \textsc{ii} $\lambda$1671}
\newcommand{\aliiil}{Al \textsc{iii} $\lambda$1857}
\newcommand{\lyal}{Ly \textsc{$\alpha$} $\lambda$1216}

\newcommand{\nvciv}{\nv/\civ}
\newcommand{\siivoivciv}{(\siiv+\oiv)/\civ}
\newcommand{\oiiialiiciv}{(\oiii+\alii)/\civ}
\newcommand{\aliiiciv}{\aliii/\civ}
\newcommand{\siiiiciv}{\siiii/\civ}
\newcommand{\ciiiciv}{\ciii/\civ}

\definecolor{lime}{HTML}{A6CE39}
\DeclareRobustCommand{\orcidicon}{%
    \begin{tikzpicture}
    \draw[lime, fill=lime] (0,0) 
    circle [radius=0.16] 
    node[white] {{\fontfamily{qag}\selectfont \tiny ID}};
    \draw[white, fill=white] (-0.0625,0.095) 
    circle [radius=0.007];
    \end{tikzpicture}
    \hspace{-2mm}
}

\newcommand{\orcidChrisO}{\href{https://orcid.org/0000-0003-0017-349X}{\orcidicon}}
\newcommand{\orcidChrisW}{\href{https://orcid.org/0000-0002-4569-016X}{\orcidicon}}
\newcommand{\orcidSamuel}{\href{https://orcid.org/0000-0001-9372-4611}{\orcidicon}}
\newcommand{\orcidFuyan}{\href{https://orcid.org/0000-0002-1620-0897}{\orcidicon}}
\newcommand{\orcidChiara}{\href{https://orcid.org/0000-0002-5941-5214}{\orcidicon}}
\newcommand{\orcidJT}{\href{https://orcid.org/0000-0002-4544-8242}{\orcidicon}}
\newcommand{\orcidOnoue}{\href{https://orcid.org/0000-0003-2984-6803}{\orcidicon}}
\newcommand{\orcidValentina}{\href{https://orcid.org/0000-0003-3693-3091}{\orcidicon}}
\newcommand{\orcidManuela}{\href{https://orcid.org/0000-0002-4314-021X}{\orcidicon}}
\newcommand{\orcidjinyi}{\href{https://orcid.org/0000-0001-5287-4242}{\orcidicon}}
\newcommand{\orcidfanxiaohui}{\href{https://orcid.org/0000-0003-3310-0131}{\orcidicon}}
\newcommand{\orcidACE}{\href{https://orcid.org/0000-0003-2895-6218}{\orcidicon}}
\newcommand{\orcidFeige}{\href{https://orcid.org/0000-0002-7633-431X}{\orcidicon}}
\newcommand{\orcidEPF}{\href{https://orcid.org/0000-0002-6822-2254}{\orcidicon}}
\newcommand{\orcidSarahBosman}{\href{https://orcid.org/
0000-0001-8582-7012}{\orcidicon}}
\newcommand{\orcidYongda}{\href{https://orcid.org/
0000-0003-3307-7525}{\orcidicon}}
\newcommand{\orcidFabian}{\href{https://orcid.org/0000-0003-4793-7880}{\orcidicon}}
\newcommand{\orcidGeorgeB}{\href{https://orcid.org/0000-0003-2344-263X}{\orcidicon}}
\newcommand{\orcidGuido}{\href{https://orcid.org/0000-0002-6830-9093}{\orcidicon}}
\newcommand{\orcidEduardo}{\href{https://orcid.org/0000-0002-2931-7824}{\orcidicon}}

\title[Quasar Chemical Abundance]{Chemical Abundance of $z\sim6$ Quasar Broad-Line Regions in the XQR-30 Sample}

\author[S. Lai et al.]{Samuel Lai \begin{CJK}{UTF8}{gbsn}(赖民希)\end{CJK},$^{1,2}$\orcidSamuel\thanks{E-mail: samuel.lai@anu.edu.au}
Fuyan Bian \begin{CJK}{UTF8}{gbsn}(边福彦)\end{CJK},$^{2}$\orcidFuyan\,
Christopher A. Onken,$^{1,3}$\orcidChrisO\,
Christian Wolf,$^{1,3}$\orcidChrisW\,
\newauthor{Chiara Mazzucchelli,$^{2}$\orcidChiara\,
Eduardo Ba\~nados,$^{4}$\orcidEduardo\,
Manuela Bischetti,$^{5}$\orcidManuela\,
Sarah E.I. Bosman,$^{4}$\orcidSarahBosman\,
}
\newauthor{George Becker,$^{6}$\orcidGeorgeB\,
Guido Cupani,$^{5,7}$\orcidGuido\,
Valentina D'Odorico,$^{5, 7, 8}$\orcidValentina\,
Anna-Christina Eilers,$^{9}$\thanks{NASA Hubble Fellow}\orcidACE\,
}
\newauthor{Xiaohui Fan,$^{10}$\orcidfanxiaohui\,
Emanuele Paolo Farina,$^{11}$\orcidEPF\,
Masafusa Onoue,$^{4,12,13}$\orcidOnoue\,
Jan-Torge Schindler,$^{14}$\orcidJT\,}
\newauthor{Fabian Walter,$^{4}$\orcidFabian\,
Feige Wang,$^{10}$\footnotemark[2]\orcidFeige\,
Jinyi Yang$^{10}$\thanks{Strittmatter Fellow}\orcidjinyi\,
and Yongda Zhu$^{6}$\orcidYongda}
\\
$^{1}$Research School of Astronomy and Astrophysics, Australian National University, Canberra, ACT 2611, Australia\\
$^{2}$European Southern Observatory, Alonso de C\'{o}rdova 3107, Casilla 19001, Vitacura, Santiago 19, Chile\\
$^{3}$Centre for Gravitational Astrophysics, Research Schools of Physics, and Astronomy and Astrophysics, Australian National University\\
$^{4}$Max-Planck-Institut für Astronomie, Königstuhl 17, D-69117 Heidelberg, Germany\\
$^{5}$INAF – Osservatorio Astronomico di Trieste, Via G. B. Tiepolo 11, I-34143 Trieste, Italy\\
$^{6}$Department of Physics \& Astronomy, University of California, Riverside, CA 92521, USA\\
$^{7}$IFPU–Institute for Fundamental Physics of the Universe, via Beirut 2, I-34151 Trieste, Italy\\
$^{8}$Scuola Normale Superiore, piazza dei Cavalieri, I-56126 Pisa, Italy\\
$^{9}$MIT Kavli Institute for Astrophysics and Space Research, 77 Massachusetts Ave., Cambridge, MA 02139, USA\\
$^{10}$Steward Observatory, University of Arizona, 933 N Cherry Ave, Tucson, AZ 85721, USA\\
$^{11}$Gemini Observatory, NSF’s NOIRLab, 670 N A’ohoku Place, Hilo, Hawai'i 96720, USA\\
$^{12}$Kavli Institute for Astronomy and Astrophysics, Peking University, Beijing 100871, China\\
$^{13}$Kavli Institute for the Physics and Mathematics of the Universe (Kavli IPMU, WPI), The University of Tokyo, 5-1-5 Kashiwanoha, Kashiwa, Chiba 277-8583, Japan\\
$^{14}$Leiden Observatory, Leiden University, PO Box 9513, 2300 RA Leiden, The Netherlands\\
}

\date{Accepted XXX. Received YYY; in original form ZZZ}

\pubyear{2021}

\begin{document}
\label{firstpage}
\pagerange{\pageref{firstpage}--\pageref{lastpage}}
\maketitle

\begin{abstract}
The elemental abundances in the broad-line regions of high-redshift quasars trace the chemical evolution in the nuclear regions of massive galaxies in the early universe. In this work, we study metallicity-sensitive broad emission-line flux ratios in rest-frame UV spectra of 25 high-redshift (5.8 < z < 7.5) quasars observed with the VLT/X-shooter and Gemini/GNIRS instruments, ranging over $\log\left({\rm{M}_{\rm{BH}}/\rm{M}_{\odot}}\right) = 8.4-9.8$ in black hole mass and $\log\left(\rm{L}_{\rm{bol}}/\rm{erg \, s}^{-1}\right) = 46.7-47.7$ in bolometric luminosity. We fit individual spectra and composites generated by binning across quasar properties: bolometric luminosity, black hole mass, and blueshift of the \civ\ line, finding no redshift evolution in the emission-line ratios by comparing our high-redshift quasars to lower-redshift (2.0 < z < 5.0) results presented in the literature. Using \texttt{Cloudy}-based locally optimally-emitting cloud photoionisation model relations between metallicity and emission-line flux ratios, we find the observable properties of the broad emission lines to be consistent with emission from gas clouds with metallicity that are at least 2-4 times solar. Our high-redshift measurements also confirm that the blueshift of the \civ\ emission line is correlated with its equivalent width, which influences line ratios normalised against \civ. When accounting for the \civ\ blueshift, we find that the rest-frame UV emission-line flux ratios do not correlate appreciably with the black hole mass or bolometric luminosity. 
\end{abstract}




\section{Introduction}
The broad-line region (BLR) of quasars contains dense ($n_{\rm{H}} \approx 10^{9-14}$ cm$^{-3}$) and high-temperature (T $\sim$ 10$^4$ K) gas \citep[e.g.][]{Peterson_2006}, which is in close proximity to the supermassive black hole and photoionised by radiation from the accretion disk. Emission lines originating from the BLR can be used as virial estimators of the black hole mass \citep[e.g][]{vestergaard_2002, Mclure_2004, Greene_2005, shen_biases_2008, Onken_2008, Vestergaard_2009}, as well as to infer chemical abundances of the gas around black holes \citep[e.g.][]{hamann1992, Hamann_1999, Hamann_2002, Dietrich_2003, Nagao_2006, Matsuoka_2011, Wang_2012, shin_2017_outflow, Xu2018, Wang_2021}. The tight correlation between the supermassive black hole (SMBH) mass and the galactic bulge mass \citep[the $\rm{M}_{\rm{BH}}-\rm{M}_{\rm{bulge}}$ relation;][]{1998AJ....115.2285M, 2003ApJ...589L..21M, 2004ApJ...604L..89H, Greene_2010} as well as the velocity dispersion of the galactic bulge \citep[the $\rm{M}_{\rm{BH}}-\sigma$ relation; ][]{Ferrarese2000, Gebhardt2000, Tremaine2002, Salviander_2013} suggests that host galaxies and their central SMBH co-evolve over cosmic time. High-redshift quasars provide an opportunity to understand the formation and evolution of the earliest galaxies and their supermassive black holes. The galaxy stellar mass - gas phase metallicity relationship (MZR) \citep[e.g.][]{Maiolino_2008, Dave_2017_mufasa, Curti_2019_Klever, Maiolino_2019_metallica, Sanders_2021_MOSDEF} combined with the M$_{\rm{BH}}$/M$_{\rm{host}}$ ratio \citep[e.g.][]{Targett_2012} and a relationship between the quasar BLR metallicity with black hole mass (Z$_{\rm{BLR}}$ - M$_{\rm{BH}}$) can be used to link the mass or metallicity of the central black hole to metallicity in the host galaxy, enabling an investigation of their co-evolution in the young (< 1 Gyr) universe \citep[e.g.][]{hamann_1993}.

The study of quasar BLR metallicity is strongly motivated by the relationship between quasar activity, host galaxy evolution, and star formation episodes \citep[e.g.][]{Hamann_1999}. Elemental abundances in the BLR are indicative of the chemical evolution in galactic nuclear material. Early investigations of BLR metallicity using highly ionised ions of C, N, and O among others, indicated solar or super-solar metallicity with high associated uncertainties \citep[e.g.][]{baldwin1978, shields1976}. Given the degeneracies involved with fitting these broad emission features, high signal-to-noise ratio (SNR) spectra or stacking spectra into high SNR composites is needed to accurately fit these lines. Considerable progress has been made since then, pushing towards higher redshift \citep[e.g.][]{Pentericci_2002} and investigating chemical enrichment history. Photoionisation models suggest that rest-frame UV line flux ratios, such as \siivoivciv\ and \nvciv, can be used to infer metallicity in the BLR \citep{Hamann_2002, Nagao_2006}. These high-ionisation lines are associated to a region closer to the nuclear engine of the black hole than low-ionisation lines as indicated by reverberation mapping experiments \citep{Collin_1986, Collin_Souffrin_1988, 1991ApJ...366...64C, Korista_1995, williams2020}. Studies utilizing these metallicity-sensitive lines found that quasar metallicity correlates with luminosity \citep[e.g.][]{hamann_1993, Dietrich_2003, Nagao_2006, Xu2018}, and outflow strength or velocity \citep[e.g.][]{Wang_2012, shin_2017_outflow, metal_density}. Additional studies found that metallicity correlates with black hole mass \citep[e.g][]{Matsuoka_2011, Xu2018, Wang_2021}, possibly pointing to a more fundamental relationship between black hole mass and the metallicity of its BLR. 

Simultaneously, there is no evidence to suggest that the same line ratios evolve with redshift \citep[e.g.][]{Pentericci_2002, metallicity_distant, Xu2018} up to redshifts as high as $z = 7.64$ \citep{Onoue_2020, yang2021probing}. Such studies consistently estimate metallicities several times the solar value $Z \sim 5 Z_{\rm{\odot}}$, up to $Z > 10 Z_{\rm{\odot}}$ in some quasars \cite[e.g.][]{metallicity_distant}. High redshift quasars with super-solar metallicities suggest rapid chemical enrichment scenarios. Under this paradigm, the nuclei of the most massive galaxies in the early universe were enriched rapidly from the host galaxy's interstellar medium within $\sim 500\, \rm{Myr}$ from the formation of the first stars. From then on, observations suggest that the metallicity of the quasar BLR did not change appreciably for a significant span of cosmic time.

It has also been suggested that the diversity of high-ionisation emission line ratios measured across a wide range of black hole masses and luminosities can be attributed, in whole or in part, to gas emission from at least two distinct regions of differing densities, illuminated by different ionizing radiation \citep{sameshima2017, metal_density}. This model does not necessarily require the metallicity in the BLR to vary across the quasar population in order to account for the observed differences in emission line properties. These studies indicate that inferences on quasar chemical enrichment history utilizing emission-line flux ratios have to account for variations in the physical conditions of the emitting gas. 

In this paper, we study spectra of 25 high-redshift (z > 5.8) quasars taken with ESO's VLT/X-shooter and Gemini-N/GNIRS. The bulk of our sample consists of high-resolution and high SNR spectra from the ESO-VLT X-shooter Large Program XQR-30 (P.I. V. D'Odorico). We fit and investigate flux ratios of metallicity-sensitive lines (primarily \nvciv\ and \siivoivciv) using individual quasar spectra and composites binned by black hole mass, bolometric luminosity, and blueshift of the \civ\ line to determine whether these parameters are correlated with metallicity in the BLR. This paper follows closely other studies at lower redshift \citep[e.g.][]{Nagao_2006, Xu2018, Shin_2019} and high redshift studies based on smaller samples \citep[e.g.][]{Jiang_2007, metallicity_distant, DeRosa_2014, JJTang_2019, Onoue_2020, Wang_2021}, many of which report measurements of the same metallicity indicators we use. Compared to a recent study of 33 $z\sim 6$ quasars observed with Gemini-N/GNIRS \citep{Wang_2021}, our sample contains higher SNR and spectral resolution spectra from X-shooter. Additionally, we consider the effects of BLR outflow on the metallicity-sensitive flux ratios, where the outflow is measured by the blueshift of the \civ\ emission line \citep[e.g.][]{Sulentic_2000, baskin_laor_2005, Vietri_2018}.

The content of this paper is organized as follows: in Section \ref{sec:sample-composites}, we describe the properties of our high-redshift quasar sample, data reduction, spectrum processing, and the methodology for generating composites. In Section \ref{sec:metallicity}, we describe our approach to fitting metallicity-sensitive emission lines and the conversion from line ratios to metallicities in the BLR. We present the results for our high redshift sample in Section \ref{sec:results}, and in Section \ref{sec:discussion}, we discuss and contextualize the results, presenting correlations found between the properties of the high redshift quasars and the metallicity of their BLRs. We summarize and conclude in Section \ref{sec:conclusion}. Throughout the paper, we adopt flat $\Lambda$CDM cosmology with H$_{0} = 70$ km s$^{-1}$ Mpc$^{-1}$ and $\left(\Omega_{\rm m}, \Omega_{\Lambda}\right) = \left(0.3, 0.7\right)$. All referenced wavelengths of emission lines are measured in vacuum.

\section{Quasar Sample and Composites} \label{sec:sample-composites}
\subsection{Sample Selection} 
The bulk of the sample originates from quasars in the ESO-VLT X-shooter Large Program XQR-30 (P.I. V. D'Odorico, program number 1103.A-0817)\footnote{Collaboration website: \href{https://xqr30.inaf.it/}{https://xqr30.inaf.it/}}. The XQR-30
program targets 30 southern hemisphere bright QSOs at $5.8 < z < 6.6$ to study the universe in its infancy. These quasars have virially estimated BH masses of $(0.8 - 6.0) \times 10^9\, \rm{M}_{\odot}$ (Mazzucchelli et al. in prep.). At lower BH masses, $0.2 - 1.0 \times 10^9\, \rm{M}_{\odot}$, we include 1 quasar spectrum from \cite{Shen_2019} and 9 spectra from \cite{yang2021probing}, all of them taken with Gemini/N GNIRS. From these other samples, we only considered spectra covering quasar properties outside the range of XQR-30 quasars with SNR > 5 per resolution element near rest-frame 1600\AA\ and 2800\AA, which are in the proximity of the emission lines of interest. Their redshifts span a similar range from $z = 6.0$ to $z = 6.8$ with one quasar at $z = 7.54$ \citep[i.e.][]{Onoue_2020}. All redshifts are measured from the peak of the best fit models to the \mgii\ emission line, where we use the complete reconstructed line profile in case of multi-component fits. We provide some details of the \mgii\ fits in Section \ref{sec:emission-fitting} and we leave the complete discussion for Mazzucchelli et al. in prep.

\begin{figure*}
	\includegraphics[width=\textwidth]{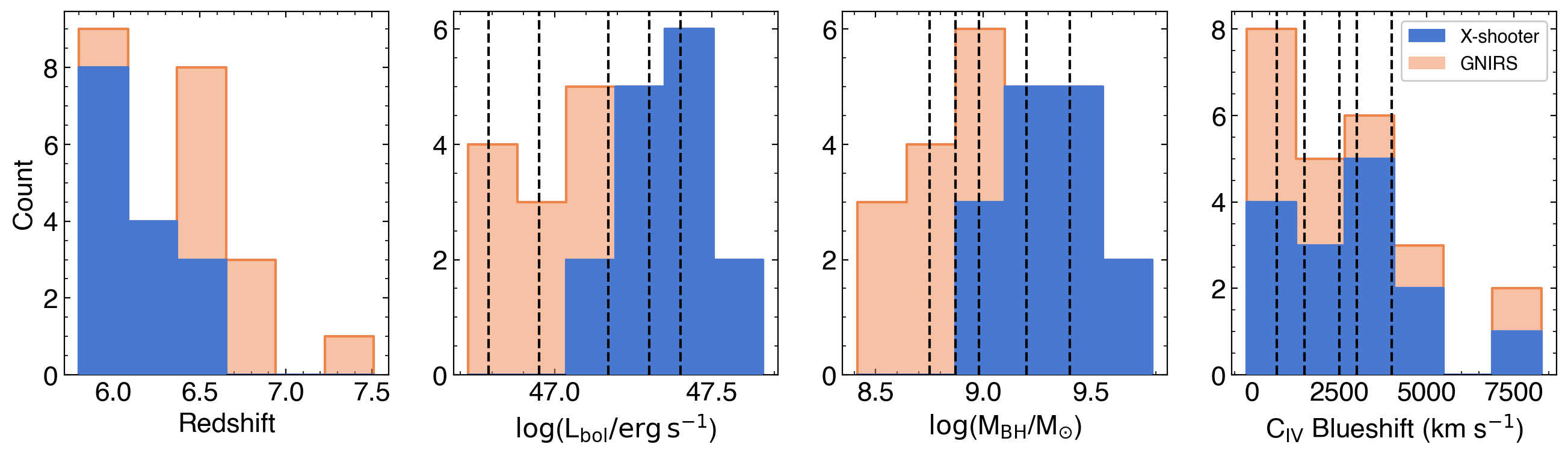}
    \caption{Stacked distribution of quasar properties in our sample from left to right: redshift, quasar bolometric luminosity, black hole mass, \civ\ blueshift. The properties of the XQR-30 sample, observed with ESO's X-shooter, are highlighted in blue while the GNIRS spectra are presented in orange. The inner bin edges of composites by quasar bolometric luminosity, black hole mass, and \civ\ blueshift are delineated by black dashed lines.}
    \label{fig:quasar_properties}
\end{figure*}

Of the 30 quasars in XQR-30, 14 are classified as quasars with broad absorption-lines (BAL) and 16 are considered non-BAL \citep{Bischetti_2022_submitted}. We exclude quasars with BAL features as they introduce additional uncertainty in the measurement of line flux. From the XQR-30 non-BAL sample, we reject J1535+1943 by visual inspection due to its dust-reddened continuum in the wavelength regions of interest \citep{yang2021probing}. Such a continuum is not well-modeled by the continuum fitting method we describe in Section \ref{sec:emission-fitting} and it would affect the continuum fit if included in composites. Combined with 10 spectra from Gemini GNIRS, a total of 25 high-redshift quasar spectra are included in this study. Figure \ref{fig:quasar_properties} and Table~\ref{tab:quasar_sample} show the distribution of the quasar sample and physical properties.

\subsection{Data Reduction and Post-Processing}
The data reduction procedure for GNIRS spectra is described in full in the source papers: \citet{Shen_2019} and \citet{yang2021probing}. \citet{Shen_2019} used a combination of the PyRAF-based \texttt{XDGNIRS} \citep{XDGNIRS_eg} and the IDL-based \texttt{XIDL} package while \citet{yang2021probing} used the Python-based spectroscopic data reduction pipeline \texttt{PypeIt} \citep{PypeIt}.

XQR-30 data \hl{are} reduced with an improved version of the flexible custom IDL-based pipeline used with data from the XQ-100 legacy survey \citep{XQ-100, Becker_2019}. The overall strategy is based on techniques described in \citet{Kelson_2003} with optimal sky subtraction, telluric absorption correction, optimal extraction, and direct combination of exposures. The pipeline is described in additional detail in \citet{Becker_2012} and it has also been used in other studies based on XQR-30 data \citep[e.g.][]{Zhu_2021}. X-shooter data \hl{are} obtained using three arms with the following wavelength ranges: UVB (300-559.5 nm), VIS (559.5-1024 nm), and NIR (1024-2480 nm). We extract data from only the VIS and NIR arms because there is no light in the UVB for our sources.

In the 50 km s$^{-1}$ rebinned quasar spectra from the XQR-30 sample, the mean SNR per pixel measured in the range 1400-1600\AA\ in the rest frame, is $\sim30$ with a minimum of 24 and maximum of 38, while the mean SNR per pixel for the GNIRS sample is $\sim13$, ranging between 6 and 30. Median pixel widths are 0.25\AA\ for the rebinned XQR-30 spectra and 0.43\AA\ for GNIRS spectra between rest-frame 1400-1600\AA. The XQR-30 SNR reported here can be different from those of other studies based on XQR-30 data because of differences in binning strategies and wavelength region over which the SNR is measured. 

After data reduction, each spectrum undergoes a common post-processing procedure described as follows:
\begin{enumerate}
    \item For every reduced spectrum in our quasar sample prior to creating composite spectra, the data \hl{are} restricted to relatively high SNR. The per-pixel SNR floor is 1 for GNIRS spectra and 5 for XQR-30 spectra. Data restricted by the SNR floor are omitted from further processing and fitting.
    \item We then apply a sigma-clip mask with a box width of \hl{30 pixels}, and a 3-$\sigma$ threshold to remove narrow absorption features and noise above 3-$\sigma$. These absorption features are not desired when fitting the intrinsic flux and profile of the broad emission lines. \hl{For the noisier and lower resolution GNIRS spectra, the sigma-clip mask has a minimal effect on the resulting spectra.}
    \item As the spectra are observed with different instruments and exhibit a diversity of redshifts, we standardize the rest-frame wavelength domain for all of the spectra, facilitating the stacking of composites later. Every spectrum is resampled using a flux-conserving algorithm into a common wavelength domain with 1 \AA\ bins in the rest-frame. The resampling calculation and error propagation are described in detail in \citet{Carnall_2017}. The number of pixels per 1 \AA\ bin in the raw spectra is wavelength-dependent, ranging from 1-3 pixels per bin for GNIRS spectra and 2-8 for X-shooter spectra. 
\end{enumerate}
To test the robustness of our measurements, we vary the details of the post-processing procedure, Among the many variations, we perform the sigma-clipping before resampling rather than after, apply an upper error threshold to restrict the maximum allowable error, and in one instance, we do not perform resampling on individual quasar spectra. In each case, we find that the majority of measurements are consistent within their uncertainties and the overall correlations and conclusions we draw from our measurements are unaffected. This gives us confidence in our results. 

\subsection{Black Hole Mass Estimate}
The black hole mass of each quasar is based on single-epoch virial mass estimates. We source the black hole masses from \citet{Shen_2019}, \citet{yang2021probing}, and Mazzucchelli et al. in prep. which span $\log\left({\rm{M}_{\rm{BH}}/\rm{M}_{\odot}}\right) = 8.4-9.8$ over the entire quasar sample. To determine the masses, these studies use the rest-frame UV \mgii\ broad emission line and the \mgii-based virial estimator, described generally by the following,
\begin{equation}
   \left(\frac{M_{\rm{BH,vir}}}{M_{\odot}}\right) = 10^{\rm{a}} \left[\frac{\lambda L_{\lambda}}{10^{44} \,\rm{erg\, s^{-1}}}\right]^{b} \left[\frac{\rm{FWHM (\mgii)}}{1000 \,\rm{km\, s^{-1}}}\right]^{2} \,,
   \label{eq:mgii-vo9}
\end{equation}
where $\lambda L_{\lambda}$ is the monochromatic luminosity of the continuum at rest frame 3000 \AA, and (a,b) are empirically calibrated against reverberation mapping experiments to the values (6.86, 0.5) in \citet{Vestergaard_2009} and (0.74, 0.62) in \citet{Shen_2011}. The masses of quasars in the XQR-30 and \citet{yang2021probing} samples are estimated using the \citet{Vestergaard_2009} calibration. The mass of the one quasar we've included from \citet{Shen_2019} is reported with the \citet{Shen_2011} calibration, but we have re-calibrated the mass with \citet{Vestergaard_2009}, resulting in a 0.1 dex difference. The continuum luminosity is estimated by fitting a power-law continuum and \feii\ emission around the \mgii\ line, as described in Section \ref{sec:emission-fitting} and the \feii\ template \cite[i.e.][]{UV_iron_template} is consistent between the different studies. The absolute fluxing of the XQR-30 spectra \hl{is based on calibrations against observed near-infrared photometry and} is described in full in Mazzucchelli et al. in prep. The \mgii\ full-width at half maximum (FWHM) is determined with single or multi-component Gaussian fits to the broad emission line and the peak of the total line profile is used to calibrate the systemic redshift of the quasar spectrum. Typical systematic errors from the virial mass estimator for the \mgii\ line can be up to 0.55 dex \citep{shen_biases_2008, Vestergaard_2009}. Bolometric luminosities are measured from the flux-calibrated spectrum using the continuum luminosity at 3000\AA\ and adopting a bolometric correction of 5.15 \citep{Shen_2011} throughout our entire sample.

The virial mass estimate is routinely applied to quasars \citep[e.g.][]{2002MNRAS.331..795M, Shen_2012} and aside from the \mgii\ line, $\rm{H}\beta$ and \civ\ emission lines have been used. Virial mass estimates using the $\rm{H}\beta$ emission line is not feasible for high redshift quasar studies prior to the James Webb Space Telescope, but the \mgii\ line width is correlated with $\rm{H}\beta$ and can be used as its substitution in single-epoch virial black hole mass estimates \citep[e.g.][]{Salviander_2007, shen_biases_2008, Wang_2009, Shen_2012}. Compared to the \civ\ emission line, the advantage of the \mgii\ line is that it is less affected by non-virial components of the black hole emission, such as the radiatively-driven BLR wind \citep[e.g.][]{Saturni_2018}. The difference between the \civ\ and \mgii\ virial mass estimates is correlated with the \civ\ blueshift \citep{Shen_2012, Coatman_2017}. 

\subsection{\civ\ Blueshift Measurement}
The \civ\ emission line is of particular interest in assessing BLR outflow strength which has also been linked to metallicity \citep[e.g.][]{Wang_2012, shin_2017_outflow}. This high-ionisation line can exhibit significant blueshifts \citep[e.g.][]{Gaskell1982, Wilkes1984, Marziani_1996, vandenberk2001, baskin_laor_2005, 2007ApJ...666..757S} and asymmetric velocity profiles \citep[e.g.][]{Sulentic_2000, baskin_laor_2005}, structure that is often interpreted as arising from a disk wind or outflow \citep[e.g.][]{2007ApJ...666..757S, Vietri_2018}. The blueshift of \civ\ is therefore an indication of the balance of emission between the outflowing ionised gas and the emission at a systematic redshift, which we call the ``wind'' and ``core'' component respectively \citep[adopting the terminology of][]{metal_density}. At high redshifts (z > 5.8), the mean and median \civ-\mgii\ velocity shifts are greater than for luminosity-matched quasars at lower redshifts, although this may potentially be biased by increased torus opacity and orientation-driven selection effects \citep{Meyer_2019, JT_2020, yang2021probing}. In this study, we define our estimate of the \civ\ blueshift as 
\begin{equation} \label{eq:blueshift}
    \frac{\rm{\civ\ blueshift}}{\rm{km \,s^{-1}}} \equiv c \times (1549.48\mbox{\normalfont\AA} - \lambda_{\rm{med}})/1549.48\mbox{\normalfont\AA}\,,
\end{equation}
where $c$ is the speed of light and $\lambda_{\rm{med}}$ is the median wavelength bisecting the total continuum-subtracted \civ\ emission line flux. The wavelength 1549.48\AA\ is the average of the \civ\ $\lambda\lambda1548.19,1550.77$ doublet. This definition is the same as in \citet{metal_density}, but their redshift is defined using a variety of low-ionization emission lines, some of which are known to exhibit velocity shifts relative to \mgii. In this study, we define our redshift using only the \mgii\ line. Due to the 1 \AA\ wavelength resolution, we prescribe a minimum precision of $\sim200$ km $\rm{s}^{-1}$ for the \civ\ blueshift, evaluated as an error of $\pm 1$\AA\ at the average wavelength of the \civ\ doublet. The overall uncertainty of the \civ\ blueshift is combined with the uncertainty from the measured redshift. 

\begin{figure}
	\includegraphics[width=\columnwidth]{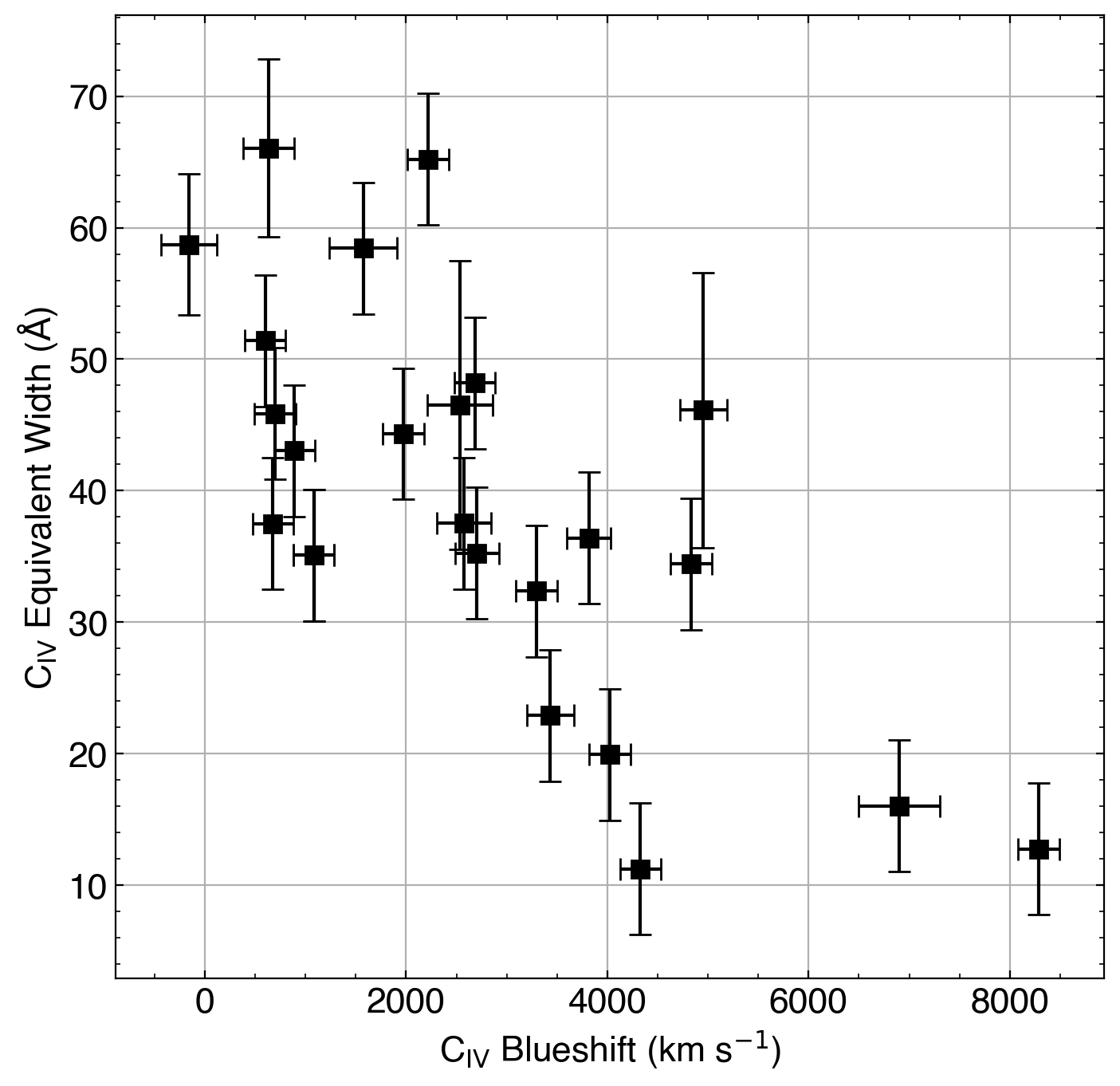}
    \caption{\civ\ equivalent width as a function of \civ\ blueshift. There is a moderate anti-correlation between these two quantities, implying that weaker \civ\ lines are more strongly blueshifted and stronger lines are less blueshifted. The outlier with high EW and blueshift is J1216+4519. Its SNR is low ($\sim 7$ per pixel) which is reflected by the large error in EW.}
    \label{fig:CIV_blueshift_ew}
\end{figure}

The \civ\ blueshift is also known to be anti-correlated with the line's equivalent width \citep[EW;][]{Leighly2004, Richards2011, Vietri_2018, Rankine2020, JT_2020, metal_density}, a relationship which is reproduced for our high-redshift quasar sample in Figure \ref{fig:CIV_blueshift_ew}. This correlation may be driven by orientation, properties of BLR winds, or the Baldwin effect linking properties of high-ionisation lines like \civ\ with the quasar luminosity \citep{Baldwin_1977_effect}. The results show highly blueshifted \civ\ lines are weak, while stronger lines are less blueshifted and more symmetric.

We note that a velocity shift relative to \cii\ of 5510$^{+240}_{-110}$ km s$^{-1}$ was measured for J1342+0928 \citep{Banados_2018, Onoue_2020} and \citet{JT_2020} also measured velocity shifts relative to \mgii\ for several XQR-30 quasars in our sample. However, due to differences in the definition of \civ\ blueshift\footnote{It is also possible to define the \civ\ blueshift by the maximum of the \civ\ line profile or the blueshift and asymmetry index (BAI) defined as the flux blueward of 1549.48\AA. Although we don't use these definition in this study, we have checked that these alternatives have little effect on the relative \civ\ blueshifts between quasars. The relationship found in Figure \ref{fig:CIV_blueshift_ew} and the correlations found in this study hold are unaffected.}, choice of \feii\ template, and sometimes the referenced line to estimate the redshift, we have re-measured the blueshifts for most quasars in our sample. However, the \civ\ blueshift could not be reliably determined for one individual quasar: J2338+2143, which, despite the minimum SNR requirement, has an overall SNR too poor to obtain a reliable fit. 

\subsection{Composite Spectra} \label{sec:composites}
The primary objective of this work is to measure flux ratios of metallicity-sensitive lines binning by quasar properties, such as bolometric luminosity, black hole mass, and \civ\ blueshift, to determine whether these parameters are correlated with metallicity in the BLR. Although most of the spectra have sufficient SNR to proceed with the emission-line fitting independently, the weak and blended emission lines of some SNR/pixel $\leq 8$ spectra in this sample could not be fit convincingly. By stacking the spectra, we are able to obtain higher SNR. Another reason to stack the spectra is to average out peculiarities of individual quasars in each bin in order to construct better comparisons to the photoionisation models referenced in Section \ref{sec:line_ratios_metallicity}. As we are interested in the average spectral properties within a binned parameter space rather than the specific individual properties, we use equivalent weighting of spectra within each composite regardless of the SNR of input spectra so that the output is not biased in favor of any contributing quasar observed with high SNR. We construct 6 bins from each of the 3 quasar properties (black hole mass, bolometric luminosity, and \civ\ blueshift), with a similar number of contributing quasar spectra in each bin. We also avoid extending the width of each bin too wide. Therefore, the average number of quasar spectra in each bin is 4, and all composites are created from 3-6 input spectra. 

The dynamic range of quasar bolometric luminosity in this sample is $\log\left(\rm{L}_{\rm{bol}}/\rm{erg \, s}^{-1}\right) = 46.7-47.7$. We split the sample into 6 luminosity bins with the following edges: 46.72, 46.79, 46.95, 47.17, 47.30, 47.40, 47.70, and a composite is created from each bin. The first three bins include all 10 GNIRS spectra and the final three bins are composed of exclusively X-shooter spectra. The total BH mass range reflected in our high-redshift quasar spectra is $\log\left({\rm{M}_{\rm{BH}}/\rm{M}_{\odot}}\right) = 8.4-9.8$. Again, we form 6 mass bins with the following edges: 8.40, 8.75, 8.87, 8.98, 9.20, 9.40, 9.80, and create a composite spectrum from each bin. We also arrange and stack all individual quasars in the sample into 6 \civ\ blueshift bins, with the following bin edges: $-$200, 680, 1500, 2500, 3000, 4000, 5000 km $\rm{s}^{-1}$. Figure \ref{fig:quasar_properties} shows the bin edges of each composite delineated by black dashed lines.

Prior to stacking, we apply an upper error threshold equal to 2 times the minimum error within box widths of 50 pixels to restrict data to where the error is reasonable. The error threshold clips wavelength bins with unusually high error and high flux that were not masked by the general post-processing procedure. The resulting spectra contains the most stable and robustly measured elements. Without the upper error threshold, the propagation of error from a small number of component spectra can create unstable composites, leading to greater uncertainty in the final flux measurements. Every spectrum is then normalised across the rest-frame 1430\AA$\sim$1450\AA\ wavelength range and the arithmetic mean of each stack is taken to generate the composite. We also take the median or geometric mean \citep[e.g.][]{vandenberk2001} of the stack and find that it does not significantly influence the result. Furthermore, we generate composites after subtracting a power-law continuum, fitted as described in Section \ref{sec:emission-fitting}, and find that the resulting measured broad emission-line fluxes are not significantly discrepant either. In all cases, the resulting line ratio measurements regardless of taking the composite arithmetic mean, geometric mean, median, or after subtracting the continuum agree to within 2.0 $\sigma$, with $\sim 75\%$ of measurements within 1.0 $\sigma$.

\begin{figure*}
	\includegraphics[width=\textwidth]{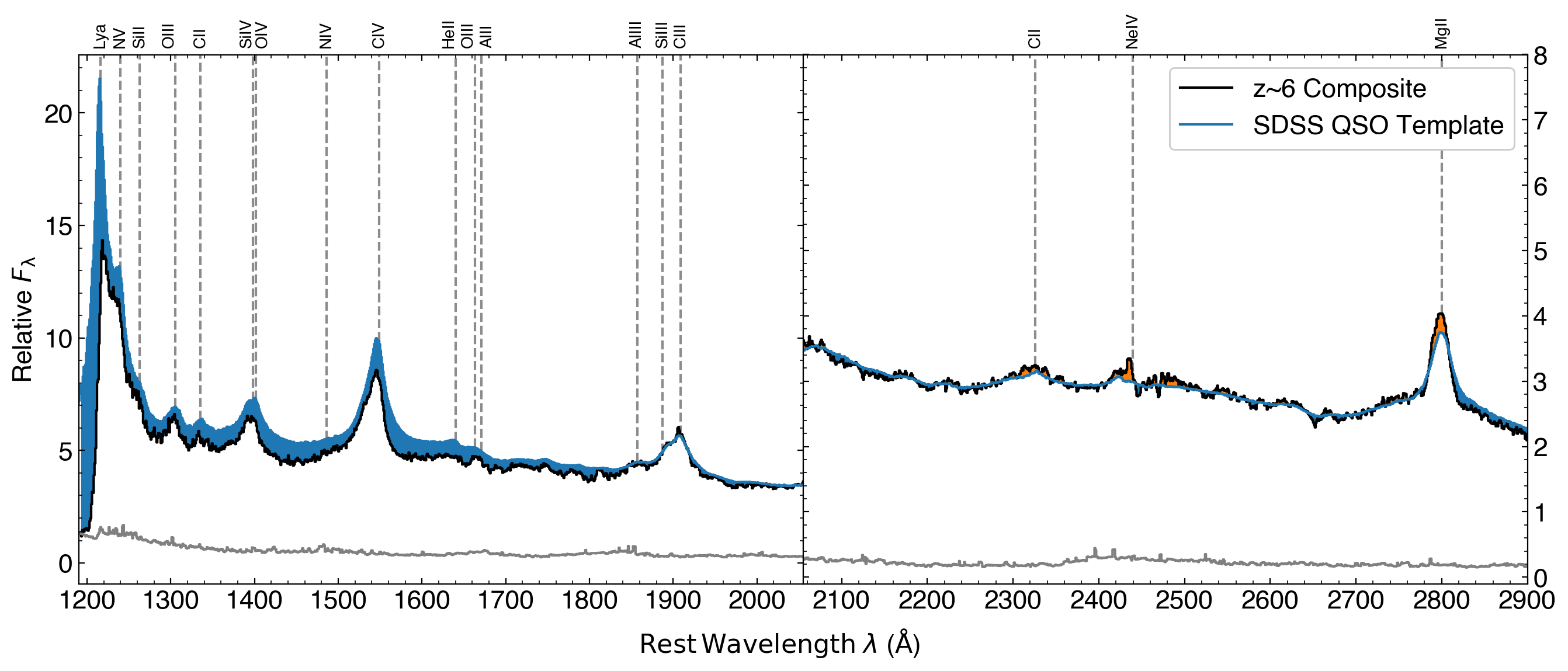}
    \caption{Composite created from all 25 high-redshift quasars in our sample compared against the SDSS quasar template comprised of 2200 spectra described in \citet{vandenberk2001}. \hl{The composite is split into two sections with independent y-axis scaling to better visualise the various spectral emission features, which are denoted with the grey dashed lines.} The composite and SDSS template are normalised together at 2200-2250\AA, \hl{and the grey spectrum is the error spectrum of the composite.} The blue and small patches of orange shaded regions indicate when the template or the \hl{composite} is in excess respectively. Our high-redshift composite has a noticeably flatter continuum slope and significant \lya\ absorption compared to the lower redshift SDSS sample, but the profiles and relative integrated fluxes of emission features other than \lya\ are comparable.}
    \label{fig:all_composite}
\end{figure*}

The uncertainty in each resolution element is composed of the error in every contributing spectrum added in quadrature, but we also estimate the systematic error in each composite by generating all of the possible composites that can be obtained if any one contributing quasar spectrum is excluded. The standard deviation in each 1\AA\ pixel from all such simulated composites is treated as the systematic error, added in quadrature to the uncertainty propagated from each contributing spectrum. This systematic error is an additional source of error which raises the uncertainty floor of the combined spectrum and reduces the relative uncertainty between each resolution element, affecting the weighting of each pixel in a least-squares fitting routine. After combining all sources of uncertainty, the resulting SNR per pixel of the composites measured between rest-frame 1400-1600\AA\ is 20-100. We show a composite constructed from all 25 quasars in our sample in Figure \ref{fig:all_composite}.

Three individual quasar spectra are treated differently for our composites. PSOJ007+04, PSOJ025-11, and J1212+0505 are affected by proximate damped \lya\ absorption (pDLA) systems \citep{Farina_2019, Banados_2019}, which causes a significant fraction of its \lya\ emission to be absorbed at the systemic redshift. We fit the emission lines of pDLA affected quasars individually, and mask their emission blueward of the \nv\ centroid from contributing to composites. 

\section{Line Fitting and Metallicity Measurement} \label{sec:metallicity}
\subsection{Emission-line Fitting} \label{sec:emission-fitting}
In this work, we study a large number of rest-frame UV emission lines: \lya, \nv, \siii, \siiv, \oiv, \niv, \civ, \heii, \oiii, \alii, \aliii, \siiii, \ciii, and \mgii. The measurement of emission-line fluxes can be tricky due to adjoining and heavily blended lines, such as \nvl \AA\ with \lyal \AA\ or \aliiil \AA, \siiiil \AA, and \ciiil \AA. Furthermore, the strong \feii\ emission biases the underlying continuum level measurement. Despite these challenges, there are two widely employed methods for fitting quasar emission-lines \citep{Nagao_2006}. One method measures the emission-line flux by integrating above a well-defined independent local continuum model \citep[e.g.][]{vandenberk2001} and the other method endeavors to fit emission lines using one or more appropriate functions, such as Gaussians or Lorentzians \citep[e.g.][]{Zheng_1997}. Both of these methods have shortcomings in measuring accurate emission-line fluxes. Defining an appropriate local continuum level below an emission-line is challenging and is sensitive to where the baseline is anchored. The additional uncertainty propagates into the resulting metallicity estimates. Regarding the function fitting approach, a single Gaussian or Lorentzian profile is insufficient for broad emission-lines with asymmetric velocity profiles \citep[e.g.][]{corbin_1997, vandenberk2001, baskin_laor_2005}. The approach utilising multiple Gaussian functions can obtain smooth realisations of the line profile, but the decomposition is not unique and a large number of free parameters is required. Modified functions such as a skewed Gaussian (defined in Appendix \ref{appendix:line-fitting-compare}) or asymmetric Lorentzian depend on fewer parameters and are arguably more physically relevant \citep[e.g.][]{mallery_2012}. With multiple reasonable approaches, there is a concern that the resulting line flux can be method-dependent. In this work, we use various appropriate functions to fit emission lines and we compare the several different methods against similar fits from existing literature in Appendix \ref{appendix:line-fitting-compare}.

We follow the general procedure from \citet{Xu2018} and define the following two line-free windows in rest-frame to fit the continuum: 1445\AA$-$1455\AA, 1973\AA$-$1983\AA. In specific circumstances, we identify two additional windows (1320\AA$-$1325\AA, and 1370\AA$-$1380\AA) to further constrain the continuum shape or we extend the blue-end of the first line-free window to 1432\AA\ in the case of a blueshifted \civ\ line. The continuum is fit with a power-law function normalised to rest-frame 3000 \AA,
\begin{equation}
    F_{\rm{pl}}(\lambda) = F_{\rm{pl, 0}} \left(\frac{\lambda}{3000 \mbox{\normalfont\AA}}\right)^{\gamma}\,,
    \label{eq:pl-cont}
\end{equation}
where $F_{\rm{pl, 0}}$ and $\gamma$ represent the normalization and power-law slope respectively. We also consider the contribution of the \feii\ pseudo-continuum spectrum using the empirical template from \citet{UV_iron_template} to cover the wavelength range from 1200 \AA\ to 3500 \AA. We convolve the template with a Gaussian broadening kernel to better fit the variety of features from the \feii\ pseudo-continuum seen across spectra in our sample,
\begin{equation}
    F_{\rm{Fe}}(\lambda) = \zeta_{\rm{0}} \,  F_{\rm{template}}|_{\lambda(1+\delta)} \circledast G(\lambda, \sigma)\,,
\end{equation}
where the free parameters of the \feii\ flux contribution include a flux scaling factor $\zeta_{\rm{0}}$, the FWHM of the broadening kernel $\sigma$, and a small wavelength shift $\delta$. The contribution from the iron continuum is more relevant at wavelengths close to the \mgiil \AA\ line and is important in obtaining the virial mass estimate. Combined, the power-law and the \feii\ template are fit to the data in the line-free windows and form the underlying continuum baseline.

Emission lines are fit with the following double power-law method adopted from \citet{Nagao_2006}, \citet{Matsuoka_2011}, and \cite{Xu2018},
\begin{equation}
  F_{\rm{em}}(\lambda) =
    \begin{cases}
      F_{\rm{0}} \times \left(\frac{\lambda}{\lambda_{\rm{0}}}\right)^{-\alpha} & \lambda > \lambda_{\rm{0}} \\
      F_{\rm{0}} \times \left(\frac{\lambda}{\lambda_{\rm{0}}}\right)^{+\beta} & \lambda < \lambda_{\rm{0}}
    \end{cases}
    \label{eq:line_fit}
\end{equation}
where the two power-law indices ($\alpha$ and $\beta$) are used to fit the red and blue sides of the emission-line profile. The peak intensity, $F_{\rm{0}}$, controls the height of the emission line and the peak wavelength, $\lambda_{\rm{0}}$, defines the location of the peak. 

Emission lines with different degrees of ionisation often show systematically varied velocity profiles \citep[e.g.][]{Gaskell1982, baskin_laor_2005}. Thus, we categorise emission lines into two distinct systems: high-ionisation lines (HILs) and low-ionisation lines (LILs). The HILs include \nv, \oiv, \niv, \civ, and \heii\ while the LILs include \siii, \siiv, \oiii, \alii, \aliii, \siiii, and \ciii\ \citep{Collin_Souffrin_1988}. The boundary separating the two main groups is an ionisation potential of 40 eV. We assume that the emission-line profiles of lines in the same category are coupled to the same line-emitting gas clouds of the BLR, sharing a common value for the $\alpha$ and $\beta$ power indices. Because we did not correct for the suppression of \lya\ from the intergalactic medium, the redder $\alpha$ index of the \lya\ line is coupled with the HILs, while the bluer $\beta$ index is left unconstrained \citep{Nagao_2006, Xu2018}. \footnote{Although the \lya\ line doesn't directly factor into the line ratios we measure or the metallicities we determine, its flux and line profile does affect the measured flux of the \nv\ line.} 

Our adopted piece-wise power-law function fit to emission lines has been compared to the double-Gaussian and modified Lorentzian methods, achieving better fits with fewer or equal number of free parameters \citep{Nagao_2006}. In cases when the piece-wise function does not produce a reasonable fit to the shape of the spectral feature, such as significantly blueshifted lines, we fit a skewed Gaussian function, where both the skew and FWHM of the Gaussian are coupled between LILs and HILs. Unlike the default piece-wise strategy, \lya\ is completely decoupled from the HIL group when fitting skewed Gaussians. We choose to fit a single skewed Gaussian because it has the same number of free parameters as the piece-wise power-law fit. Changing the fitting strategy is also motivated by the reduction in the minimum chi-square value even when the fits produce similar emission-line flux ratios. A more complete description of the comparison between these two fitting methods and the definition of the skewed Gaussian are provided in Appendix \ref{appendix:line-fitting-compare}. 

When LILs and HILs are not coupled, some local continuum methods can produce emission-line profiles with very different widths and skewness \citep{vandenberk2001}. We assume, as several similar other studies do \citep[e.g.][]{Nagao_2006, Matsuoka_2011, Xu2018}, that emission lines with similar ionising potentials originate from similar line-emitting regions in the BLR. The coupling of power indices in Equation \ref{eq:line_fit} provides a crucial constraint in ensuring that the kinematics of line-emitting clouds are preserved within the LIL and HIL groups. Furthermore, without the coupling of HILs, the decomposition of the \lya\ and \nv\ emission profile is not unique. The coupling of the \nv\ profile and the red wing of \lya\ to HILs provides a way to obtain a unique solution that disentangles their line profiles and fluxes.

\begin{figure*}
	\includegraphics[width=\textwidth]{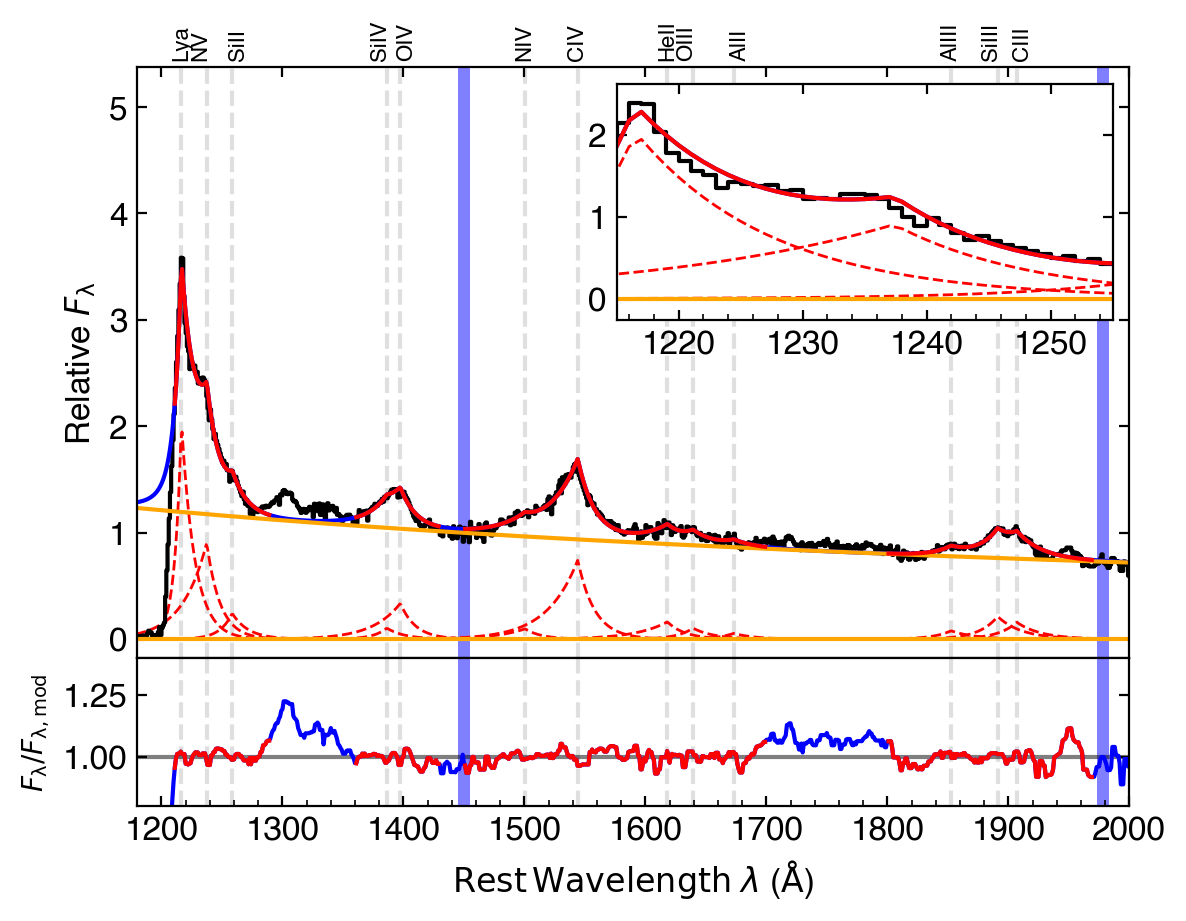}
    \caption{Example fit to ATLASJ029-36 observed using VLT/X-shooter with a mean SNR per 1\AA\ pixel of 27.35 between 1400-1600\AA. The top plot shows the spectrum after the post-processing techniques and the bottom plot shows the residuals. The inset plot provides a closer look at the line profiles of \lya\ and \nv. \hl{The vertical blue bars indicate the continuum fitting windows, which are fit by the power-law continuum denoted by the orange line.} The red lines indicate the emission line fits as well as the extent of the individual line-fitting windows. All fitted emission lines are labeled and their individual line profiles are shown.}
    \label{fig:example-fit-ATLASJ029-36}
\end{figure*}

The emission lines \lyal, \nvl, \siiil, \siivl, \oivl, \nivl, \civl, \heiil, \oiiil, \aliil, \aliiil, \siiiil, and \ciiil\ are all fit simultaneously. The line-fitting regions generally are 1214-1290, 1360-1430, 1450-1700, and 1800-1970\AA\ with some flexibility depending on the width and kinematics of the spectral features. Each line is allowed an independent $\pm 25$\AA\ shift in central wavelength, $\lambda_{\rm{0}}$, with respect to the rest-frame vacuum wavelength. Whether we used the piece-wise power-law or skewed Gaussian approach, a single emission-line is fit with only four parameters. Figure \ref{fig:example-fit-ATLASJ029-36} shows an example fit to ATLASJ029-36 which was observed with VLT/X-shooter. In this example, all of the lines from \lya\ at 1216 \AA\ to \ciii\ at 1909 \AA\ have been fit simultaneously, with coupled LILs and HILs. 

We fit the \civ\ emission line to estimate the blueshift according to Equation \ref{eq:blueshift}. When \civ\ blueshifts < 4000 km $\rm{s}^{-1}$, we adopt the piecewise power-law fit and at higher blueshifts, we use the skewed Gaussian function. We find both methods produce consistent results at lower \civ\ blueshifts as shown in the comparison described in Appendix \ref{appendix:line-fitting-compare}.

\begin{figure}
	\includegraphics[width=\columnwidth]{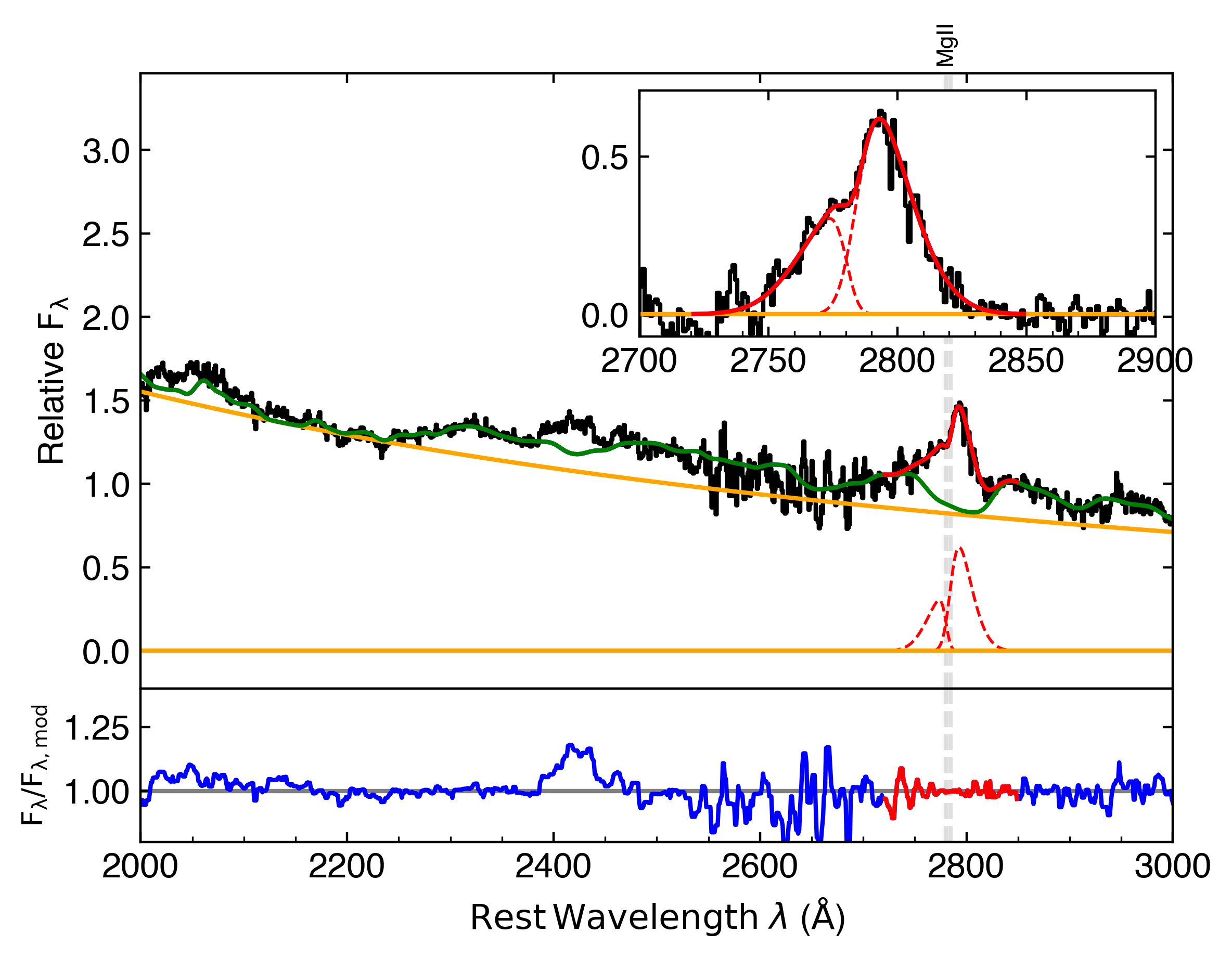}
    \caption{Example two-Gaussian fit to the \mgii\ emission line of PSOJ217-16, observed using VLT/X-shooter. The top plot shows the spectrum after the post-processing techniques and the bottom plot shows the residuals. The inset plot provides a closer look at the \mgii\ line profile after continuum subtraction. The orange line is the power-law and the green line is full baseline continuum with the \feii\ template. The red line marks the fit to the emission line and the extent of the fitting windows. These fits are presented in greater detail in Mazzucchelli et al. in prep.}
    \label{fig:mgii_fit_example}
\end{figure}

We also fit the \mgii\ line independently with one skewed Gaussian or two symmetric Gaussians, using the following line-free windows: 1770-1810, 2060-2340, 2600-2740, 2840-3100\AA\ to measure the continuum. \hl{The continuum is measured independently for the} \mgii\ line fit because the contribution from the \feii\ emission \hl{is} much more significant at these longer wavelengths. \hl{Although the fit parameters are not always consistent between the two wavelength ranges, we find the power-law to be an adequate local approximation of the accretion disk emission}. We use the \mgii\ FWHM, Equation \ref{eq:mgii-vo9}, and calibration from \citet{Vestergaard_2009} to determine the black hole mass and find good agreement with Mazzucchelli et al. in prep. We calibrate the spectrum against the observed quasar AB magnitude in the J bandpass, where flux from the spectrum integrated over the filter transmission profile is scaled appropriately to the observed value. The peak flux wavelength of the total \mgii\ line profile is used to determine the redshift and its error. Figure \ref{fig:mgii_fit_example} shows an example of a multiple Gaussian fit to the \mgii\ emission line along with the combined power-law and \feii\ continuum.

For the line flux error estimation, we adopt a Monte-Carlo approach used in similar studies of high-redshift quasar spectra \citep[e.g.][]{Shen_2019, 2020z7.5, 2021_luminous7.6}. We create 50 mock spectra for each individual spectrum and stacked composite, where the flux at each pixel is resampled from a symmetric distribution with a standard deviation equivalent to the pixel spectral error. We assume the spectrum noise to follow a Gaussian distribution in this case. The same fitting procedure is applied to every mock spectrum generated in this way and we filter out outlier fits by sigma-clipping the line flux measurements using a 3-$\sigma$ threshold. The final line flux of each metallicity-sensitive line is the median of all remaining fits and the final uncertainty is its standard deviation.

\hl{Using other empirical templates of the} \feii\ \hl{emission} \citep[e.g.][]{Tsuzuki_2006, Bruhweiler_2008, Mejia_2016} \hl{can result in a maximum discrepancy of 20\% in the} \mgii\ \hl{FWHM, and 0.2\% in the estimated redshift. Although this can modify the significance of the black hole mass correlation presented in} Section \ref{sec:results}, \hl{it has no effect on the metallicity estimates or the conclusions of this study, which depends primarily on the} \civ\ \hl{profile.}

Figure \ref{fig:quasar_properties} presents the distribution of redshift, bolometric luminosity, black hole mass, and \civ\ blueshift in our sample. The completed sample covers redshifts $z = 5.8-7.5$, spanning a range in bolometric luminosity $\log\left(\rm{L}_{\rm{bol}}/\rm{erg \, s}^{-1}\right) = 46.7-47.7$ and a range in black hole mass $\log\left({\rm{M}_{\rm{BH}}/\rm{M}_{\odot}}\right) = 8.4-9.8$. Quasars with luminosities $\log\left(\rm{L}_{\rm{bol}}/\rm{erg \, s}^{-1}\right) \leq 47.0$ and black hole masses $\log\left({\rm{M}_{\rm{BH}}/\rm{M}_{\odot}}\right) \leq 8.9$ are all observed with GNIRS. We measure \civ\ blueshifts spanning over 5000 km $\rm{s}^{-1}$ in our sample, from values consistent with no detectable blueshift to the most extreme outflow-dominated spectrum in PSOJ065-25. None of the targets exhibit a significantly redshifted \civ\ emission line. We list all of the measured quasar properties in Table \ref{tab:quasar_sample}.

\subsection{Line Ratios and Metallicity} \label{sec:line_ratios_metallicity}
In order to interpret the results of the fitting, it is useful to compare the measured high-ionisation line ratios against predictions from photoionisation models. It's well-known that single-zone photoionisation models are unable to fully reproduce the observed emission from the BLR, because the gas clouds span a wide range of densities and degrees of ionisation \citep[e.g.][]{Davidson_1977, Collin_Souffrin_1988}. Multi-zone models incorporate emission from gas with a wide range of physical properties and are shown to be consistent with observation \citep[e.g.][]{Rees_1989, Hamann_1998}.

The flux ratio-metallicity relation is sensitive to the density of line-emitting clouds, spectral energy distribution (SED) of the ionizing continuum, and microturbulence, but under the locally optimally-emitting cloud model \citep[LOC;][]{Baldwin_1995_LOC}, the net emission spectrum can be reproduced by integrating across a wide range of physical conditions. Therefore, the characteristics of the observable spectrum originate from an amalgamation of emitters, where each emission line is formed in a region that is optimally suited to emit the targeted line. This model consistently reproduces properties of both low and high-ionisation emission lines observed in quasar spectra \citep[e.g.][]{Korista_2000_LOC, Hamann_2002, Nagao_2006}.

We primarily utilise two broad emission-line flux ratios (\nvciv, \siivoivciv) because the relevant lines are easier to detect and more commonly studied in the existing literature. However, we also present results for additional line ratios (\oiiialiiciv, \aliiiciv, \siiiiciv, \ciiiciv). The emission from these other lines are substantially more difficult to measure and can only be detected in a robust manner in high SNR spectra, such as our XQR-30 sample of high-redshift quasars. We convert all of the line ratios into metallicity estimates using relations derived from \texttt{Cloudy} photoionisation simulations described in \citet{Hamann_2002} and \citet{Nagao_2006}. Both models utilise the LOC model \citep{Ferland_1998_cloudy}. \citet{Nagao_2006} predicts line flux ratios for all listed line ratios with two models for the ionizing continuum: one with a strong UV thermal bump matching results from \citet{large_uv_bump} and one with a weak UV thermal bump similar to Hubble Space Telescope quasar templates \citep{Zheng_1997, telfer2002}. These two SEDs are thought to be extreme and opposite cases for the actual ionising continuum \citep{Nagao_2006}, which gives us the full range of possible inferred metallicities. \citet{Hamann_2002} predicts line flux ratios of only \nvciv\ for three mock incident spectra: that of \citet{mf87}; a single hard power-law with index $\alpha = -1.0$ ($f_{\rm{\nu}} \propto \nu^{\rm{\alpha}}$); and a segmented power-law with indices $\alpha = [-0.9, -1.6, -0.6]$ for 0.25\AA\ to 12\AA, 12\AA\ to 912\AA, and 912\AA\ to 1 $\mu$m respectively. The segmented power-law continuum approximates data gathered from observations \citep[e.g.][]{Laor_1997}. For \citet{Hamann_2002}, the \citet{mf87} incident spectrum predicts the highest metallicities for the same line ratio and the $\alpha = -1.0$ spectrum produces the lowest. The spread in metallicities predicted for the same line ratio is incorporated into our uncertainties. For \nvciv\ the results from both publications are largely consistent with minor differences arising from the SED of the ionizing continuum, integration ranges of gas density ($n_{\rm{H}}$) or ionizing flux ($\Phi_{\rm{H}}$), cloud column density, and the version of \texttt{Cloudy} used. 

Line ratios which imply metallicities over 10 $Z_{\odot}$ extend beyond the parameter space probed by \citet{Hamann_2002} or \citet{Nagao_2006}. This occurs when the measured line ratio exceeds 0.84 for \nvciv\ or 0.45 for \siivoivciv. For the other line ratios, this occurs at (0.39, 0.16, 0.36, 0.57) for (\oiiialiiciv, \aliiiciv, \siiiiciv, \ciiiciv). In order to investigate inferred metallicities for higher line ratios, we assume that the observed line flux ratio-metallicity relationship maintains a linear trend in log-space and linearly extrapolate beyond the parameter space probed by the simulations. Super-solar metallicites over 10 $Z_{\odot}$ have not been calibrated against \texttt{Cloudy} simulations.

We consider all of the photionisation calculations with different ionizing SEDs in our metallicity estimate. The effect of the ionizing continuum SED is responsible for up to a factor of two difference in the resulting metallicity predictions from \nvciv. For \nvciv, the uncertainty from the metallicity calibration based on the various photoionisation models is dominant over the observational uncertainty. \siivoivciv\ is a more robust metallicity indicator than \nvciv\ because it is not as sensitive to differences in the ionizing continuum or assumed weighting functions \citep[e.g.][]{Nagao_2006, Matsuoka_2011, Maiolino_2019_metallica}, and it is not affected by bias propagating from a poor fit to the highly absorbed \lya\ emission line. However, in this study, we offer no discussion on the discrepancy between the metallicity indicators. Instead, we present the inferred metallicities separately and use the spread of results from different assumed ionizing SEDs as the uncertainty of each individual measurement. A comparison between inferred metallicities from \nvciv\ and \siivoivciv\ is provided in the Appendix.

\section{Results} \label{sec:results}
We present all quasars and their measured properties in Table \ref{tab:quasar_sample}. In addition to fitting the composites described in Section \ref{sec:composites}, the emission-lines of nearly all of the quasar spectra can be fit individually. We fit the \nvciv\ line ratio for 16 of the 25 individual quasars, wherever the \nv\ emission can be separated from the \lya\ emission. We show the \nvciv\ and \siivoivciv\ line ratio results for individual fits in Table \ref{tab:quasar_sample} and provide 6 example fits to individual quasar spectra in the Appendix, covering the lowest and highest quasar bolometric luminosity, black hole mass, and \civ\ blueshift. Also available are figure sets which show sample fits to bolometric luminosity composites, black hole mass composites, and \civ\ blueshift composites. We provide all line fluxes measured from our composites normalised against \civ\ in Tables \ref{tab:composite_results}, \ref{tab:composite_mass}, and \ref{tab:composite_blueshift}.

\begingroup
\begin{table*}
\caption {\label{tab:quasar_sample} Properties of the quasars and spectra included in this study and their measured emission line flux ratios. The redshift is determined from the \mgii\ line with an uncertainty floor of 0.001. The \civ\ blueshift is measured using the median wavelength of the \civ\ fit and the mean SNR is measured in the rest-frame wavelength range 1400-1600\AA. We prescribe a minimum error of the \civ\ blueshift equivalent to $\sim$200 km s$^{-1}$, based on the 1\AA\ resolution of the resampled grid. All quasars listed above the horizontal divider are observed with GNIRS and all quasars listed below the divider are observed with X-shooter. The last column indicates our source for the black hole mass and bolometric luminosity. The bolometric correction used to measure the luminosity, single-epoch virial mass calibration, and \feii\ template used to measure \mgii\ are all consistent throughout the sample.}  
\begin{tabular}{lcccccccc}
\hline \hline
 Name & R.A. & Decl. & \mgii\ Redshift & \civ\ Blueshift & SNR & \nvciv & \siivoivciv & M$_{\rm{BH}}$/L$_{\rm{bol}}$ Ref\\
 & (J2000) & (J2000) & & (km s$^{-1}$) & & & & \\
 \hline
J0024+3913 & 00:24:29.77 & 39:13:19.00 & 6.620 $\pm$ 0.004 & 635 $\pm$ 255 & 9.45 & 0.77 $\pm$ 0.15 & 0.27 $\pm$ 0.06 & 1 \\ 
J0829+4117 & 08:29:31.97 & 41:17:40.40 & 6.773 $\pm$ 0.007 & 1574 $\pm$ 336 & 16.33 & 0.49 $\pm$ 0.09 & 0.12 $\pm$ 0.05 & 1 \\ 
J0837+4929 & 08:37:37.84 & 49:29:00.40 & 6.702 $\pm$ 0.001 & 600 $\pm$ 204 & 30.91 & 1.49 $\pm$ 0.08 & 0.66 $\pm$ 0.10 & 1 \\ 
J0910+1656 & 09:10:13.63 & 16:56:29.80 & 6.719 $\pm$ 0.005 & $-$159 $\pm$ 279 & 9.42 & 0.50 $\pm$ 0.08 & 0.19 $\pm$ 0.09 & 1 \\ 
J0921+0007 & 09:21:20.56 & 00:07:22.90 & 6.565 $\pm$ 0.001 & 678 $\pm$ 204 & 9.66 & 0.49 $\pm$ 0.13 & 0.16 $\pm$ 0.05 & 1 \\ 
J1216+4519 & 12:16:27.58 & 45:19:10.70 & 6.648 $\pm$ 0.003 & 4955 $\pm$ 232 & 7.63 & --- & 0.50 $\pm$ 0.33 & 1 \\ 
J1342+0928 & 13:42:08.10 & 09:28:38.60 & 7.510 $\pm$ 0.010 & 6900 $\pm$ 405 & 24.79 & --- & 0.78 $\pm$ 0.28 & 1 \\ 
J2102-1458 & 21:02:19.22 & $-$14:58:54.00 & 6.652 $\pm$ 0.003 & 3433 $\pm$ 232 & 11.47 & 1.82 $\pm$ 0.44 & 0.57 $\pm$ 0.23 & 1 \\ 
P333+26 & 22:15:56.63 & 26:06:29.40 & 6.027 $\pm$ 0.006 & 2534 $\pm$ 325 & 5.73 & 0.80 $\pm$ 0.47 & 0.27 $\pm$ 0.19 & 2 \\ 
J2338+2143 & 23:38:07.03 & 21:43:58.20 & 6.565 $\pm$ 0.009 & --- & 7.07 & --- & --- & 1 \\ \hline
PSOJ007+04 & 00:28:06.56 & 04:57:25.64 & 6.001 $\pm$ 0.002 & 3816 $\pm$ 218 & 24.58 & --- & 0.45 $\pm$ 0.20 & 3 \\ 
PSOJ025-11 & 01:40:57.03 & $-$11:40:59.48 & 5.816 $\pm$ 0.004 & 2575 $\pm$ 266 & 26.54 & --- & 0.70 $\pm$ 0.13 & 3 \\ 
PSOJ029-29 & 01:58:04.14 & $-$29:05:19.25 & 5.976 $\pm$ 0.001 & 3295 $\pm$ 205 & 27.39 & 1.14 $\pm$ 0.13 & 0.64 $\pm$ 0.15 & 3 \\ 
ATLASJ029-36 & 01:59:57.97 & $-$36:33:56.60 & 6.020 $\pm$ 0.002 & 2705 $\pm$ 217 & 27.35 & 1.07 $\pm$ 0.11 & 0.49 $\pm$ 0.13 & 3 \\ 
VDESJ0224-4711 & 02:24:26.54 & $-$47:11:29.40 & 6.528 $\pm$ 0.001 & 2217 $\pm$ 204 & 28.68 & 0.79 $\pm$ 0.04 & 0.34 $\pm$ 0.11 & 3 \\ 
PSOJ060+24 & 04:02:12.69 & 24:51:24.42 & 6.170 $\pm$ 0.001 & 1082 $\pm$ 204 & 30.73 & 0.76 $\pm$ 0.05 & 0.10 $\pm$ 0.05 & 3 \\ 
PSOJ065-26 & 04:21:38.05 & $-$26:57:15.60 & 6.188 $\pm$ 0.001 & 8288 $\pm$ 204 & 36.57 & --- & 0.78 $\pm$ 0.65 & 3 \\ 
PSOJ108+08 & 07:13:46.31 & 08:55:32.65 & 5.945 $\pm$ 0.001 & 4832 $\pm$ 205 & 37.27 & --- & 0.78 $\pm$ 0.27 & 3 \\ 
PSOJ158-14 & 10:34:46.50 & $-$14:25:15.58 & 6.068 $\pm$ 0.001 & 2683 $\pm$ 204 & 31.67 & 0.81 $\pm$ 0.07 & 0.28 $\pm$ 0.09 & 3 \\ 
J1212+0505 & 12:12:26.98 & 05:05:33.49 & 6.439 $\pm$ 0.001 & 4329 $\pm$ 204 & 31.27 & --- & 0.93 $\pm$ 0.37 & 3 \\ 
PSOJ217-16 & 14:28:21.39 & $-$16:02:43.30 & 6.150 $\pm$ 0.001 & 4023 $\pm$ 204 & 34.57 & --- & 0.30 $\pm$ 0.13 & 3 \\ 
PSOJ242-12 & 16:09:45.53 & $-$12:58:54.11 & 5.830 $\pm$ 0.001 & 891 $\pm$ 205 & 15.55 & 0.87 $\pm$ 0.11 & 0.34 $\pm$ 0.22 & 3 \\ 
PSOJ308-27 & 20:33:55.91 & $-$27:38:54.60 & 5.799 $\pm$ 0.001 & 1971 $\pm$ 205 & 33.18 & 0.98 $\pm$ 0.06 & 0.95 $\pm$ 0.10 & 3 \\ 
PSOJ323+12 & 21:32:33.19 & 12:17:55.26 & 6.586 $\pm$ 0.001 & 697 $\pm$ 204 & 31.34 & 0.73 $\pm$ 0.05 & 0.36 $\pm$ 0.08 & 3 \\ 
PSOJ359-06 & 23:56:32.45 & $-$06:22:59.26 & 6.172 $\pm$ 0.001 & 1082 $\pm$ 204 & 35.36 & 0.96 $\pm$ 0.13 & 0.11 $\pm$ 0.04 & 3 \\ 
\hline \hline
\multicolumn{6}{l}{\footnotesize
$^{1}$ \citet{yang2021probing}
$^{2}$ \citet{Shen_2019}
$^{3}$ Mazzucchelli et al. in prep.}
\end{tabular}
\end{table*}
\endgroup

\begin{figure*}
\begin{tabular}{cc}
  \includegraphics[width=90mm]{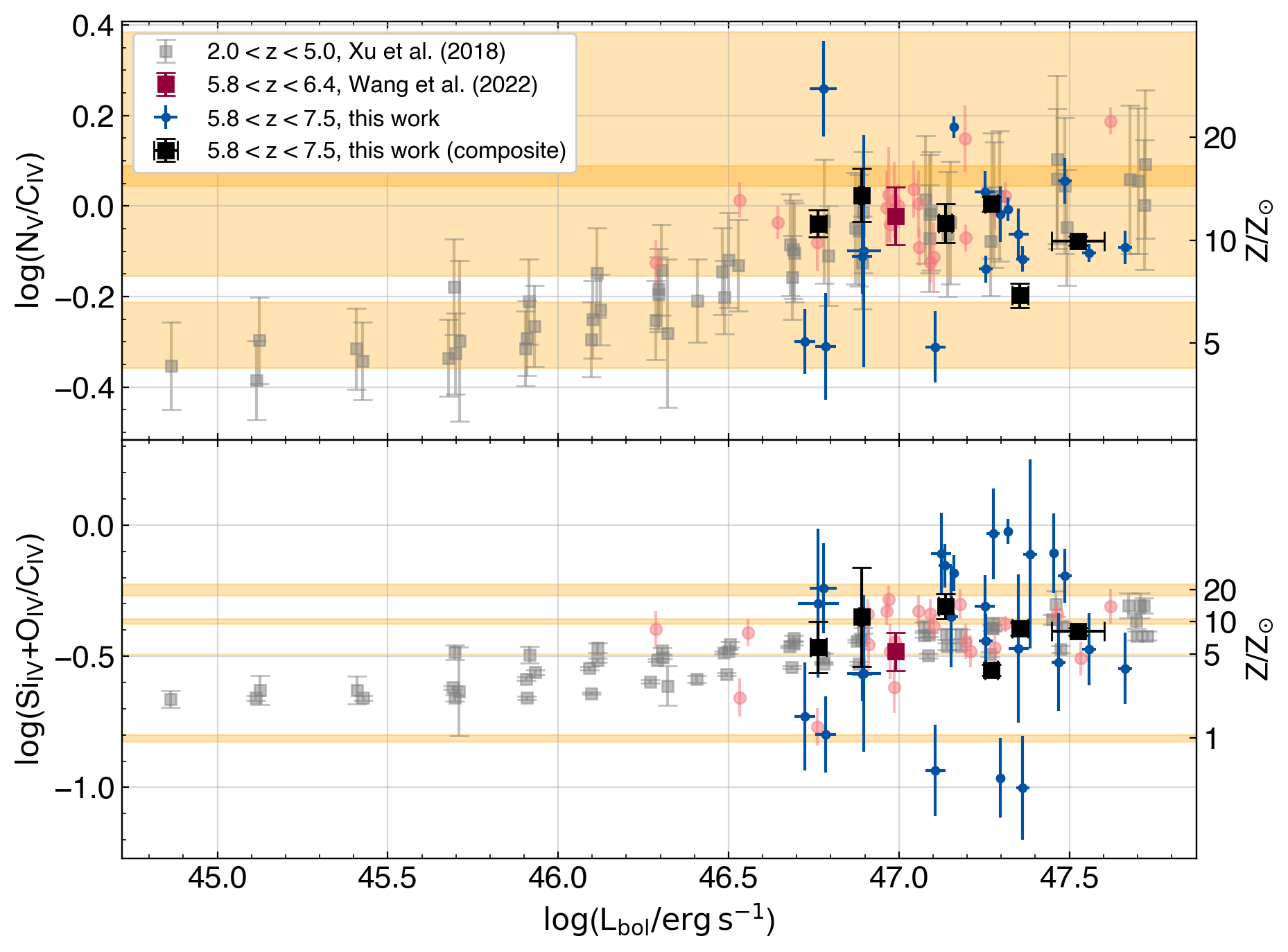} &   \includegraphics[width=90mm]{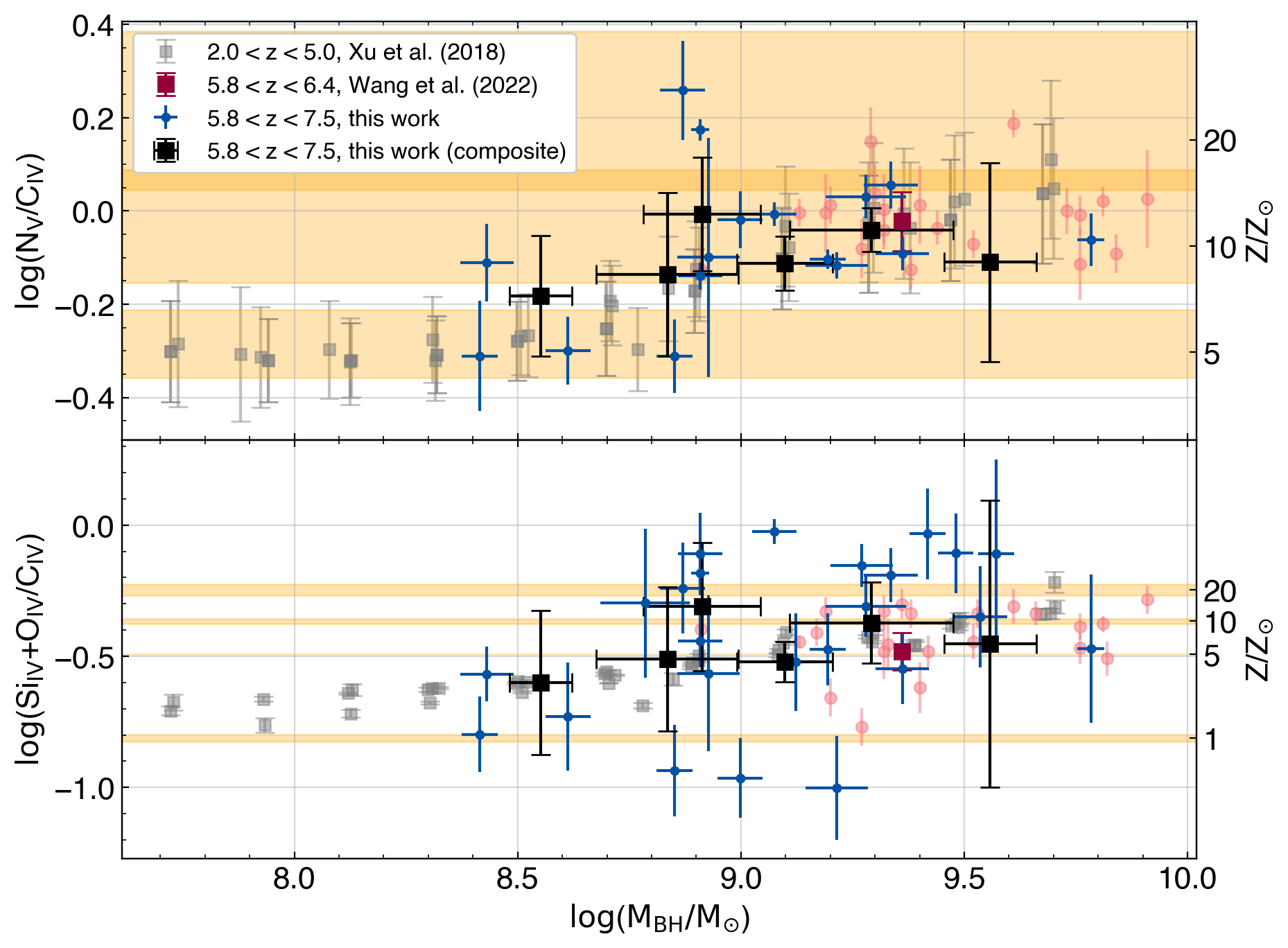} 
\end{tabular}
\caption{\nvciv\ and \siivoivciv\ flux ratios as a function of the quasar bolometric luminosity (left) and virially estimated black hole mass (right). The low-redshift sample (2.0 < $z$ < 5.0) indicated in grey is from \citet{Xu2018} while another higher-redshift comparison sample indicated in red is sourced from \citet{Wang_2021}. Our sample is presented in blue and black. Square points with capped error bars indicate composites while circular points indicate individual fits. Not all individual quasars involved in the composites are plotted. The single red square denotes the composite from \citet{Wang_2021}. The black hole masses in this study and in \citet{Wang_2021} are estimated with single-epoch virial estimates using the \mgii\ emission line, while the \citet{Xu2018} study uses the \civ\ emission line. \hl{The orange shaded space indicates a range of line ratios which are consistent with the metallicity indicated in the secondary axis based on photoionisation calculations with different ionizing SEDs}. The overlapping region in the \nvciv\ plot indicates a range of line ratios which is consistent with both $\rm{Z} = 10\,\rm{Z}_{\odot}$ and $\rm{Z} = 20\,\rm{Z}_{\odot}$ \citep[e.g.][]{Hamann_2002, Nagao_2006}. Metallicity values larger than 10 Z$_{\odot}$ are extrapolated.} \label{fig:line_ratio_Lbol_Mbh}
\end{figure*}

\begin{figure*}
	\includegraphics[width=\textwidth]{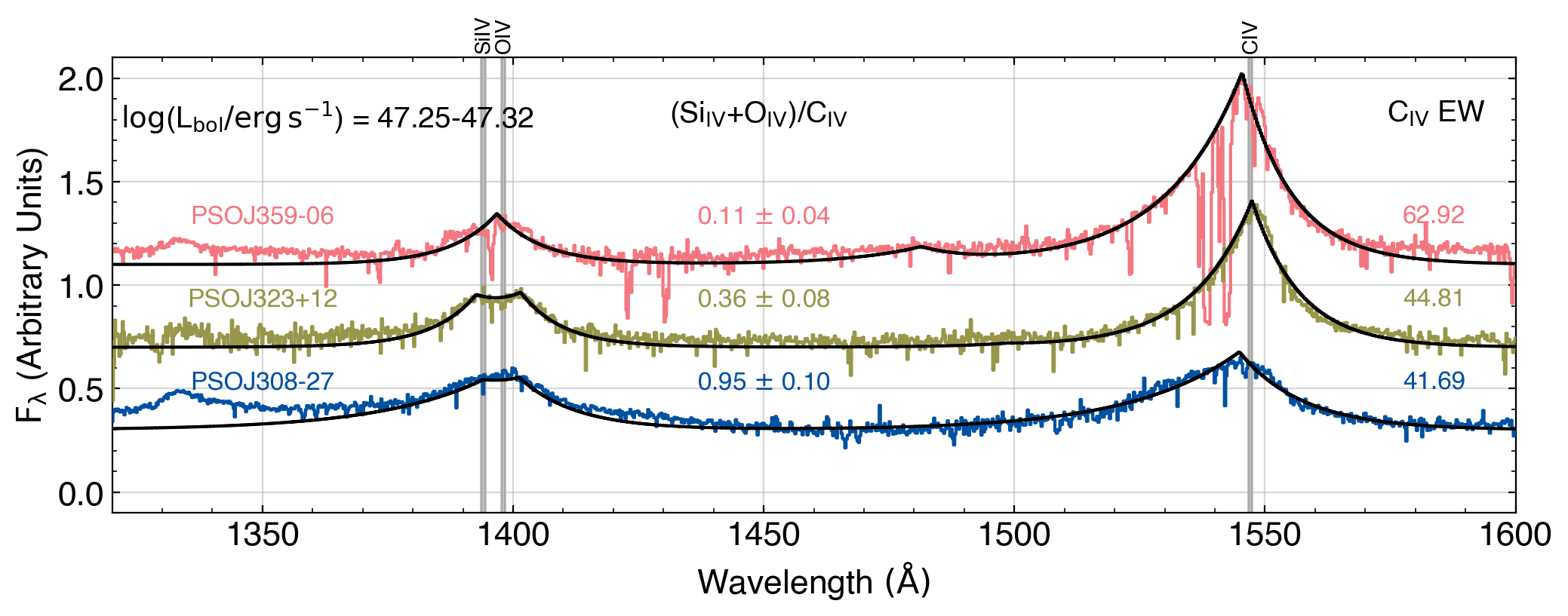}
    \caption{Example spectra of three quasars in the sample (PSOJ242-12, PSOJ308-27, and PSOJ359-06) which exhibit similar bolometric luminosities. The raw spectrum is shown in this plot, but many of the significant absorption features (e.g. in PSOJ359-06) are masked by the post-processing procedure. Overall line fits are shown in black. A large scatter in the \siivoivciv\ emission line flux ratio is observed between quasar spectra with significantly different \civ\ equivalent width, which is inversely correlated with the \civ\ blueshift.}
    \label{fig:line_ratio_Lbolscatter}
\end{figure*}

We present the measured line ratios of all individual and composite fits of bolometric luminosity and black hole mass in Figure \ref{fig:line_ratio_Lbol_Mbh}. For comparison, we show SDSS low-redshift composites reported in \citet{Xu2018} and high-redshift ($z\sim6$) quasars observed with GNIRS from \citet{Wang_2021}. Square symbols indicate measurements from fits of composites while circular points indicate fits of individual spectra. Our data, indicated in blue and black, have the highest SNR and spectral resolution of the data represented in the figure. Measurements of both metallicity-sensitive line ratios show a large scatter between individual quasar fits even when controlling for quasar luminosity or black hole mass. Particularly at a bolometric luminosity range of $\log\left(L_{\rm{bol}}/\rm{erg \, s}^{-1}\right) = 47.30-47.35$, we see over a factor of 8 difference between the individual quasar measured with the highest and lowest \siivoivciv\ line ratio, as seen in Figure \ref{fig:line_ratio_Lbolscatter}. The composites suppress the high variance that we observe in the individual measurements, which can be attributed to varied \civ\ blueshifts. The associated uncertainty of the bolometric luminosity and black hole mass for each composite is determined by the mean and standard deviation of the input spectra. We note that in \citet{Xu2018}, the black hole masses are estimated using the \civ\ emission line which can be biased by its blueshift. 

The line flux ratio measurements of our high-redshift quasar sample are essentially indistinguishable from the lower-redshift results of comparable luminosity and black hole mass sourced from \citet{Xu2018}. Therefore, we do not observe appreciable evolution with redshift. However, our high-redshift sample does not show a statistically appreciable correlation between observed line ratios and the bolometric luminosity, as evidenced in previous work \citep{hamann_1993, Dietrich_2003, Nagao_2006, Xu2018}. This may be because this sample covers a restricted luminosity range compared to the lower redshift sample. Deeper observations of quasars with high redshift and lower luminosity \citep[e.g.][]{Matsuoka_2016} are needed to verify any trend in emission-line ratio with quasar bolometric luminosity. On the other hand, we recover the positive correlation between \siivoivciv\ and the estimated black-hole mass as shown in Figure \ref{fig:line_ratio_Lbol_Mbh}. The measurements of individual quasars exhibit large scatter for very similar quasar properties. We present the line ratio dependence on the \civ\ blueshift in the following section to explain this variance.

\subsection{\civ\ Blueshift and Line Flux Ratios} \label{sec:blueshift}
\begin{figure}
	\includegraphics[width=\columnwidth]{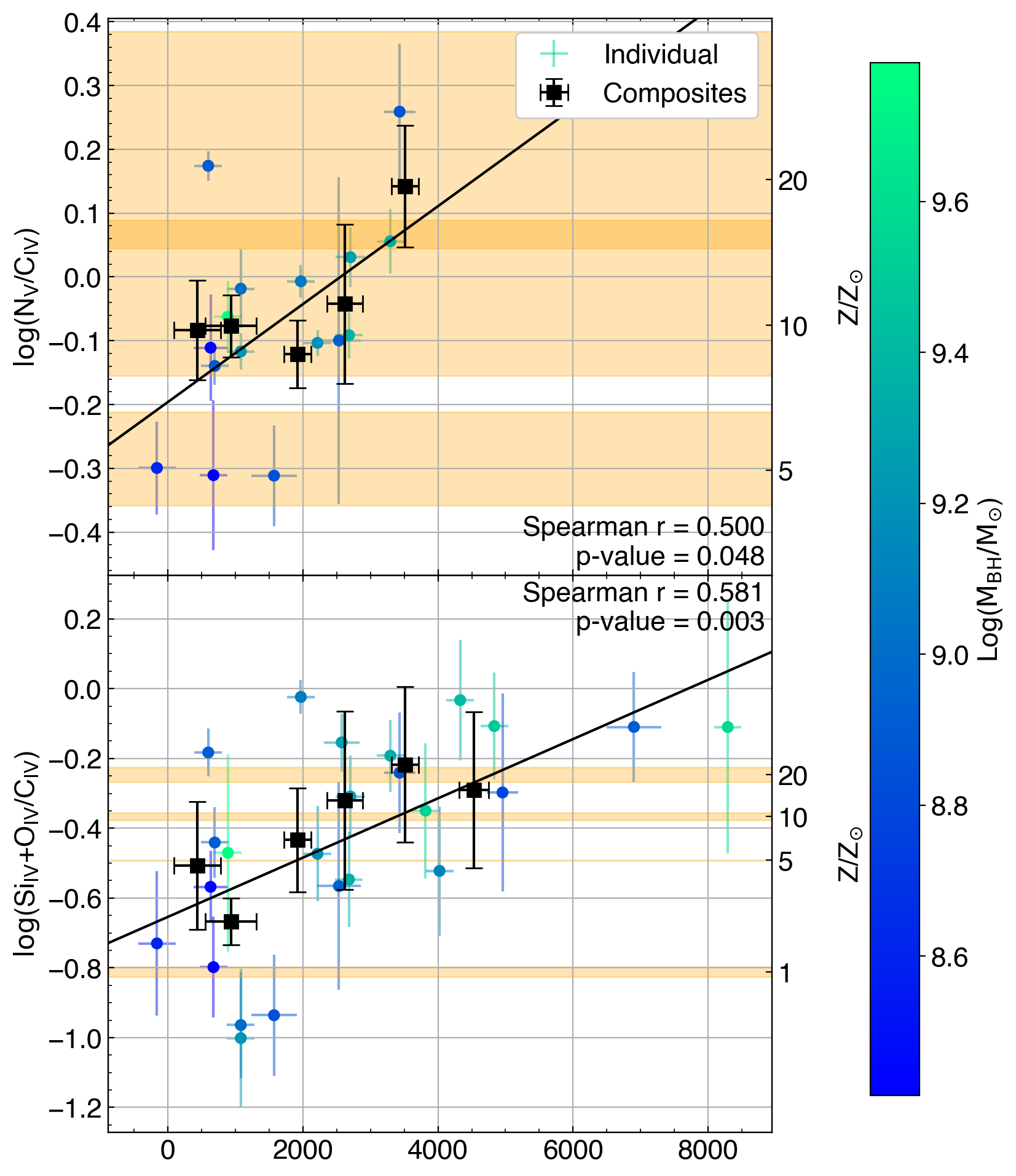}
    \caption{\nvciv\ and \siivoivciv\ flux ratios as a function of the \civ\ blueshift of quasars in the sample. Blueshift composites are presented in black while individual fits are mapped onto a blue-green gradient scaled to the black hole mass. The minimum \civ\ blueshift error is 200 km s$^{-1}$ based on the 1\AA\ wavelength bins, but the contribution from the systemic redshift error is added in quadrature for an average total of 230 km s$^{-1}$ uncertainty. The Spearman correlation coefficients and p-values are derived from the fits to individual quasars. \hl{The orange shaded space indicates a range of line ratios which are consistent with the metallicity indicated in the secondary axis based on photoionisation calculations with different ionizing SEDs}. The overlapping region in the \nvciv\ plot indicates a sub-space of parameters which is consistent with both $\rm{Z} = 10\,\rm{Z}_{\odot}$ and $\rm{Z} = 20\,\rm{Z}_{\odot}$. The line ratio correlation with the \civ\ blueshift is more significant than the correlation with black hole mass or bolometric luminosity.}
    \label{fig:blueshift_line_ratio_results}
\end{figure}

Figure \ref{fig:blueshift_line_ratio_results} plots emission line flux ratios against the \civ\ blueshift along with the estimated black hole mass as a third axis, represented by the blue-green color scale. The \civ\ blueshift and uncertainty of the composite spectra is obtained from the mean and standard deviation of the input spectra. We prescribe an uncertainty floor of the \civ\ blueshift equivalent to 200 km s$^{-1}$ based on the 1 \AA\ wavelength grid, but the total uncertainty for measurements of individual quasars is composed also of the redshift error added in quadrature. On average, the \civ\ blueshift error is 230 km s$^{-1}$. Measurements of the \nv\ emission line becomes more challenging to deblend from the \lya\ flux at high blueshifts, especially for individual lower SNR spectra, thus the high \civ\ blueshift parameter space for \nvciv\ is only sparsely explored.

The results from Figures \ref{fig:line_ratio_Lbolscatter} and \ref{fig:blueshift_line_ratio_results} demonstrate that the measured emission line ratios are strongly correlated with the \civ\ blueshift. Controlling for the \civ\ spectral shape in Figure \ref{fig:blueshift_line_ratio_results}, the relationship between the two emission line ratios with black hole mass is no longer clear. The \civ\ lines of the highest mass quasars are typically more blueshifted, but the most massive quasars do not necessarily have the highest line ratios among other quasars with similar blueshifts. It could also be seen that quasars with moderate ($\sim$1000 km s$^{-1}$) \civ\ blueshifts can have a very large scatter in observable line ratios whereas quasars with higher blueshifts consistently possess some of the highest line ratios observed in our sample. At high \civ\ blueshifts, the lower flux of the \civ\ line, as evidenced by its correlation with narrower EWs shown in Figure \ref{fig:CIV_blueshift_ew}, drives the \civ-normalised flux ratio measurements higher. The responses of the \nv, \siiv, and \oiv\ equivalent widths are not proportionate to that of the \civ\ line as the \civ\ blueshift rises. We see similar trends on other metallicity-sensitive line ratios that depend on the \civ\ flux as shown in Appendix Figure \ref{fig:other_line_ratios}.

\begin{figure}
	\includegraphics[width=\columnwidth]{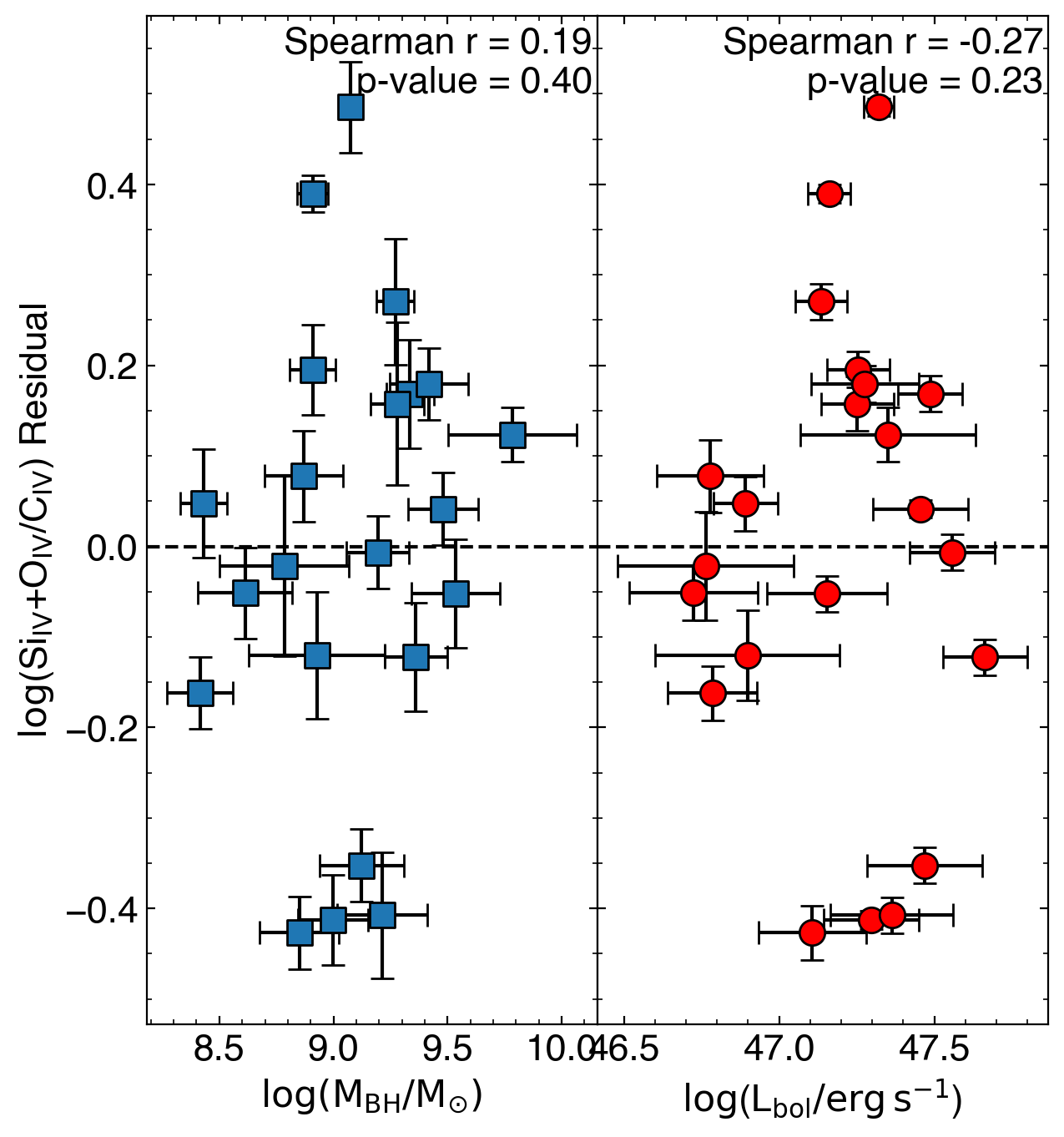}
    \caption{\siivoivciv\ residual as a function \hl{of} black hole mass (left) and bolometric luminosity (right). The residual is measured by subtracting the correlation found between composites of \civ\ blueshift and the \siivoivciv\ line ratio from the measured line ratios of individual quasars. The scatter is large around zero with no apparent systematic deviation with black hole mass or bolometric luminosity. This shows that higher mass or more luminous quasars are not more likely to exhibit higher line ratios than their smaller or fainter counterparts. However, we caution the reader when interpreting this relationship, because the \civ\ blueshift and quasar properties are not independent.}
    \label{fig:line_ratio_residual}
\end{figure}

Figure \ref{fig:blueshift_mbh_lbol} presents the correlations found in our sample between the \civ\ blueshift with the virially estimated black hole mass, quasar bolometric luminosity, and Eddington ratio. The \civ\ blueshift is not significantly correlated with the quasar bolometric luminosity, but there is a moderate relationship between the \civ\ blueshift with the Eddington ratio and estimated black hole mass, with the magnitude of Spearman correlation coefficients greater than 0.4 and at least 5\% significance. We show in Figure \ref{fig:line_ratio_residual} the residuals calculated by subtracting the correlation found between composites of \civ\ blueshift and the \siivoivciv\ line ratio from the measured line ratios of individual quasars. The results show that higher mass and more luminous quasars are not more likely to lie above this relationship, indicating no strong correlation with black hole mass or luminosity when controlling for blueshift \citep[also see][]{metal_density}. The stronger and more significant correlation between line ratios and the \civ\ blueshift \hl{could be attributed to the fact that the} \civ\ blueshift \hl{is a more direct observable compared to the black hole mass or luminosity, which are estimated using calibrations with large associated uncertainties. When not controlled, the} \civ\ blueshift \hl{can bias other correlations found between quasar properties and the metallicity in the BLR, such as the apparent} mass correlation in Figure \ref{fig:line_ratio_Lbol_Mbh}. It's important to note that the \civ\ blueshift and quasar properties are not independent for our sample. This suggests that studies measuring metallicity-sensitive line flux ratios dependent on the \civ\ flux should consider the \civ\ spectral shape before interpreting the diversity of emission-line ratios as an indication of evolution in the BLR metallicity. 

\begin{figure*}
	\includegraphics[width=\textwidth]{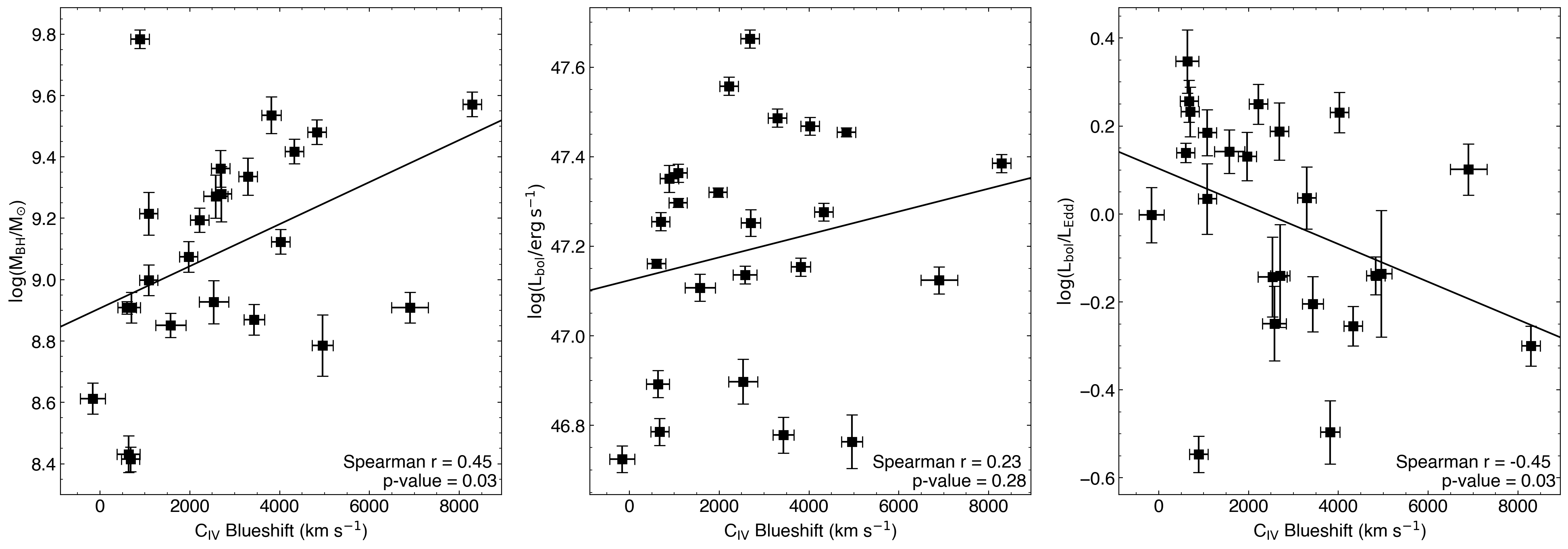}
    \caption{The quasar black hole mass (left), bolometric luminosity (middle), and Eddington ratio (right) are plotted against the measured \civ\ blueshift for our sample. The least-squares linear fits are shown along with the Spearman r-coefficients and their significance. There is a moderate, but significant, correlation between the quasar outflow indicator, i.e. the \civ\ blueshift, and the black hole mass. The correlation between the \civ\ blueshift with the Eddington ratio is weaker and less significant, while no significant correlation was found with the bolometric luminosity.}
    \label{fig:blueshift_mbh_lbol}
\end{figure*}

\subsection{Inferred Metallicity in the Quasar BLR} \label{sec:metallicity_result}
Using models derived from the photoionisation code \texttt{Cloudy}, we convert the measured \nvciv\ and \siivoivciv\ line ratios into metallicity estimates in the BLR. Figures \ref{fig:line_ratio_Lbol_Mbh} and \ref{fig:blueshift_line_ratio_results} show the range of line ratios consistent with 5, 10, and 20 Z$_{\odot}$ for \nvciv\ and  1, 5, 10, and 20 Z$_{\odot}$ for \siivoivciv. The minimum and maximum bounds of each metallicity estimate are determined by the variations on the assumed ionizing SED used in \texttt{Cloudy} photoionisation LOC models presented in \citet{Hamann_2002} and \citet{Nagao_2006}. The central tick is determined by the median of all relevant models. Larger variations in metallicity can be seen for the \nvciv\ line ratio, indicating that \siivoivciv\ is less dependent on the shape of the ionizing flux SED \citep[e.g.][]{Nagao_2006, Matsuoka_2011, Maiolino_2019_metallica}. The overlapping region in the \nvciv\ plot shows a range in the line ratio which is consistent with both $\rm{Z} = 10\,\rm{Z}_{\odot}$ and $\rm{Z} = 20\,\rm{Z}_{\odot}$ depending on the referenced photoionisation model, implying a factor of two uncertainty. The results for the other line ratios (\oiiialiiciv, \aliiiciv, \siiiiciv, and \ciiiciv) plotted against quasar bolometric luminosity, estimated black hole mass, and \civ\ blueshift are presented in Figure \ref{fig:other_line_ratios} in the Appendix. Generally, these other line ratios predict metallicities similar to \siivoivciv.

Spectra with high \civ\ blueshift are dominated by emission from an outflowing BLR wind, which is correlated with high X/\civ\ line flux ratios. Photoionisation models suggest that the wind emission originates from higher density gas clouds closer in to the accretion disk, illuminated by high ionising fluxes, while the core emission is composed of emission from clouds with a broad range of physical properties, as in the LOC model \citep{metal_density}. We therefore consider the metallicity results from spectra where the \civ\ blueshift < 1500 km s$^{-1}$, minimizing the contribution from the wind emission. The two composites satisfying this requirement yield (\nvciv, \siivoivciv) line ratios of ($0.83 \pm 0.15$, $0.31 \pm 0.13$) and ($0.84 \pm 0.09$, $0.22 \pm 0.03$) for \civ\ blueshifts from $-$200-680 and 680-1500 km s$^{-1}$ respectively. We do not consider the other line ratios in this discussion as they are substantially more difficult to measure. The \nvciv\ line ratio typically predicts higher metallicities with greater corresponding uncertainty than the \siivoivciv\ line ratio. Figure \ref{fig:blueshift_line_ratio_results} also plots the line ratios for each \civ\ blueshift composite and the inferred metallicity in the secondary axis. Using the aforementioned reference photoionisation models, the measured \nvciv\ line ratio is consistent with being produced by gas clouds with Z$_{-\rm{200-680}}$ = $9.77 \pm 2.35$ Z$_{\odot}$ and Z$_{\rm{680-1500}}$ = $9.96 \pm 2.42$ Z$_{\odot}$ for the two lowest \civ\ blueshift composites. For \siivoivciv, it is Z$_{-\rm{200-680}}$ = $4.61 \pm 0.01$ Z$_{\odot}$ and Z$_{\rm{680-1500}}$ = $2.01 \pm 0.01$ Z$_{\odot}$. Both metallicity indicators individually suggest super-solar metallicities with high significance ($\gtrsim$ 4-$\sigma$). The absolute measured metallicity differs by a factor of 2-4 between the indicators although the \siivoivciv\ line provides more robust results and is less affected by the model chosen for the ionising flux. Using the \siivoivciv\ line ratio, we can see that the metallicity in the quasar BLR at $z\sim6$ is at least 2-4 times super-solar. Inferred metallicities from spectra observed with high \civ\ blueshifts range from ${\rm{Z}}_{\rm{>1500}}$ = 8 Z$_{\odot}$ to as high as ${\rm{Z}}_{\rm{>3000}}$ = 20 Z$_{\odot}$

\section{Discussion} \label{sec:discussion}
Previous studies of chemical abundances in the BLR have suggested metallicities that are several times solar across a wide range of redshifts (2.0 < $z$ < 7.5) \citep[e.g.][]{hamann1992, Dietrich_2003, metallicity_distant, Xu2018, Onoue_2020}, consistent with some galactic chemical evolution models \citep{Tinsley_1980, Arimoto_1987, hamann_1993, Hamann_1999}. Complementary probes targeting quasar narrow absorption features also suggest super-solar (Z > 2 Z$_{\odot}$) metallicites \citep[e.g.][]{Hamann_1999, dodorico_2004_absorption, Jiang_2018, Maiolino_2019_metallica}. The lack of apparent redshift evolution up to $z\sim6$ stands in contrast to studies of metallicity in star-forming galaxies and Lyman-break galaxies up to $z\sim3.5$, which show evolution in the mass-metallicity relationship and an overall decrease in metallicity with redshift \citep[e.g.][]{Maiolino_2008, Mannucci_2009_LSD}. It's possible to infer the host galaxy metallicity using the mass of the central black hole using the tight (0.1 dex) galaxy stellar mass - gas phase metallicity relationship (MZR) \citep[e.g.][]{Maiolino_2008, Dave_2017_mufasa, Curti_2019_Klever, Maiolino_2019_metallica, Sanders_2021_MOSDEF} combined with the M$_{\rm{BH}}$/M$_{\rm{host}}$ ratios \citep[e.g.][]{Targett_2012}. The results from comparisons between quasar BLR and host galaxy metallicities at redshifts $2.25 < z < 5.25$ suggest that the BLR is enriched in excess of the inferred metallicities of the host galaxies, which are approximately solar \citep{Xu2018}. This discrepancy has been attributed to the black hole mass-metallicity relationship and selection effects, where only the most massive and enriched high-redshift quasars are selectively observed in a magnitude-limited survey \citep[e.g.][]{metallicity_distant, Maiolino_2019_metallica}. However,  \citet{Xu2018}, \citet{Wang_2021}, and this paper study samples with comparable quasar properties as shown in Figure \ref{fig:line_ratio_Lbol_Mbh}. The \citet{Xu2018} composites include hundreds of SDSS DR12 quasar spectra in the redshift range $2.25 < z < 5.25$, whereas \citet{Wang_2021} utilises a higher-redshift sample with 33 $z\sim6$ quasars. These quasars occupy a similar black hole mass and luminosity range, and show very similar line ratios within the scatter of the data, suggesting that a selection bias is not sufficient to explain the apparent lack of redshift evolution. 

Additionally, we note that the \civ\ blueshift, a signature of quasar outflows, is a significant factor correlated with the measured \civ\ flux (see Figure \ref{fig:CIV_blueshift_ew}). According to a study of 34 low-redshift quasars spanning nearly 3 dex in black hole mass and bolometric luminosity, outflow indicators are not correlated with black hole mass and only marginally correlated with luminosity and Eddington ratio \citep{shin_2017_outflow}. Although our study covers a smaller range of quasar parameters, our results in Figure \ref{fig:blueshift_mbh_lbol} show a moderate, but significant, correlation between the \civ\ blueshift and the black hole mass. There is also a similar negative correlation between the \civ\ blueshift and the Eddington ratio, and no significant correlation with quasar bolometric luminosity. There are several methodical differences between the measurements in our study and those in \citet{shin_2017_outflow}. We measure black hole virial masses based on \mgii\ instead of H$\beta$, and calculate systemic redshifts from the \mgii\ line rather than from a combination of low-ionisation narrow lines (\siii, \oii, \oi, H$\beta$). We also use different outflow indicators: our \civ\ blueshift is defined in Equation \ref{eq:blueshift}, in contrast to the ``velocity shift index'' (VSI) and ``blueshift and asymmetry index'' (BAI) defined in Equations 2 and 3 of \citet{shin_2017_outflow}. According to our result, a correlation between the black hole mass or bolometric luminosity with the \civ\ blueshift implies that the \civ\ flux is anti-correlated with the \civ\ blueshift by extension (see Figure \ref{fig:CIV_blueshift_ew}). As the \civ\ blueshift is found to be correlated with the metallicity-sensitive rest-frame UV line ratios and quasar properties, this has the potential to bias correlations between metallicity and black hole mass or luminosity. The relationship between the \civ\ blueshift and these line ratios can be explained by increased gas opacity with metallicity, leading to larger absorption and increased acceleration \citep[e.g.][]{Wang_2012}. However, the extreme metallicities ($\sim 20\, \rm{Z}_{\odot}$) seen in the most blueshifted high redshift ($z > 6.0$) quasars in our sample suggest that while the relationship between the \civ\ blueshift and the line ratio is real, the comparison to the simple photoionisation models is no longer appropriate as emission from the BLR outflow dominates the observed spectrum. The dynamics, density, and geometry of the BLR wind is not the same as for symmetric core emission \citep{metal_density}. An alternative explanation is that the \civ\ blueshift relationship with the rest-frame UV line ratios is driven primarily by the weakening of the symmetric \civ\ core emission and enhanced emission toward the line-of-sight from quasar orientation \citep[e.g.][]{Yong_2020}. Further studies on quasar properties and their emission-line flux ratios will need to account for indications of quasar outflows or avoid using emission lines that are strongly affected by BLR outflow.

It has also been suggested that the observed diversity of line ratios (\nvciv\ and \siivoivciv\ among others) can be attributed to the variation of density of the emitting gas and the incident ionizing flux instead of metallicity. \citet{metal_density} proposes a model with two kinematically distinct regions, the core and the wind, that can reproduce the range of observed broad emission-line flux ratios under solar metallicities, as long as the spatial density distribution of the emitting gas clouds is adjusted accordingly. Such multiple zone photoionisation models have been used to great effect in reproducing quasar and AGN spectra \citep[e.g.][]{Rees_1989, Peterson_1993, Baldwin_1996, Hamann_1998, Korista_2000_LOC}. The locally optimally emitting cloud (LOC) model, proposed in \citet{Baldwin_1995_LOC}, is a natural extension of multi-zone models. The advantage of the LOC model is that the total line emission is composed of an integration over the density ($n_{\rm{H}}$) and ionizing flux ($\Phi_{\rm{H}}$) parameter space assuming certain empirically motivated distribution functions \citep{Nagao_2006}, thereby bypassing the need for specific knowledge of either $n_{\rm{H}}$ or $\Phi_{\rm{H}}$. The properties of the emission lines are then dominated by the emitters that are optimally suited to emit the targeted line \citep{Baldwin_1995_LOC}. It is well-documented that the line ratios depend sensitively on $n_{\rm{H}}$ and $\Phi_{\rm{H}}$ \citep[e.g.][]{Hamann_2002, Nagao_2006, metal_density}, but the LOC results represent the average properties of diverse quasar samples. We note that the high-density wind component ($n_{\rm{H}} \approx 10^{13-14}$ cm$^{-3}$, $\Phi_{\rm{H}} \approx 10^{22-24}$ cm$^{-2}$ s$^{-1}$) and the range of typically assumed BLR properties ($n_{\rm{H}} \approx 10^{9-12}$ cm$^{-3}$, $\Phi_{\rm{H}} \approx 10^{18-21}$ cm$^{-2}$ s$^{-1}$) suggested in \citet{metal_density} are parameter ranges which are also covered by the LOC photoionisation models used in this study ($n_{\rm{H}} \approx 10^{7-14}$ cm$^{-3}$, $\Phi_{\rm{H}} \approx 10^{17-24}$ cm$^{-2}$ s$^{-1}$) \citep{Hamann_2002, Nagao_2006}. However, if the assumed cloud distribution functions used in photoionisation models are inaccurate, the absolute metallicity inferred from line ratios is subject to change. For example, photoionisation models using emission from clumpy disk winds can produce spectra resembling that of quasars \citep{Dannen_2020_clumpy, Matthews_2020_clumpy}, showing that there are viable alternatives to the LOC models we have referenced for the conversions between line ratio and metallicity.

The results from \citet{metal_density} further motivated us to use a quasar outflow indicator, the \civ\ blueshift, as a control to limit the effect of the BLR wind. Even for composites of low \civ\ blueshift where the assumed contribution to the overall emission from the wind is low, we observe in Figure \ref{fig:blueshift_line_ratio_results} that the average emission-line properties are comparable to emission from gas clouds with metallicity several times solar (e.g. Z$_{-\rm{200-680}}$ = $4.61 \pm 0.01$ Z$_{\odot}$ using \siivoivciv) under the LOC model.

The super-solar metallicities in the quasar BLR inferred from low \civ\ blueshift composites imply rapid enrichment scenarios that are not unrealistic under normal galactic chemical evolution scenarios in the cores of massive galaxies \citep[e.g.][]{Gnedin_1997, Dietrich_2003}. The BLR is a small nuclear region of the galaxy (<1 pc) with higher densities entailing shorter dynamical timescales \citep[e.g.][]{Gnedin_1997, Cen_1999, Kauffman_2000, Granato_2004}. The total mass of the BLR is on the order of 10$^{4}\, \rm{M}_{\odot}$ \citep{Baldwin_2003}, and it can be enriched rapidly to super-solar metallicities within 10$^{8}$ yrs by a single supernova explosion every 10$^{4}$ yrs \citep{metallicity_distant}. Under some multi-zone chemical evolution models, massive star formation in the galactic central regions and subsequent metal enrichment via supernovae can predict super-solar metallicities (up to 10 Z$_{\odot}$) within 0.5 - 0.8 Gyrs \citep[e.g.][]{hamann_1993, Friaca_1998, romano_2002}. This rapid enrichment scenario means that the properties of the BLRs do not necessarily trace the chemical properties of their host galaxies \citep[e.g.][]{Suganuma_2006, Matsuoka_2018}. This is supported by studies presenting estimates of metallicity in the quasar narrow-line region (NLR) which represent a region over 1000 pc in size \citep[e.g.][]{Bennert_2006_NLRsize}. The metallicity in the NLR was found to be 2-3 times lower than the BLR, following similar MZR trends as star-forming galaxies \citep[e.g.][]{Dors_2019}. However, in addition to high-metallicity BLRs, there is now mounting evidence that entire host galaxies can be highly enriched to solar values in early cosmic time as evidenced by measurements of C, N, and O ions \citep[e.g.][]{Walter_2003, Venemans_2017, Novak_2019, Pensabene_2021}.

More exotic enrichment scenarios such as enhanced supernova rates in central star clusters \citep[e.g.][]{artymowicz_1993, Shields_1996}, star formation inside quasar accretion disks \citep[e.g.][]{Collin_1999, Goodman_2004, Toyouchi_2021}, or nucleosynthesis without stars \citep[e.g.][]{Chakrabarti_1999, Hu_2008, Datta_2019} are also capable of producing highly enriched BLRs in a short time. However, we do not consider these to be strictly necessary to explain the metallicities in the $z\sim6$ redshift quasars inferred from observations in this study. Pushing metallicity estimates to even higher redshifts $z>8$ when the universe is only 0.6 Gyr old would place more stringent constraints on the metal enrichment timescales from the era of re-ionisation of the universe, where such rapid enrichment scenarios could be required to produce super-solar metallicities \citep[e.g.][]{Friaca_1998}.

\section{Conclusions} \label{sec:conclusion}
In this study, we examined a sample of 25 high-redshift (z > 5.8) quasars, 15 of which were observed with X-shooter during the XQR-30 programme and 10 of which were observed with Gemini North's GNIRS sourced from \citet{Shen_2019} and \citet{yang2021probing}. The sample from XQR-30 contains the highest-quality spectra covering the rest-frame UV emission lines observed in quasars in this redshift range. The bolometric luminosity of the quasars in this sample covers $\log\left(\rm{L}_{\rm{bol}}/\rm{erg \, s}^{-1}\right) = 46.7-47.7$ assuming a bolometric correction factor of 5.15 from the continuum luminosity at 3000\AA. The black hole mass range in the sample is $(0.2 - 6.0) \times 10^9\, \rm{M}_{\odot}$, measured with single-epoch virial mass estimates utilizing the FWHM of the \mgii\ emission line. We measured the blueshift of the \civ\ line in most of the quasars in this sample and created composites by quasar luminosity, black hole mass, and \civ\ blueshift. We then measured broad rest-frame UV emission-line flux ratios in individual quasar spectra and all composites. The main results are as follows:
\begin{itemize}
    \item Due to the relationship between the \civ\ blueshift and its equivalent width, the metallicity-sensitive broad emission-line ratios correlate with the \civ\ blueshift, which is an indicator of the projected BLR outflow velocity. If not accounted for, this correlation biases studies of quasar metallicity and its relationship with black hole mass and luminosity. The correlation between the metallicity-sensitive emission line flux ratios and the \civ\ blueshift is stronger and more significant than for the quasar bolometric luminosity or black hole mass. 
    \item Comparing against \texttt{Cloudy}-based photoionisation models, the metallicity inferred from line ratios of the high-redshift ($z\sim6$) quasars in this study is several (at least 2-4) times super-solar, consistent with studies of much larger samples at lower redshifts and similar studies at comparable redshifts. We also find no strong evidence of redshift evolution in the BLR metallicity, indicating that the BLR is already highly enriched at $z\sim6$. The metallicity-sensitive emission-line flux ratios are sensitive to the density $n_{\rm{H}}$ of gas clouds and the incident ionizing flux $\Phi_{\rm{H}}$, but we use locally optimally-emitting cloud photoionisation models to draw conclusions based on the average properties of diverse samples of quasars. Our low \civ\ blueshift composites are good probes of metallicity at this redshift as they minimise the effects of the BLR wind.
    \item The lack of redshift evolution in the BLR metallicity is contrary to studies of metallicity in star-forming and Lyman-break galaxies, which show a significant redshift dependence. Furthermore, estimates of host galaxy properties based on black hole mass suggest metallicities that are approximately solar. We find that selection effects are not sufficient to explain the apparent lack of redshift evolution and the discrepancy between the BLR metallicity and host galaxy metallicity. However, given the small scale of the BLR, rapid enrichment scenarios make it a poor tracer of host galaxy metallicity.
    \item The super-solar metallicity inferred for BLRs at $z\sim6$ provides stringent constraints on the timescales of star formation and metal enrichment in the vicinity of some of the earliest supermassive black holes. Rapid metal enrichment scenarios of the BLR are not unrealistic under normal galactic chemical evolution models and more exotic explanations, such as nucleosynthesis or star formation inside the accretion disk, are not strictly necessary. 
\end{itemize}

Intrinsic absorption lines could, in principle, provide more straightforward estimates of the BLR metallicity \citep[e.g.][]{Hamann_1999, dodorico_2004_absorption, Maiolino_2019_metallica}. In the past, such studies were not possible due to low SNR of quasar spectra at $z \sim 6$, but the high-quality data of XQR-30 enables this type of investigation, which will be explored in a future study. More precise metallicity diagnostics would solidify and refine these results, especially for individual quasars.

\section*{Acknowledgements}

We thank Matthew Temple for the helpful discussion \hl{and the reviewer, Yoshiki Matsuoka, for the thoughtful comments and suggestions which have improved this work.}

The results of this research is based on observations collected at the European Organisation for Astronomical Research in the Southern Hemisphere under ESO programme 1103.A-0817. 

This work is also based, in part, on observations obtained at the international Gemini Observatory, a program of NSF's NOIRLab, which is managed by the Association of Universities for Research in Astronomy (AURA) under a cooperative agreement with the National Science Foundation on behalf of the Gemini Observatory partnership: the National Science Foundation (United States), National Research Council (Canada), Agencia Nacional de Investigaci\'{o}n y Desarrollo (Chile), Ministerio de Ciencia, Tecnolog\'{i}a e Innovaci\'{o}n (Argentina), Minist\'{e}rio da Ciencia, Tecnologia, Inova\c{c}\~{o}es e Comunica\c{c}\~{o}es (Brazil), and Korea Astronomy and Space Science Institute (Republic of Korea).

S.L. is grateful to the Australian National University Research School of Astronomy \& Astrophysics (ANU/RSAA) for funding his Ph.D. studentship and the European Southern Observatory for the research internship.

CAO was supported by the Australian Research Council (ARC) through Discovery Project DP190100252.

M.B. acknowledges support from PRIN MIUR project ``Black Hole winds and the Baryon Life Cycle of Galaxies: the stone-guest at the galaxy evolution supper'', contract \#2017PH3WAT. 

ACE acknowledges support by NASA through the NASA Hubble Fellowship grant $\#$HF2-51434 awarded by the Space Telescope Science Institute, which is operated by the Association of Universities for Research in Astronomy, Inc., for NASA, under contract NAS5-26555. 

SEIB acknowledges funding from the European Research Council (ERC) under the European Union’s Horizon 2020 research and innovation programme (grant
agreement No. 740246 “Cosmic Gas”).

JTS acknowledges funding from the European Research Council (ERC) under the European Union’s Horizon 2020 research and innovation programme (grant agreement No. 885301 ``Quasar Chronicles'').

\section*{Data Availability}
The data underlying this article will be shared on reasonable request to the corresponding author. 



\bibliographystyle{mnras}
\bibliography{bibliography} 




\appendix
\newpage
\section{Comparison of emission-line fitting methods} \label{appendix:line-fitting-compare}

In this study, we use a piece-wise power-law function to fit emission features. This approach to emission-line fitting is different from other widely adopted functions such as multiple Gaussians and modified Lorentzians. Nevertheless, we demonstrate here that the resulting flux ratios are consistent between different emission-line fitting functions. For the comparison, we used the skewed Gaussian distribution because it is an alternative which, like the piece-wise power-law, also fits each emission-line with four free parameters. The probability density function (pdf) of a skewed Gaussian distribution is described by the following formula
\begin{equation}
    f(\lambda) = \frac{2}{\omega} F_{\rm{0}} \, \phi\left(\frac{\lambda - \lambda_{0}}{\omega}\right) \Phi\left(\alpha\left(\frac{\lambda - \lambda_{0}}{\omega}\right)\right) \,,
\end{equation}
where $\phi(x)$ is the standard Gaussian pdf,
\begin{equation}
    \phi(x) = \frac{1}{\sqrt{2\pi}} e^{-\frac{x^2}{2}}\,,
\end{equation}
and $\Phi(x)$ is its cumulative distribution function given by,
\begin{equation}
    \Phi(x) = \int_{-\infty}^{x} \phi(t) dt \,,
\end{equation}
and $\omega$ controls the scale of the distribution while $\alpha$ determines the skewness. The normalisation, $F_{\rm{0}}$, is proportional to the peak of the emission line at the peak wavelength, $\lambda_{\rm{0}}$. Analogous to the coupling of power indicies of LIL and HIL lines, we couple the skewness and scale of lines in each category, with the exception of the parameters of \lya\ which can vary independently. The scale of the skewed Gaussian distribution is coupled in velocity space, which preserves the FWHM and kinematic status of line-emitting clouds. 

\begin{figure}
	\includegraphics[width=\columnwidth]{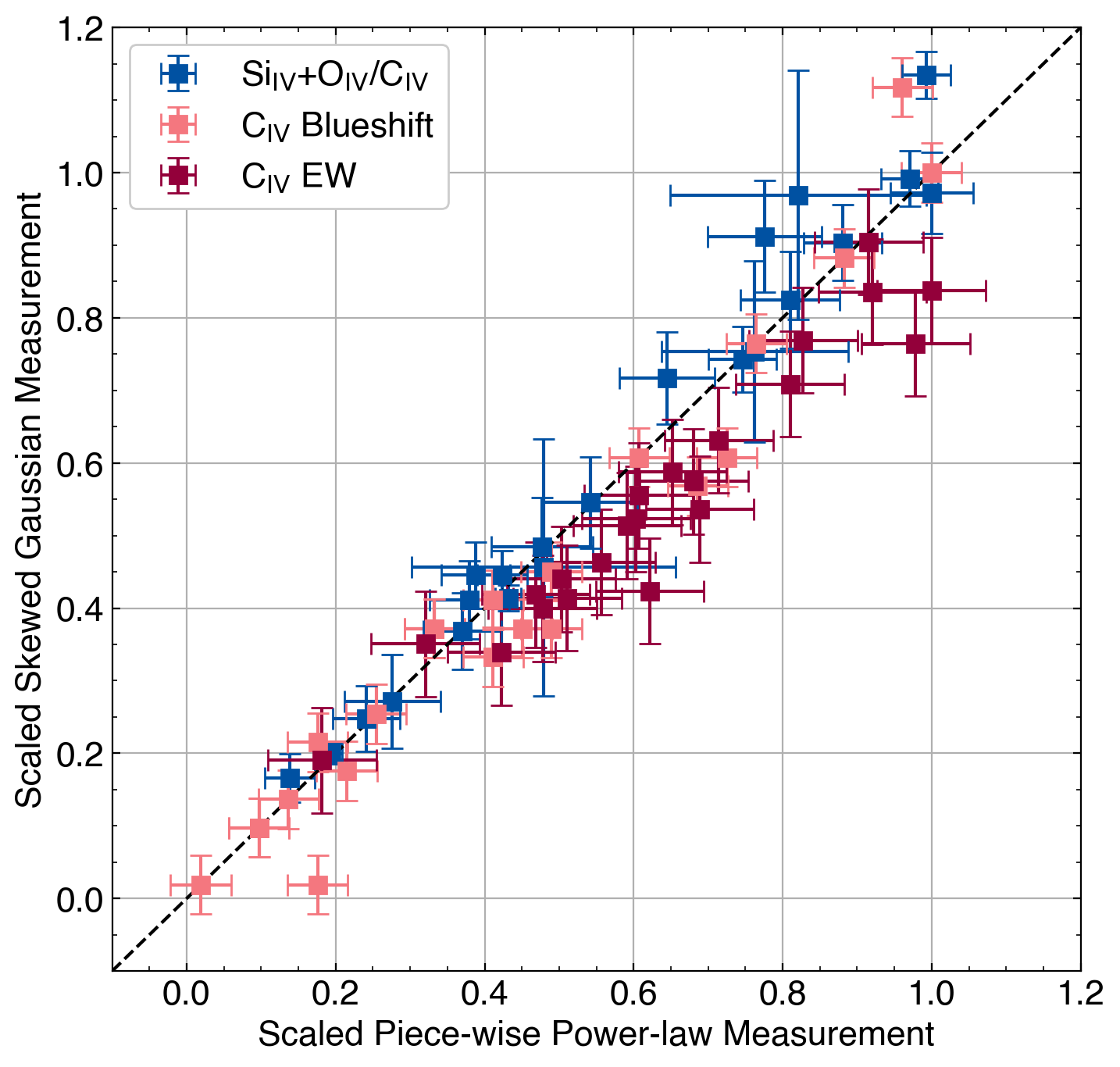}
    \caption{Comparison between measurements obtained with the piece-wise power-law fit and the skewed Gaussian fit. The three quantities plotted are the \siivoivciv\ emission-line flux ratio, the \civ\ blueshift, and the \civ\ equivalent width. Each of these measurements are scaled to the maximum quantity produced by the piece-wise power-law fitting method. We find no systematic difference with the \siivoivciv\ flux ratio, whereas the \civ\ blueshift and equivalent width show a slight measurement bias where the piece-wise power-law method produces comparatively larger values.}
    \label{fig:PL_SN_compare}
\end{figure}

Because of the blended \lya\ and \nv\ line profile, and the absence of an analogous approach to coupling the red wing of \lya\ to HIL lines with this method, we compare the flux ratio of \siivoivciv\ as well as the \civ\ equivalent width and blueshift. Figure \ref{fig:PL_SN_compare} shows a comparison between these quantities, scaled to the maximum value produced by the power-law fitting method. We find no systematic difference with the \siivoivciv\ flux ratio, but the \civ\ equivalent width and blueshift show a slight measurement bias where the piece-wise power-law fit produces larger values ($\sim 1 \sigma$) than the skewed Gaussian approach. On average, the scatter in the \siivoivciv\ emission-line ratio is 0.7-$\sigma_{\rm{SD}}$, while the scatter in the \civ\ equivalent width and blueshift are 1.2-$\sigma_{\rm{SD}}$ and 1.1-$\sigma_{\rm{SD}}$ respectively. In general, the skewed Gaussian approach under-fits the peak of \civ\ emission lines, especially in the lower \civ\ blueshift and higher equivalent width regime. This indicates that the skewed Gaussian approach should be used when the \civ\ emission is more blueshifted, which is the strategy we have adopted in this study. 

We also compare our results against a similar study of metallicity at redshift $z \sim 6$ \citep{Wang_2021}. Our data overlap consists of 1 quasar (P333+26) out of the 24 considered in this study, for which \citet{Wang_2021} does not report measurements for either \nvciv\ nor \siivoivciv. Nevertheless, we obtained 2 of their spectra (J0008-0626 and J1250+3130) from their parent sample, \citet{Shen_2019} which are not used in this study because they are BAL quasars. In \citet{Wang_2021}, a multiple-Gaussian approach was used for emission-line fitting and each individual quasar spectrum was fit independently. In addition, emission lines were not coupled together into groups of LILs and HILs. Though both spectra were of BAL quasars and despite the different methods used, we find less than 10\% difference in \nvciv\ and less than 5\% difference in \siivoivciv. These comparisons show that the results and correlations presented in this study can be reproduced with a variety of continuum and line-fitting methods. This gives us confidence that our measurements are robust.

\section{Additional Tables and Figures}
\begin{figure*}
	\includegraphics[width=\textwidth]{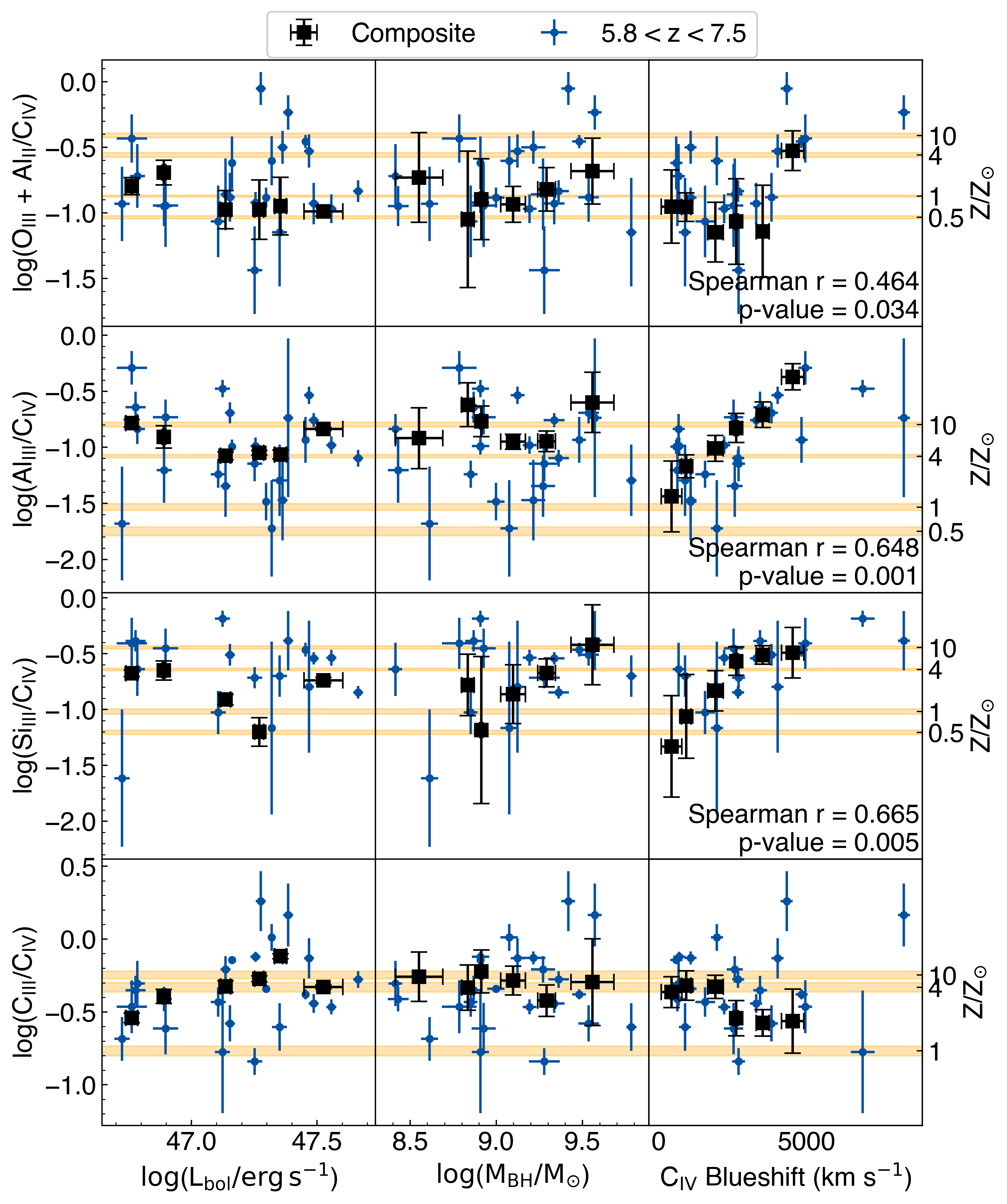}
    \caption{Line flux ratios of \oiiialiiciv, \aliiiciv, \siiiiciv, and \ciiiciv\ as functions of the bolometric luminosity, black hole mass, and \civ\ blueshift of quasars in the sample. Composites are plotted as black squares while individual fits are represented by blue points. The orange shaded space indicates a range of line ratios which are consistent with the metallicity indicated in the secondary axis based solely on varying the assumed ionizing SED in the various photoionisation calculations. We present the Spearman correlation coefficient and p-value for correlations based on fits of individual quasars only if the p-value $\leq 0.1$.}
    \label{fig:other_line_ratios}
\end{figure*}

\begin{figure}
	\includegraphics[width=\columnwidth]{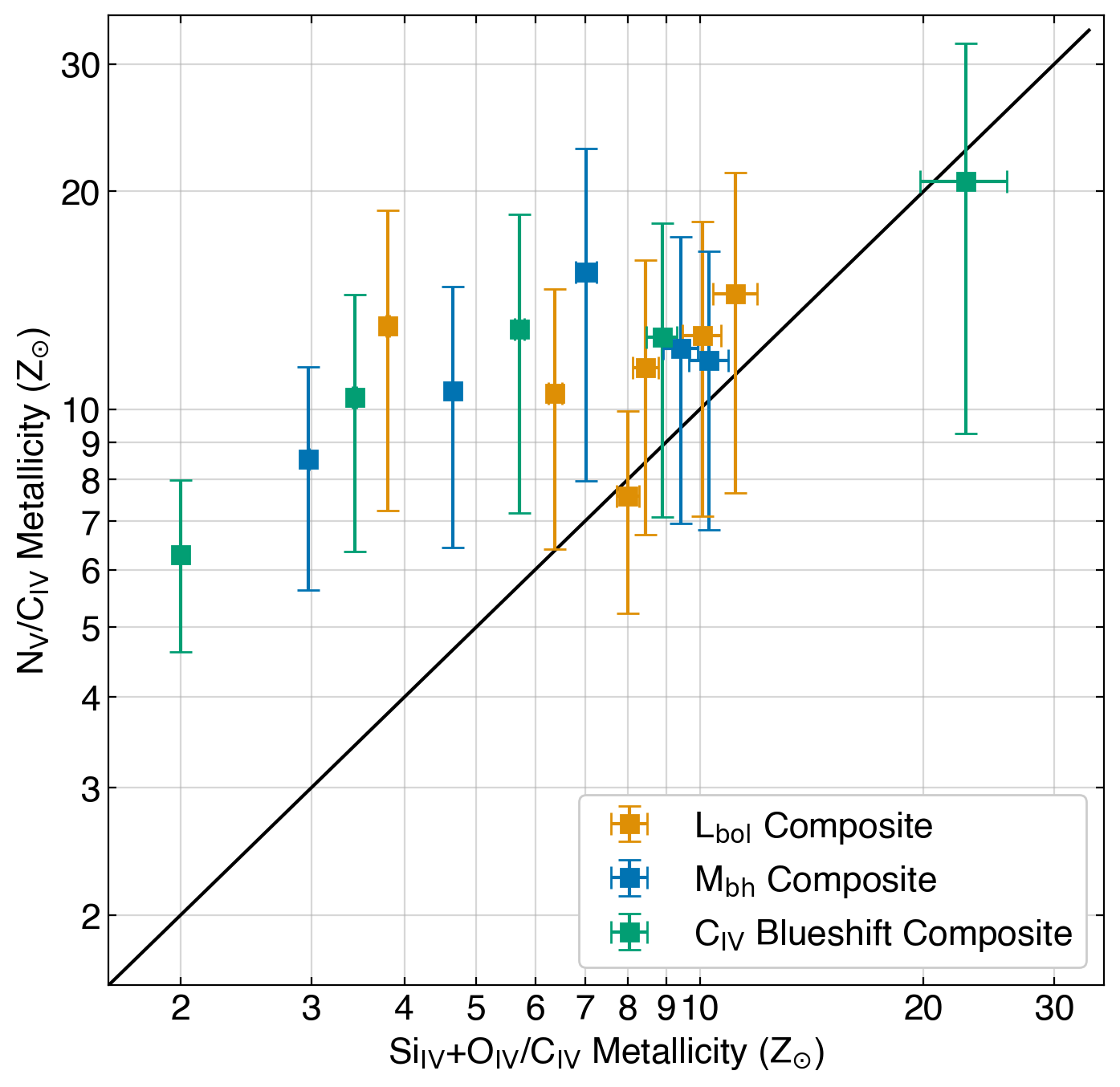}
    \caption{Comparison between inferred metallicities from \nvciv\ and \siivoivciv\ in log-space and units of solar metallicity. We show values for all of the quasar bolometric luminosity, black hole mass, and \civ\ blueshift composites if their \nvciv\ line ratio could be measured. In this plot, the uncertainty is determined by the maximum and minimum values between all of the photoionisation calculations from \citet{Hamann_2002} and \citet{Nagao_2006}, assuming different ionizing SEDs. The metallicity inferred from \nvciv\ is typically greater than from \siivoivciv.}
    \label{fig:NV_SiIV_Zcompare}
\end{figure}

\begin{figure*}
\begin{tabular}{cc}
(a) J0910+1656: Lowest L$_{\rm{bol}}$ \& Lowest \civ\ Blueshift & 
(b) PSOJ158-14: Highest L$_{\rm{bol}}$ \\
  \includegraphics[width=80mm]{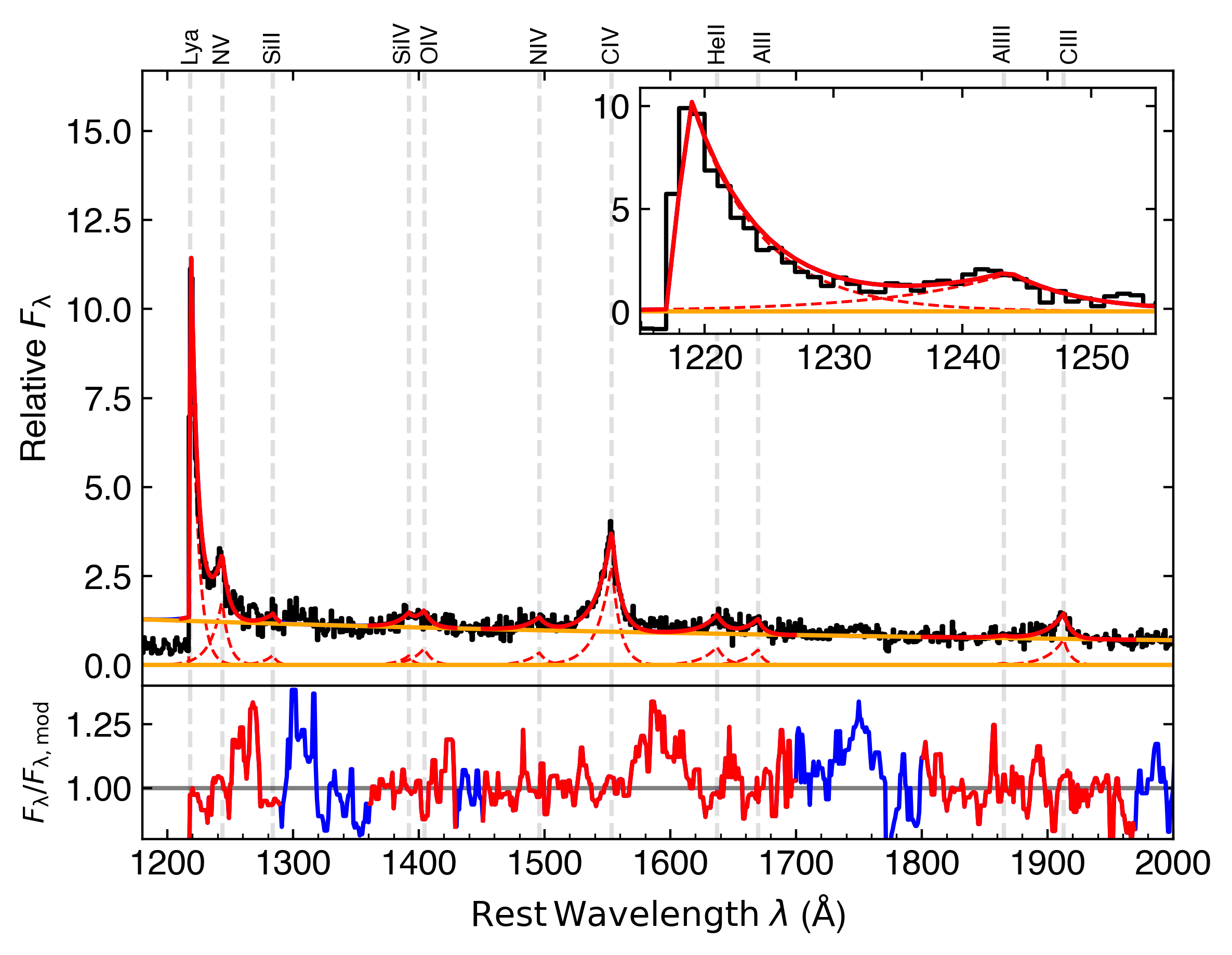} &   \includegraphics[width=80mm]{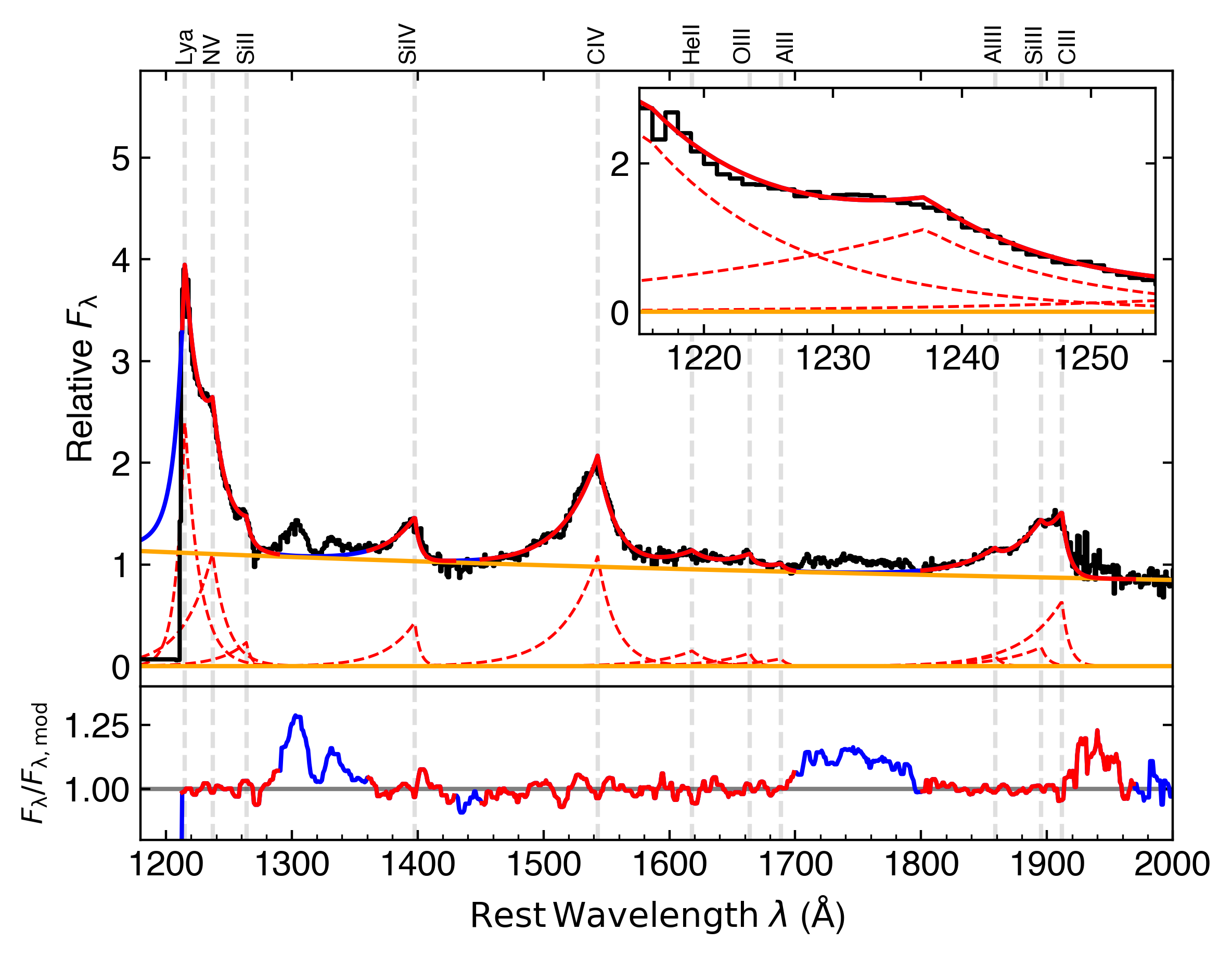} \\ [6pt]
(c) J0921+0007: Lowest M$_{\rm{BH}}$ & 
(d) PSOJ242-12: Highest M$_{\rm{BH}}$ \\
 \includegraphics[width=80mm]{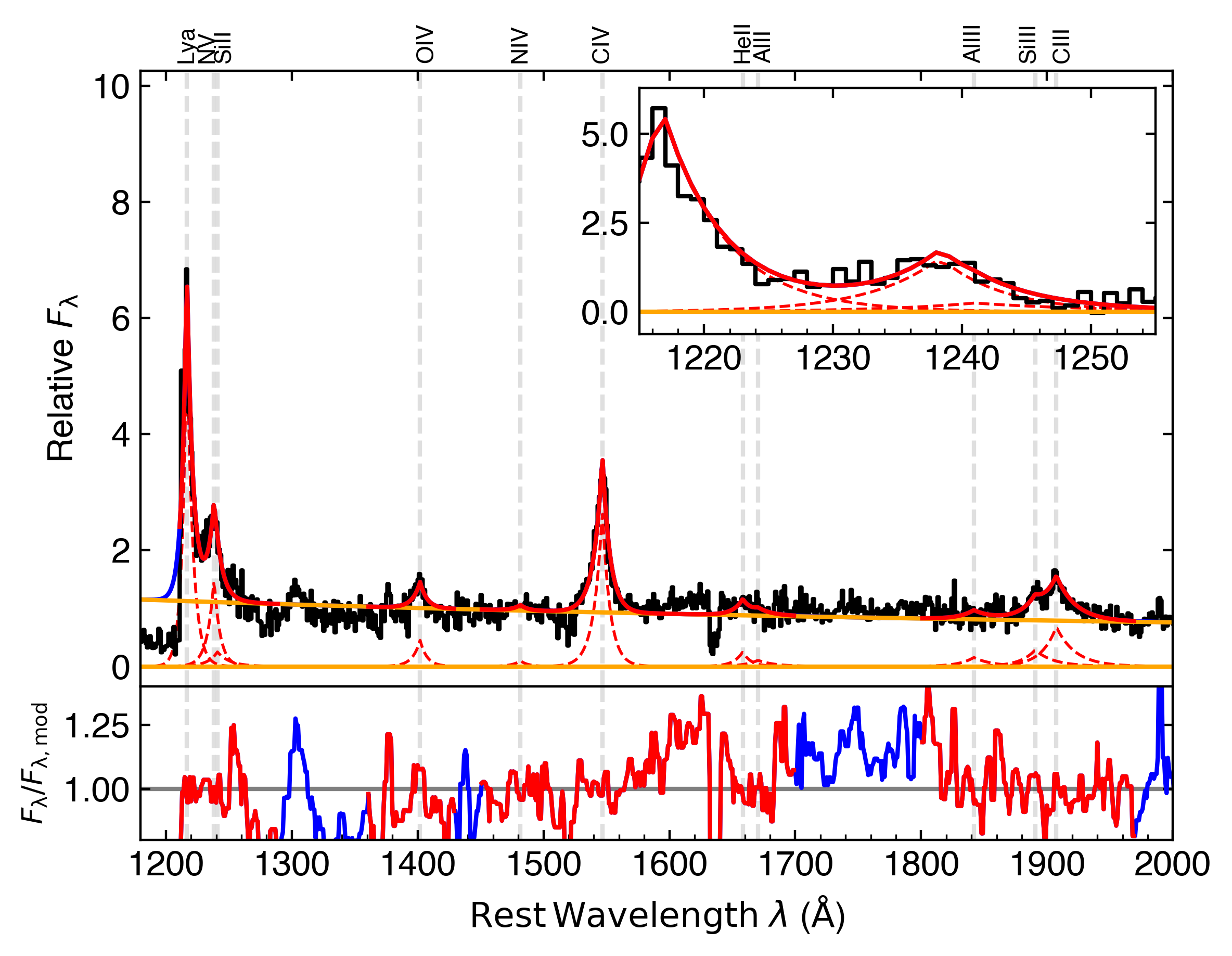} &   \includegraphics[width=80mm]{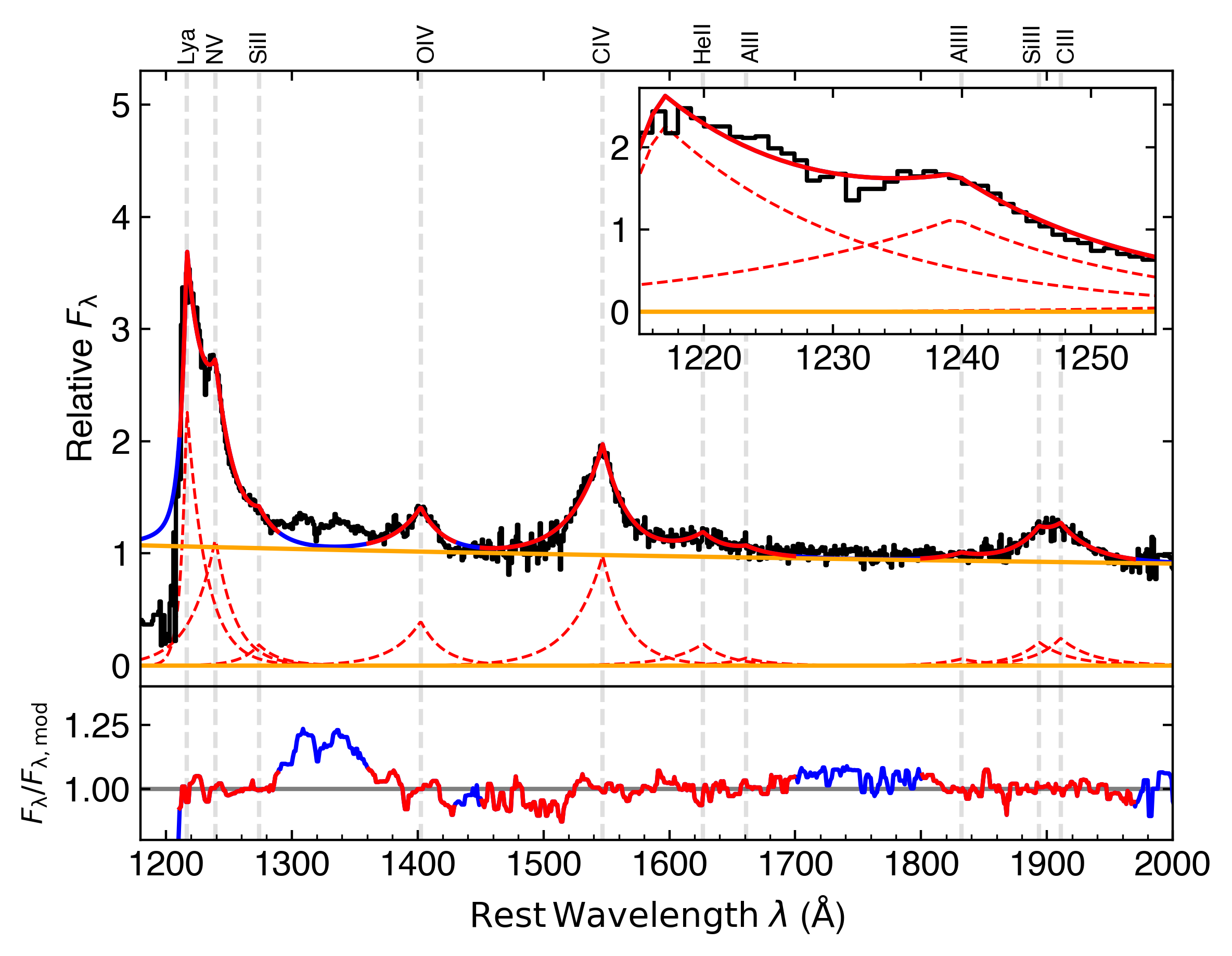} \\ [6pt]
(e) J0837+4929: 2$^{\rm{nd}}$ Lowest \civ\ Blueshift  & 
(f) PSOJ065-25: Highest \civ\ Blueshift \\
 \includegraphics[width=80mm]{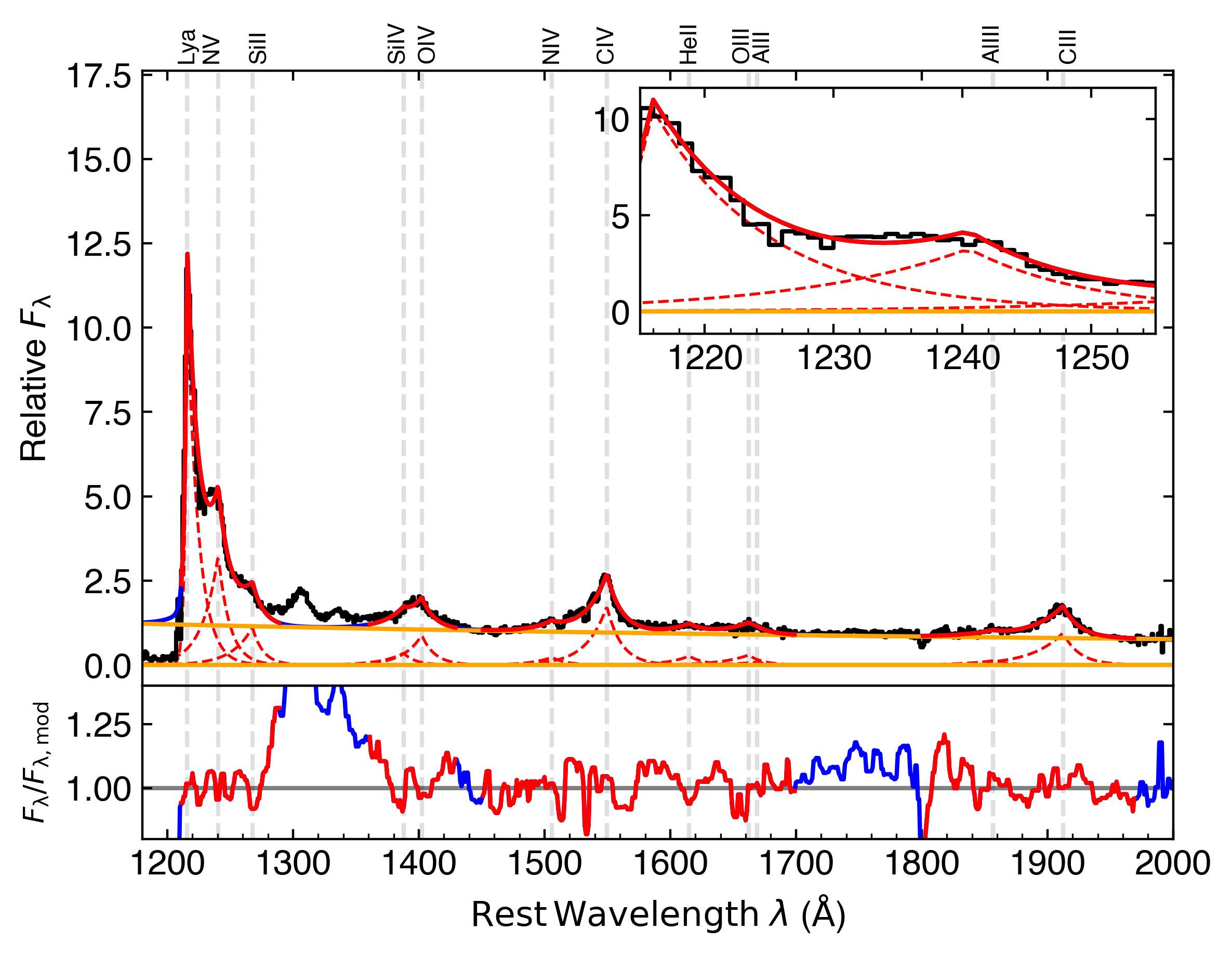} &   \includegraphics[width=80mm]{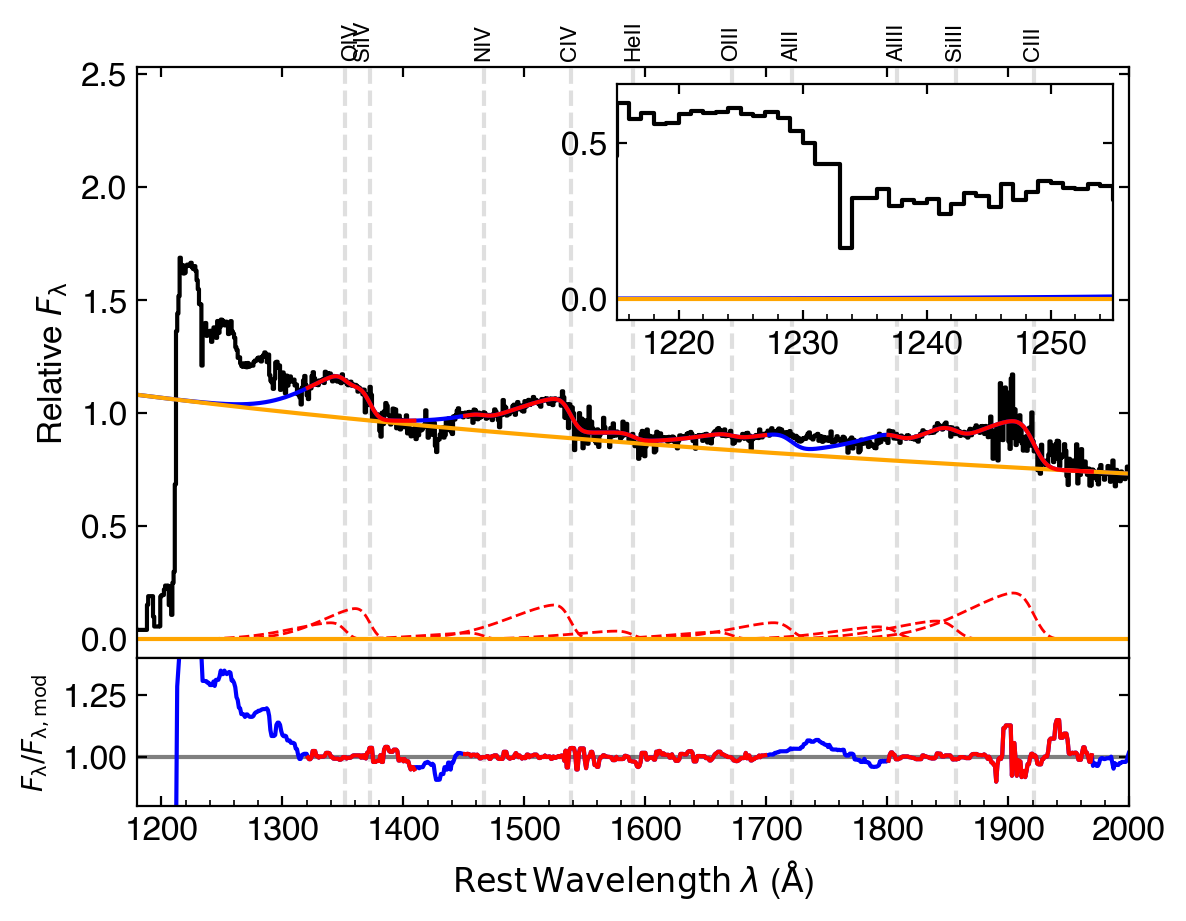} \\ [6pt]
\end{tabular}
\caption{Example fits to individual quasar spectra based on the indicated quasar properties. Line fitting and continuum fitting windows have been changed slightly to obtain the best fits. If the \lya+\nv\ complex was particularly noisy with low SNR or if the \civ\ blueshift is high, then emission lines blueward of \siiv\ are not included in the fit.} \label{fig:set-individual-fits}
\end{figure*}

\begin{figure*} 
\begin{tabular}{cc}
(a) $\log\left(\rm{L}_{\rm{bol}}/\rm{erg \, s}^{-1}\right) = 46.72-46.79$ & (b) $\log\left(\rm{L}_{\rm{bol}}/\rm{erg \, s}^{-1}\right) = 46.79-46.95$ \\
  \includegraphics[width=80mm]{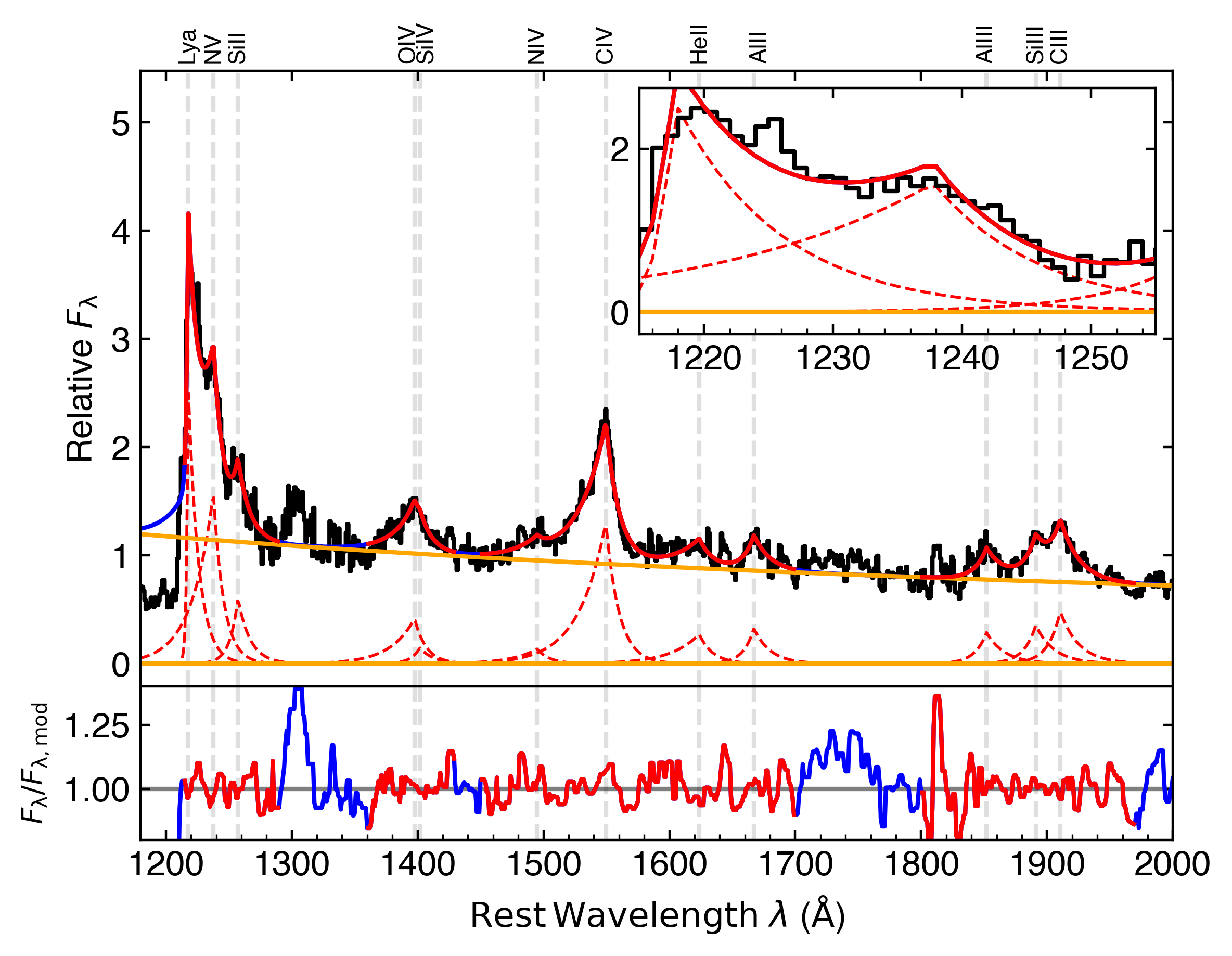} &   \includegraphics[width=80mm]{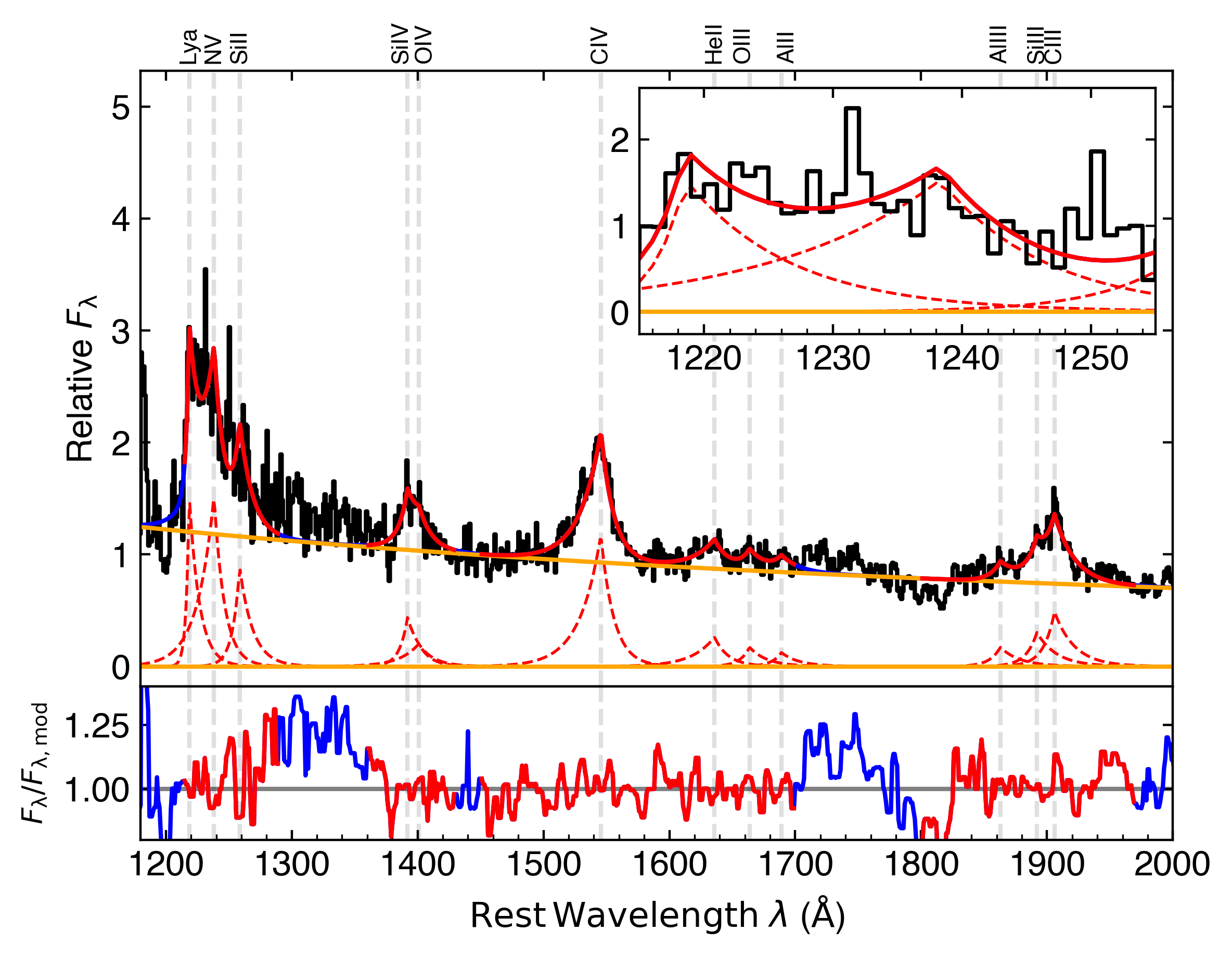} \\ [6pt]
(c) $\log\left(\rm{L}_{\rm{bol}}/\rm{erg \, s}^{-1}\right) = 46.95-47.17$ & (d) $\log\left(\rm{L}_{\rm{bol}}/\rm{erg \, s}^{-1}\right) = 47.17-47.30$ \\
 \includegraphics[width=80mm]{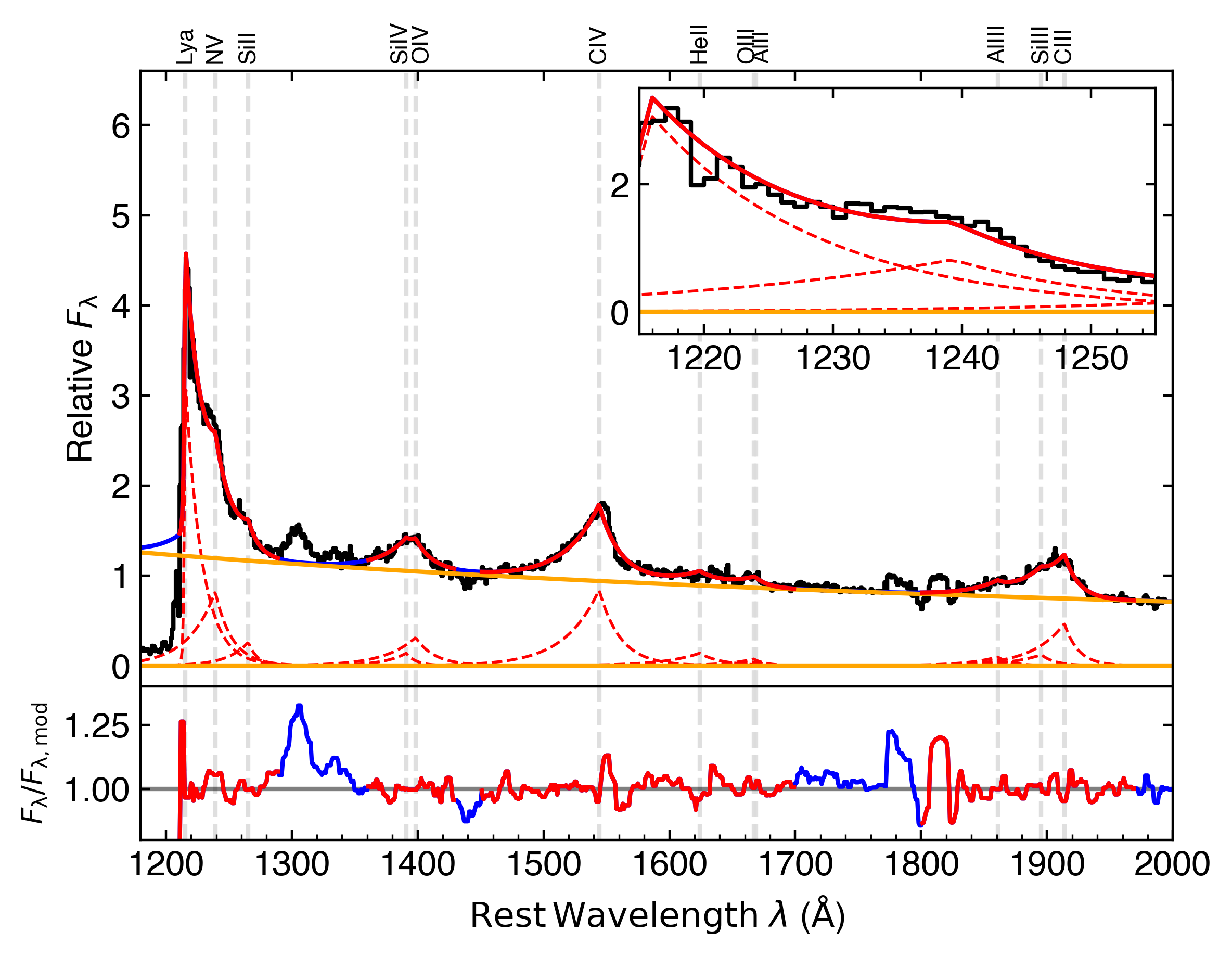} &   \includegraphics[width=80mm]{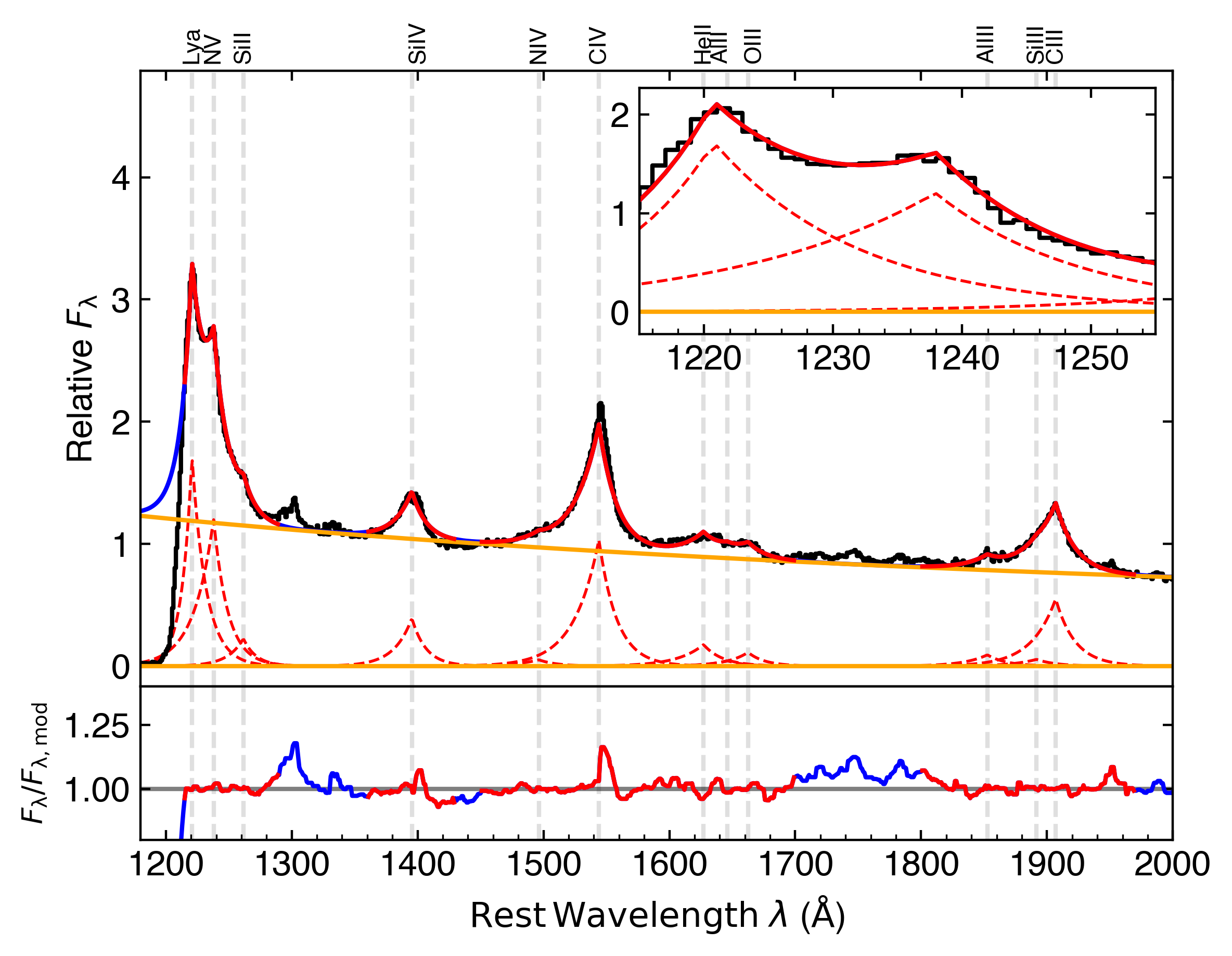} \\ [6pt]
(e) $\log\left(\rm{L}_{\rm{bol}}/\rm{erg \, s}^{-1}\right) = 47.30-47.40$ & (f) $\log\left(\rm{L}_{\rm{bol}}/\rm{erg \, s}^{-1}\right) = 47.40-47.70$ \\
 \includegraphics[width=80mm]{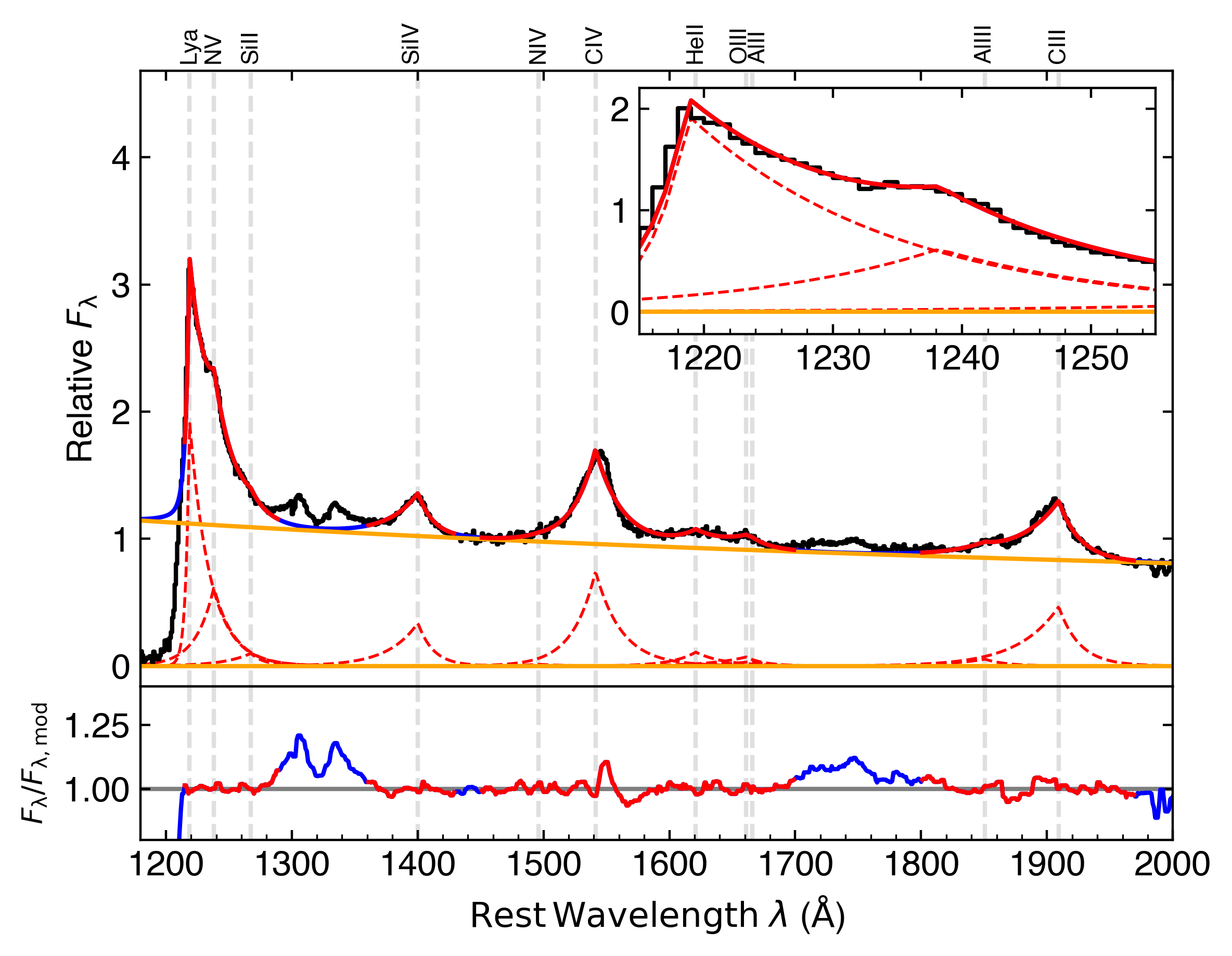} &   \includegraphics[width=80mm]{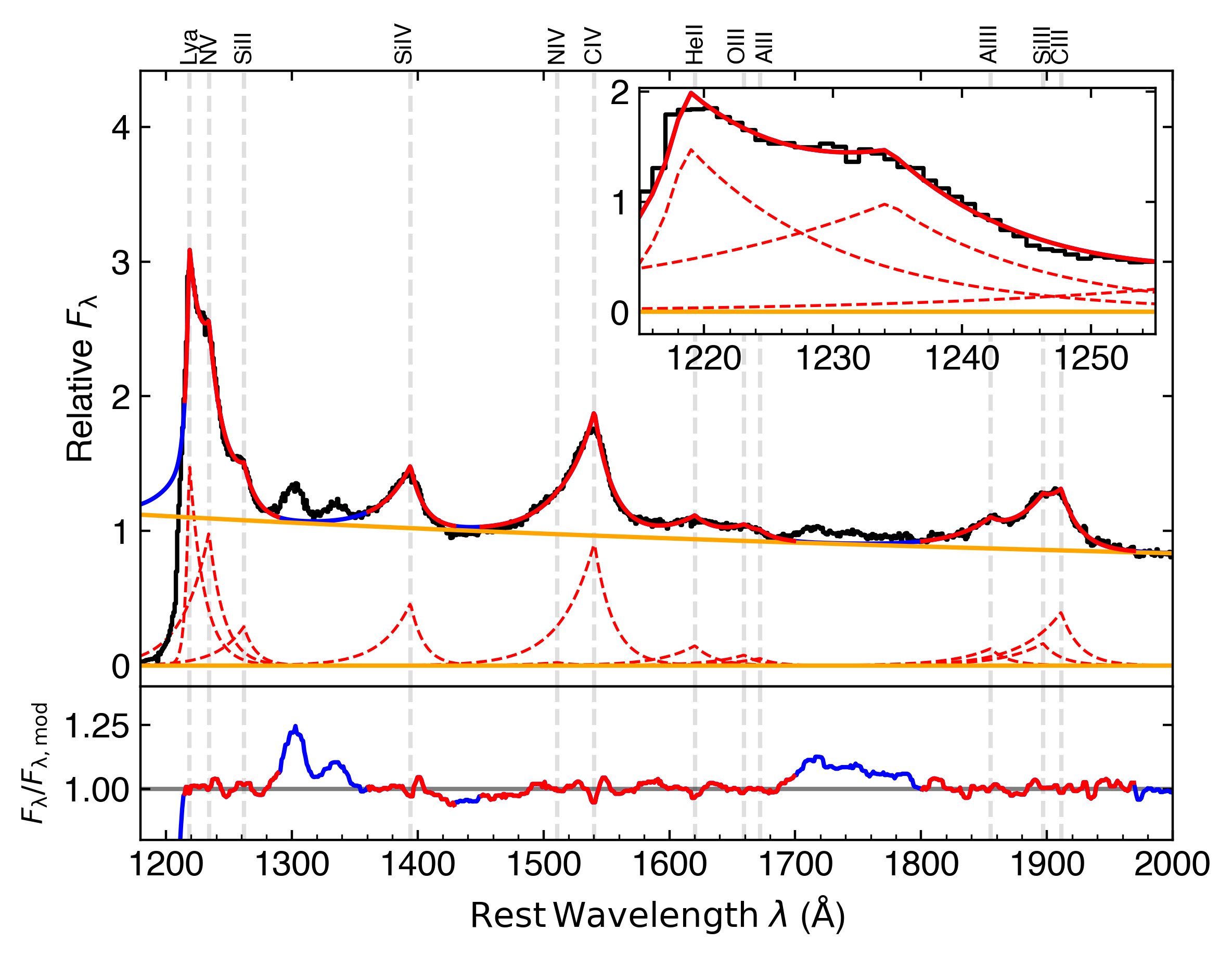} \\ [6pt]
\end{tabular}
\caption{Example fits to composites stacked in bins of bolometric luminosity ordered by increasing luminosity from panels a-f. Line fitting and continuum fitting windows have been changed slightly between fits. The red lines indicate the emission line fits as well as the extent of the individual line-fitting windows. All fitted emission lines are labeled and their individual line profiles are shown. For the top-right panel, the \nv\ emission line could not be convincingly fit independent of the \lya\ emission line.} \label{fig:set-lbol-fits}
\end{figure*}

\begin{figure*} 
\begin{tabular}{cc}
(a) $\log\left({\rm{M}_{\rm{BH}}/\rm{M}_{\odot}}\right) = 8.40-8.75$ & (b) $\log\left({\rm{M}_{\rm{BH}}/\rm{M}_{\odot}}\right) = 8.75-8.87$ \\
  \includegraphics[width=80mm]{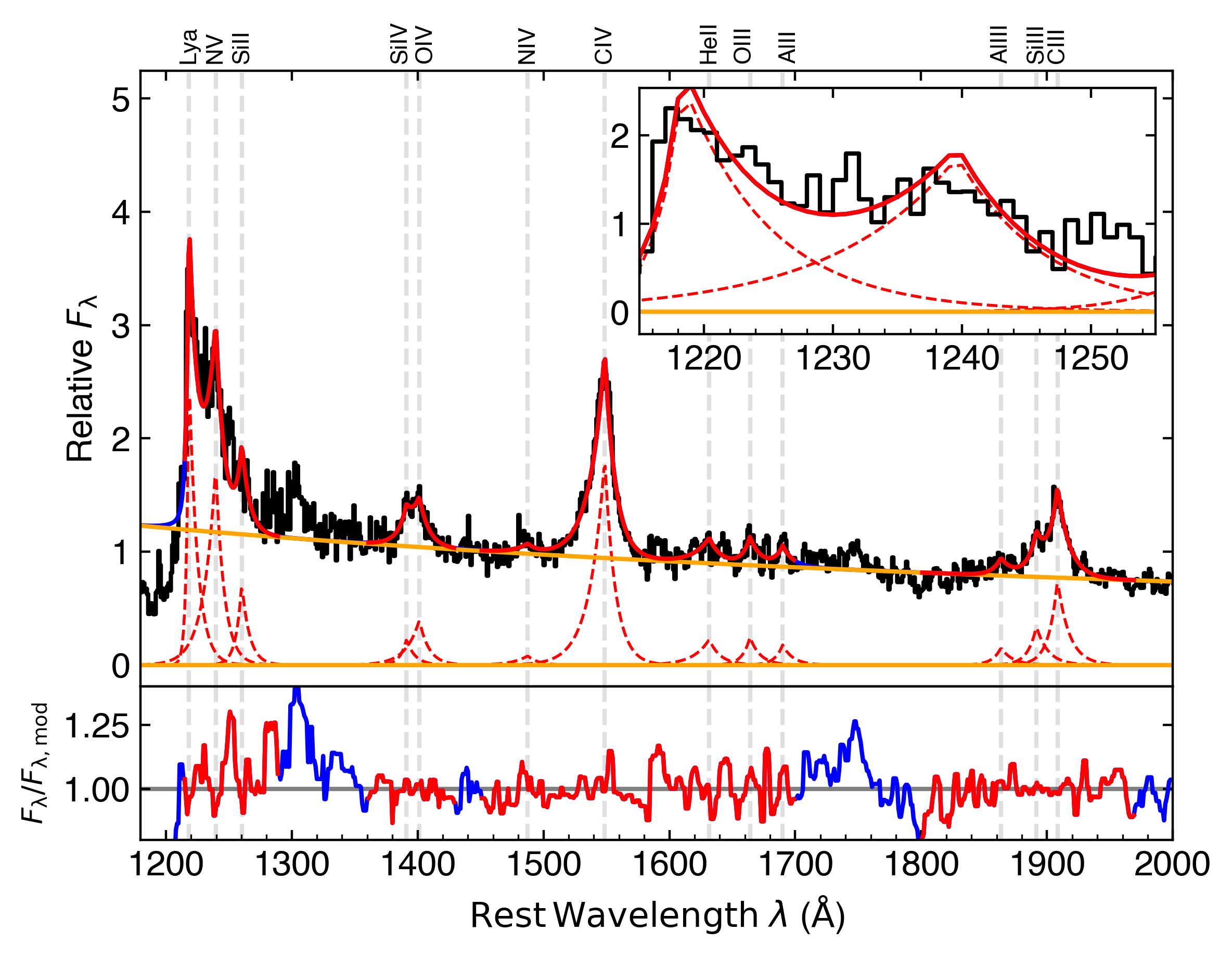} &   \includegraphics[width=80mm]{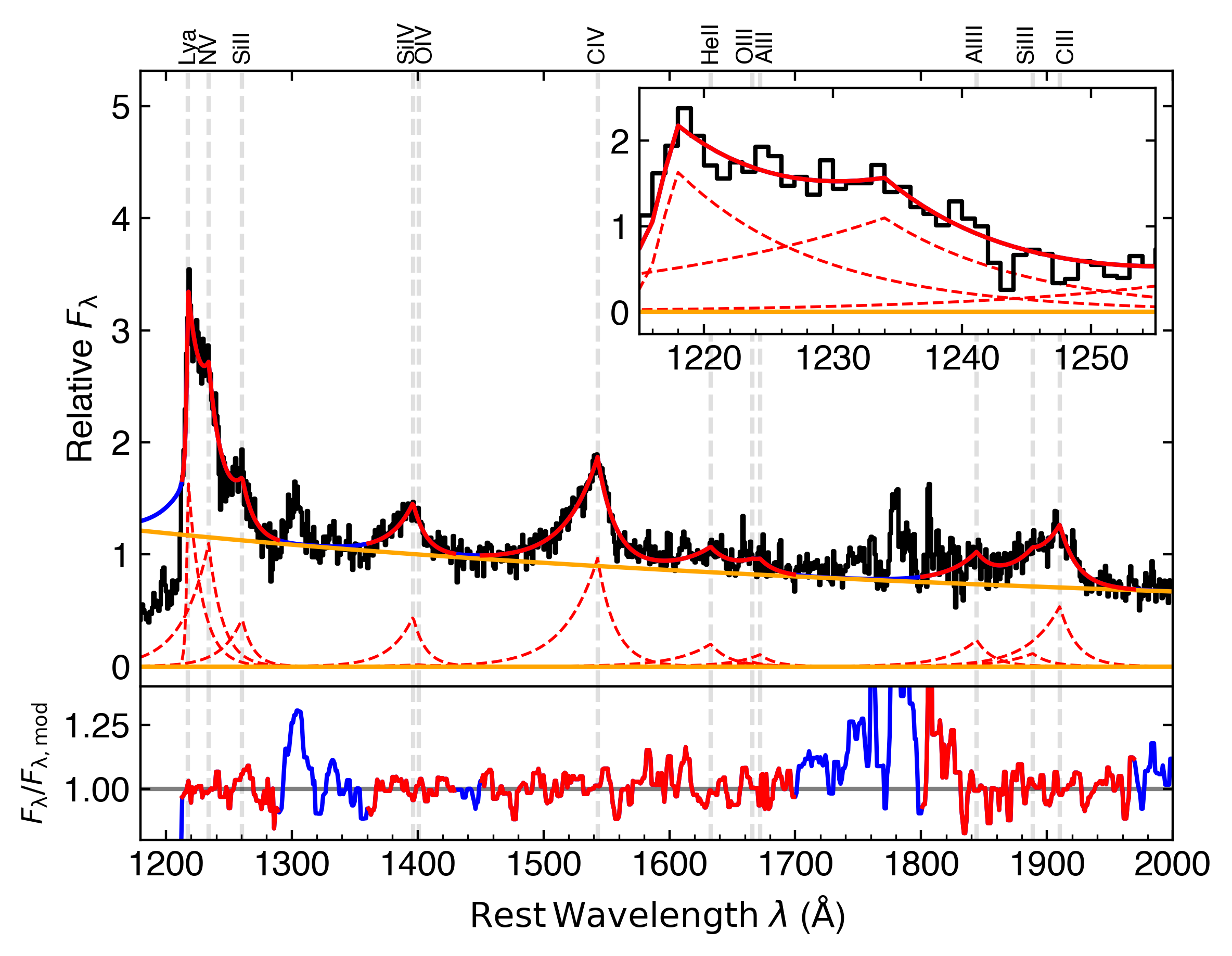} \\ [6pt]
(c) $\log\left({\rm{M}_{\rm{BH}}/\rm{M}_{\odot}}\right) = 8.87-8.98$ & (d) $\log\left({\rm{M}_{\rm{BH}}/\rm{M}_{\odot}}\right) = 8.98-9.20$ \\
 \includegraphics[width=80mm]{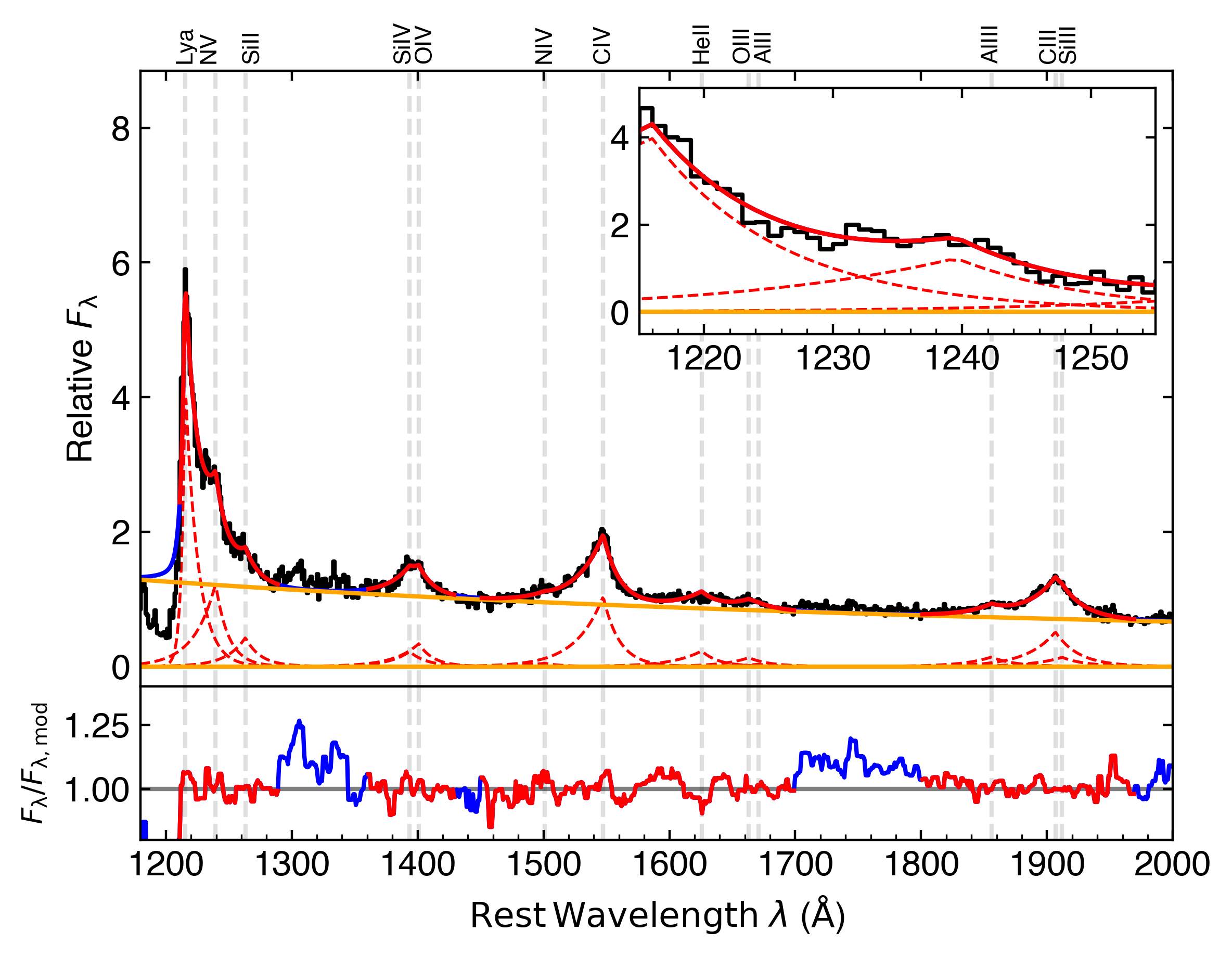} &   \includegraphics[width=80mm]{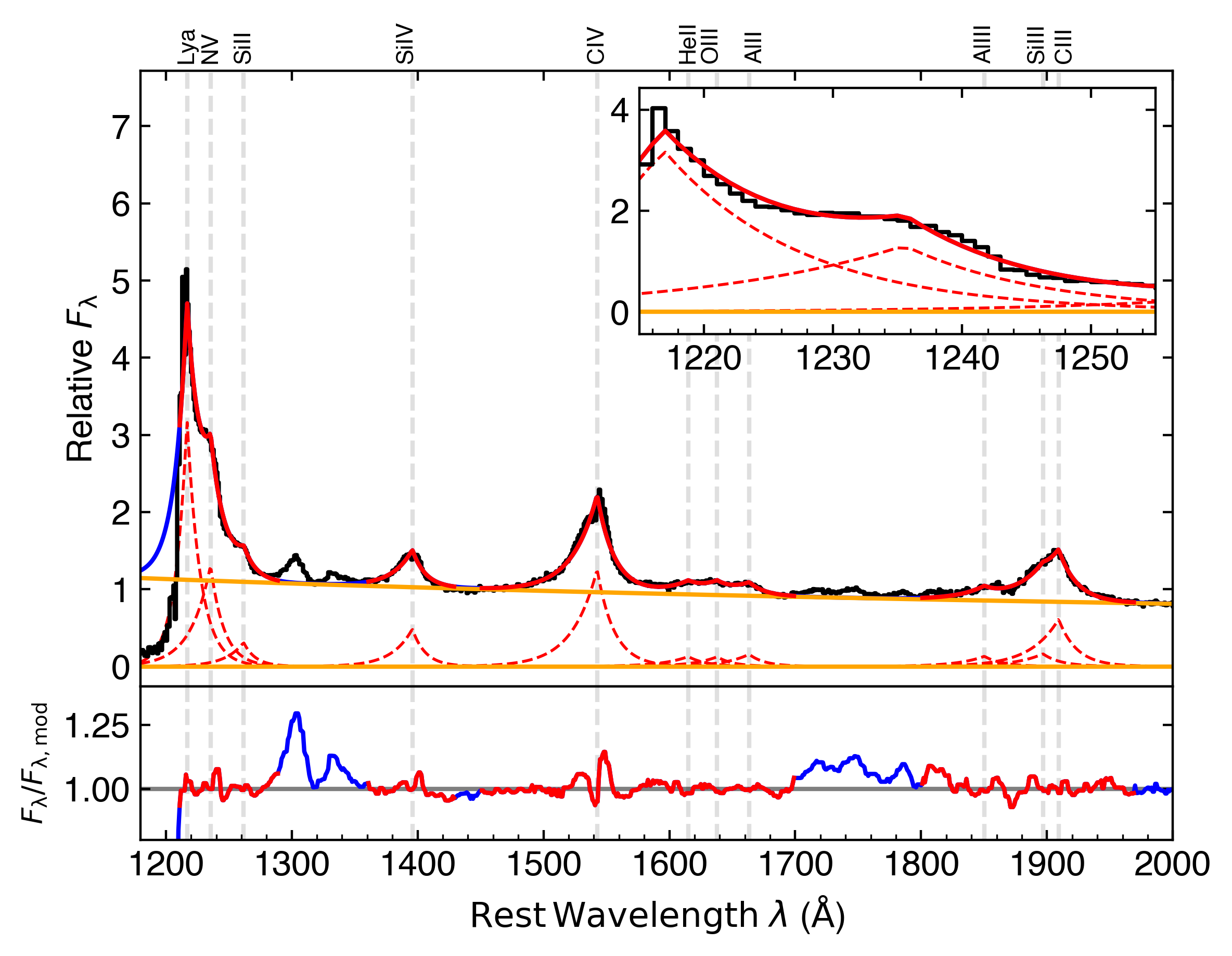} \\ [6pt]
(e) $\log\left({\rm{M}_{\rm{BH}}/\rm{M}_{\odot}}\right) = 9.20-9.40$ & (f) $\log\left({\rm{M}_{\rm{BH}}/\rm{M}_{\odot}}\right) = 9.40-9.80$ \\
 \includegraphics[width=80mm]{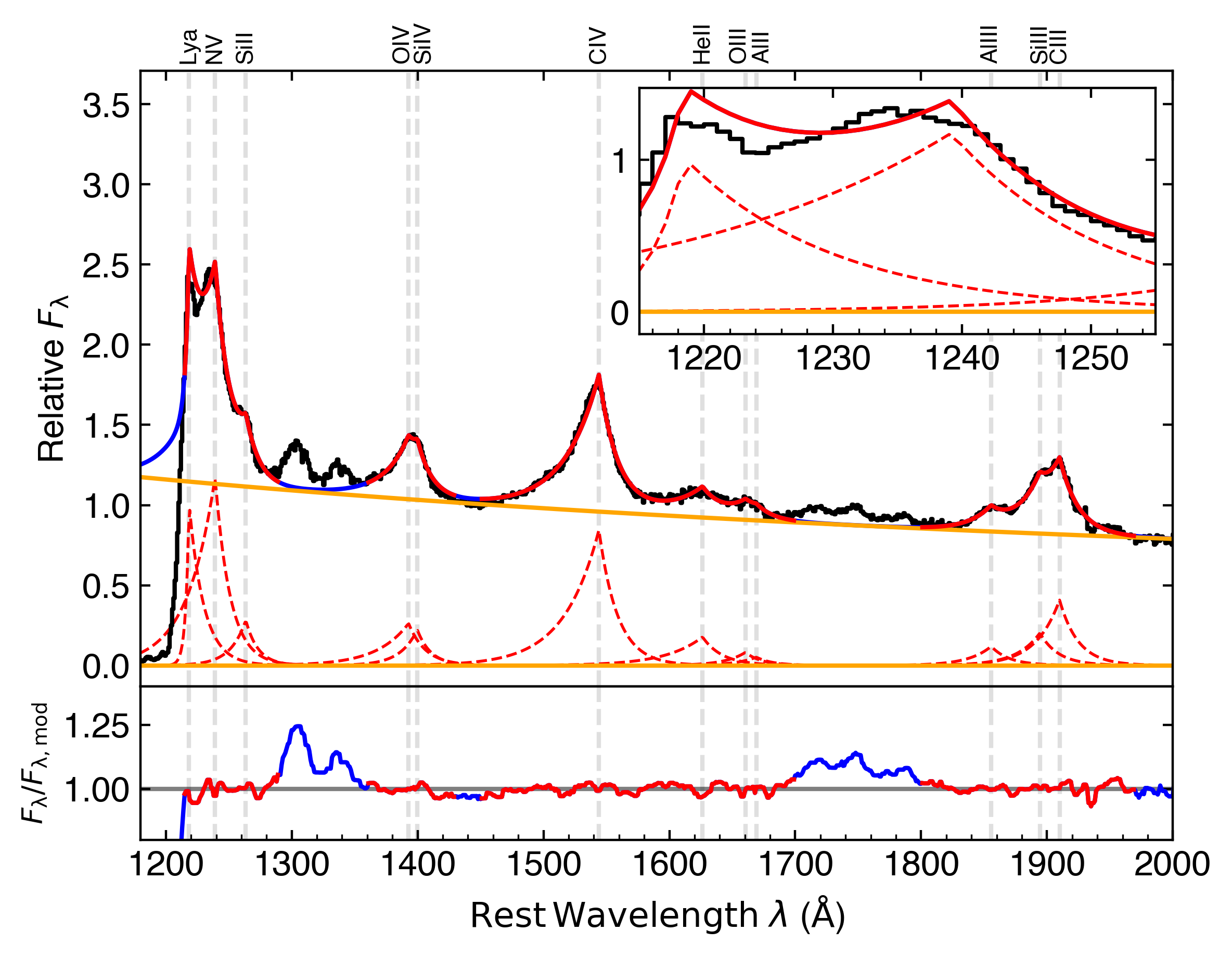} &   \includegraphics[width=80mm]{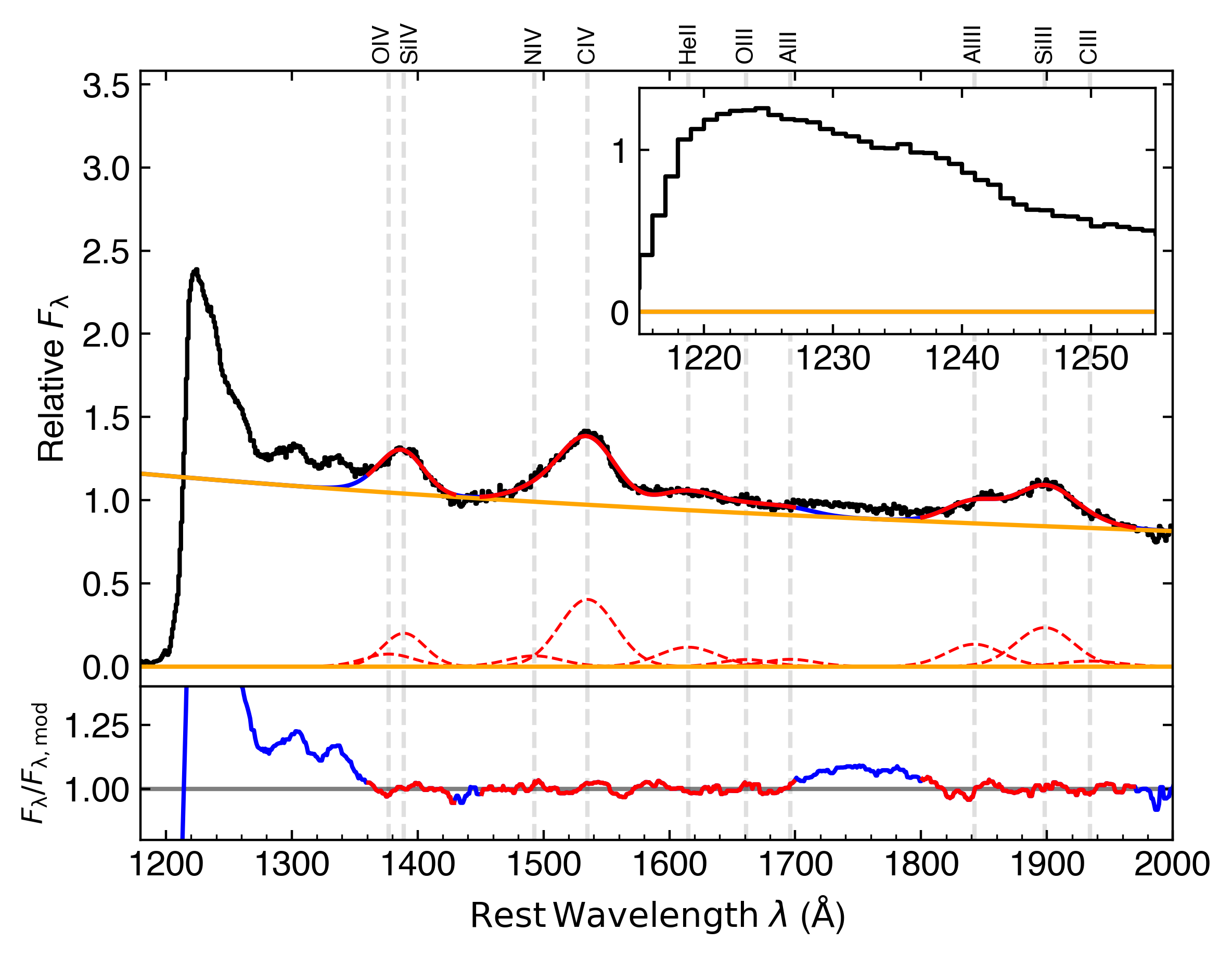} \\ [6pt]
\end{tabular}
\caption{Example fits to composites stacked in bins of virially-estimated black hole mass ordered by increasing mass from panels a-f. Line fitting and continuum fitting windows have been changed slightly between fits. The red lines indicate the emission line fits as well as the extent of the individual line-fitting windows. All fitted emission lines are labeled and their individual line profiles are shown. In the bottom-right panel, \nv\ was difficult to disentangle from the \lya\ emission, so only the \siivoivciv\ line ratio is fit. A skewed Gaussian was used to fit emission lines because of the highly blueshifted \civ\ spectral feature. } \label{fig:set-mbh-fits}
\end{figure*}

\begin{figure*}
\begin{tabular}{cc}
(a) \civ\ blueshift = $-$200-680 km s$^{-1}$ & 
(b) \civ\ blueshift = 680-1500 km s$^{-1}$ \\
  \includegraphics[width=80mm]{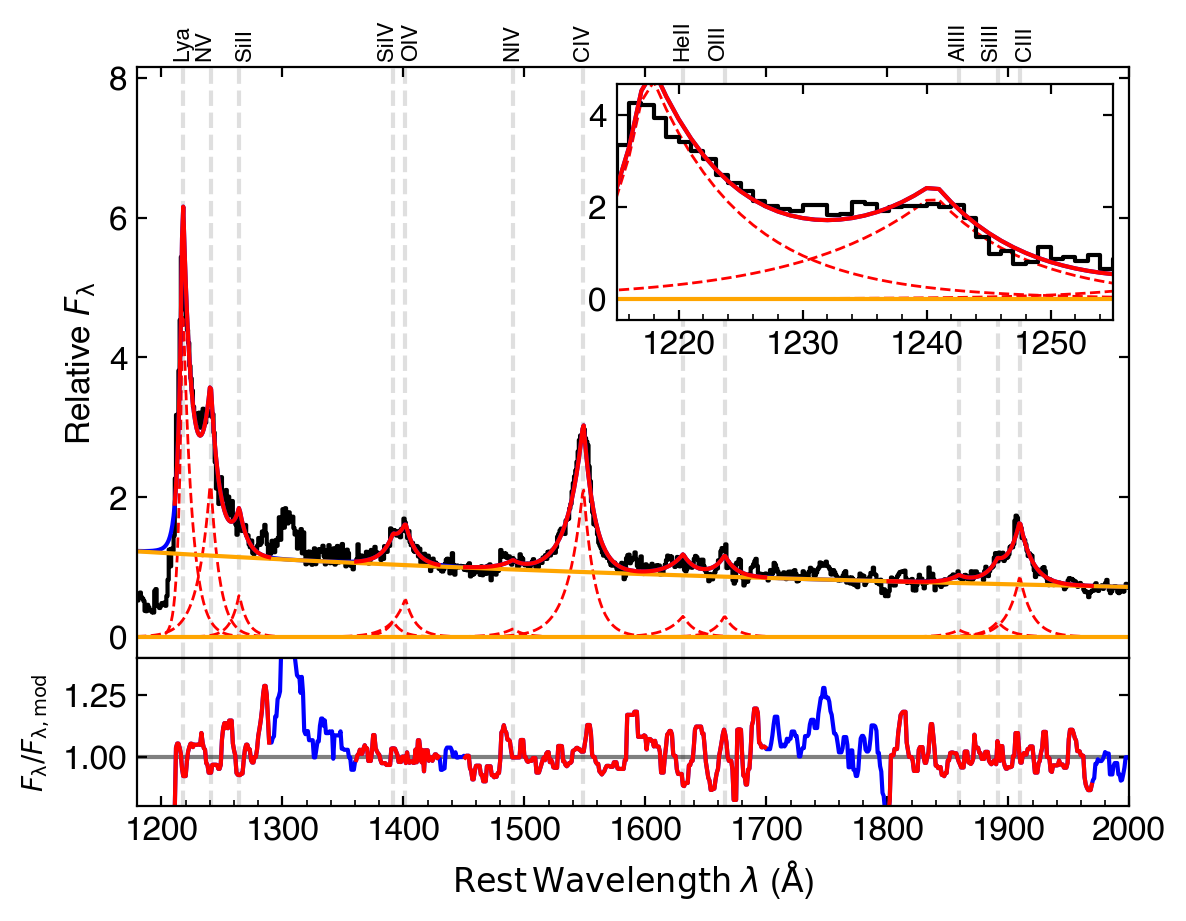} &   \includegraphics[width=80mm]{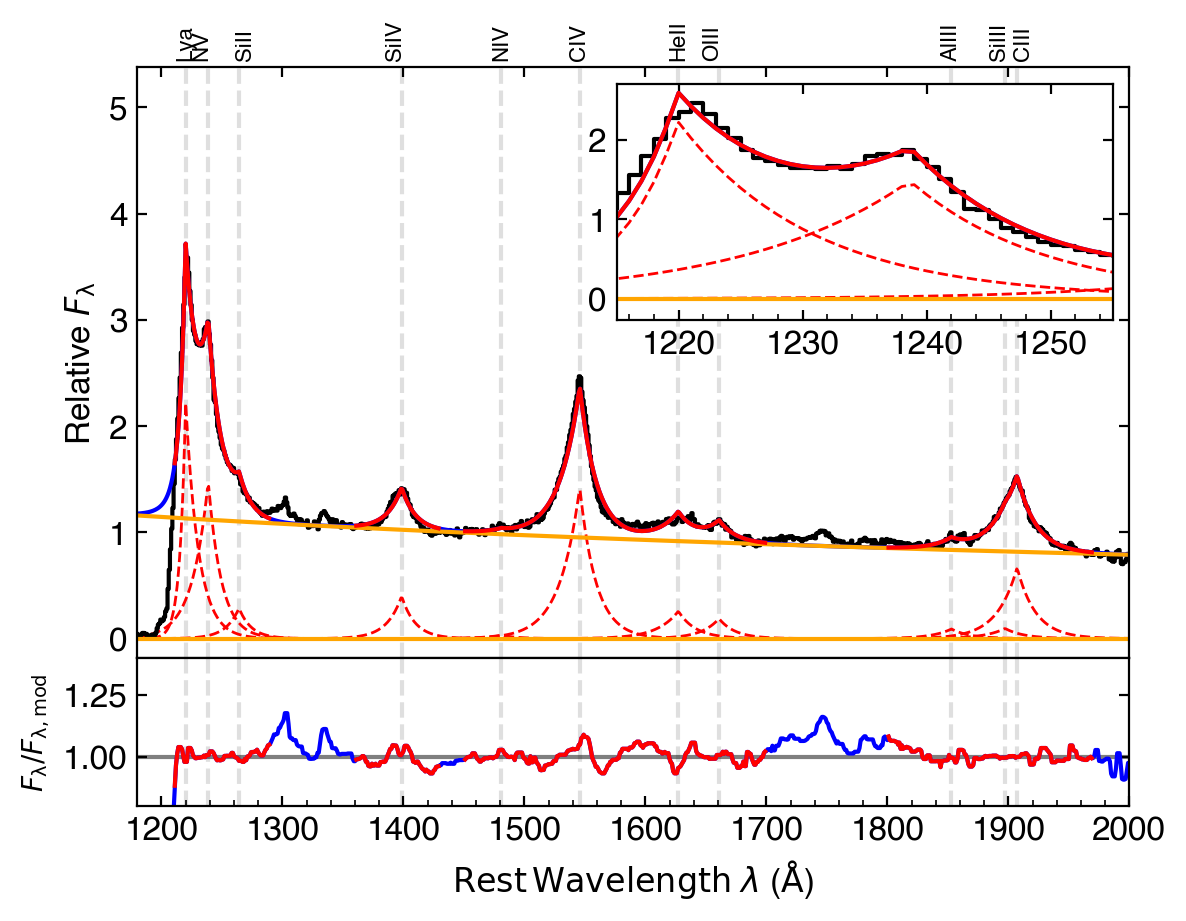} \\ [6pt]
(c) \civ\ blueshift = 1500-2500 km s$^{-1}$ & 
(d) \civ\ blueshift = 2500-3000 km s$^{-1}$ \\
 \includegraphics[width=80mm]{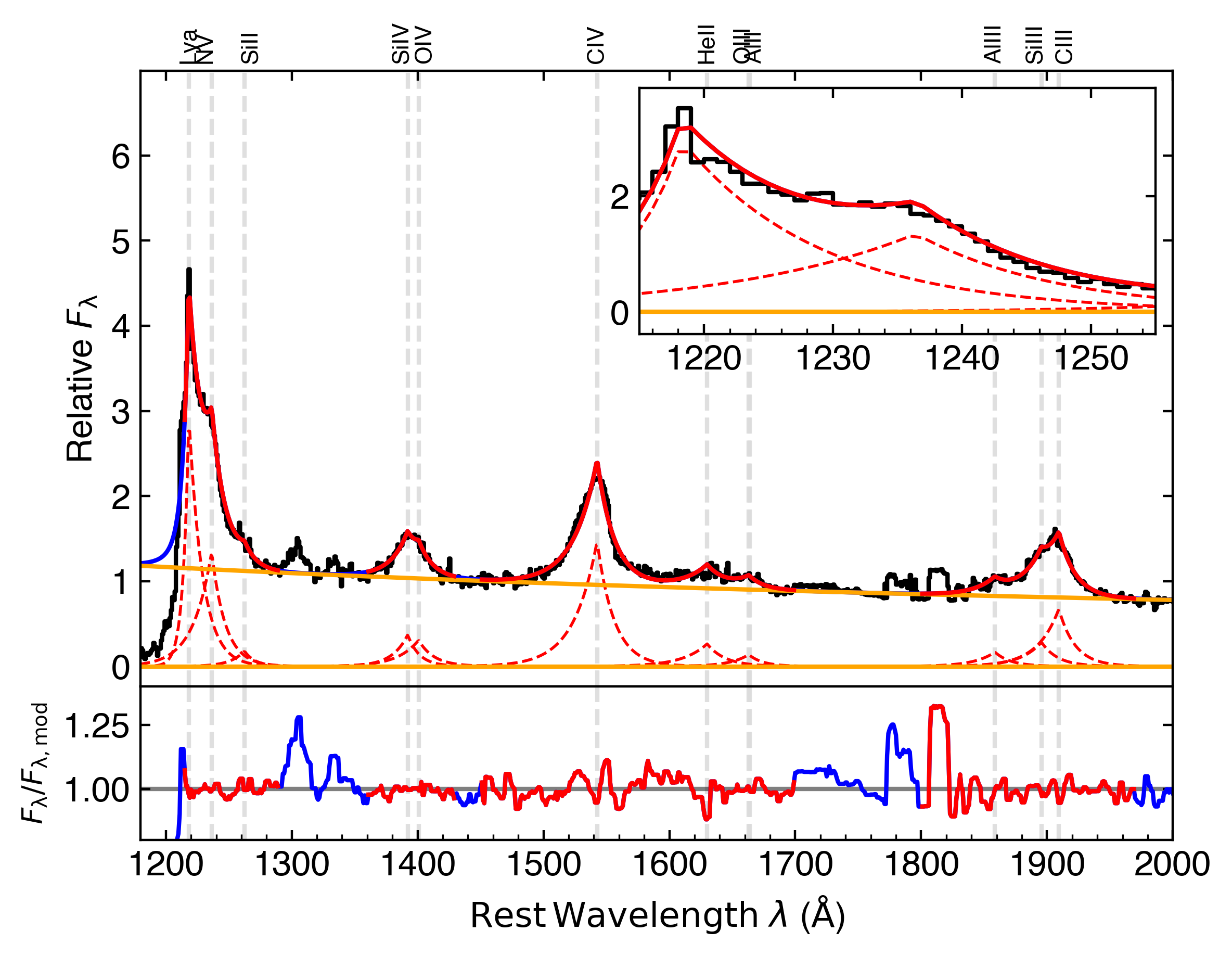} &   \includegraphics[width=80mm]{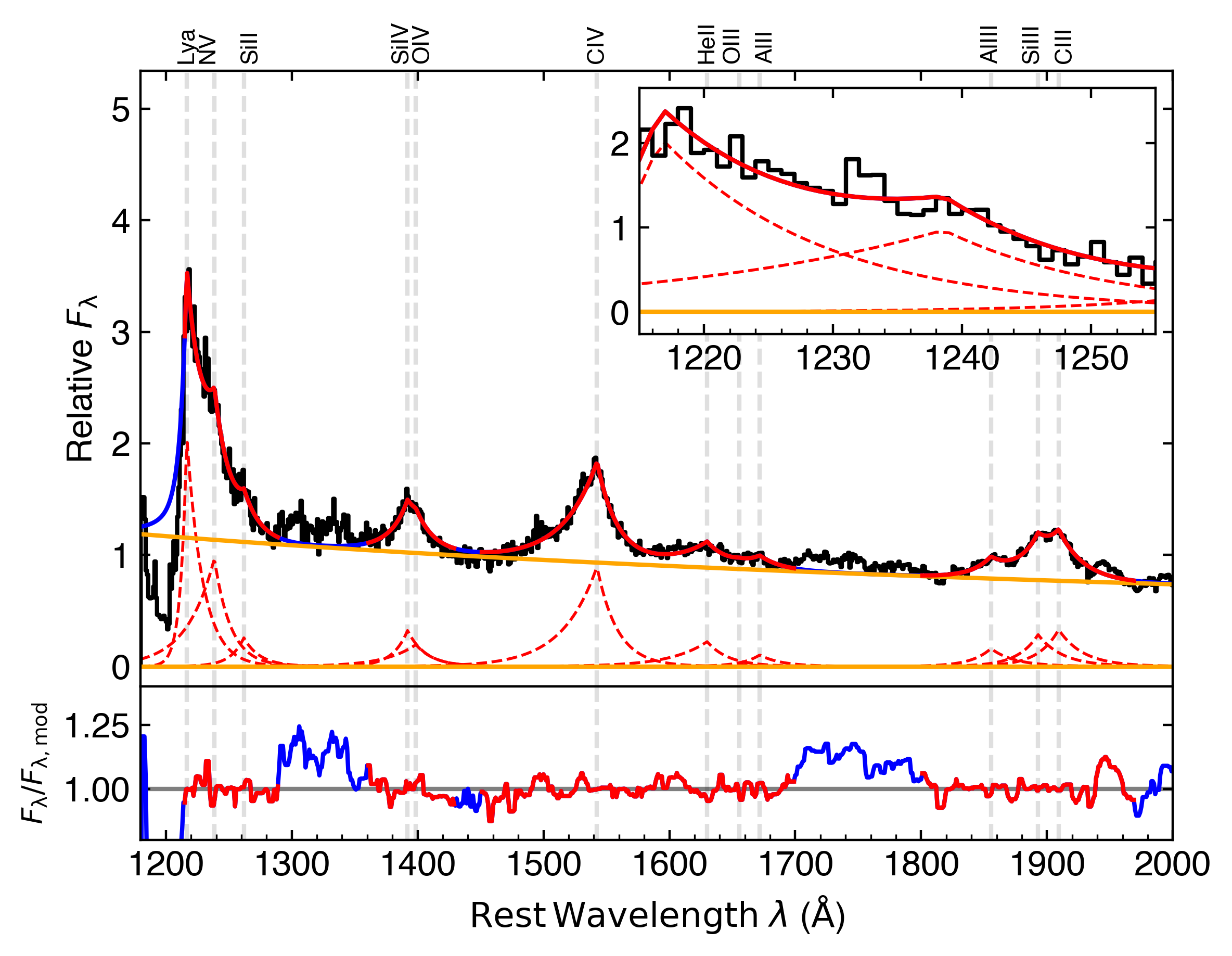} \\ [6pt]
(e) \civ\ blueshift = 3000-4000 km s$^{-1}$ & 
(f) \civ\ blueshift = 4000-5000 km s$^{-1}$ \\
 \includegraphics[width=80mm]{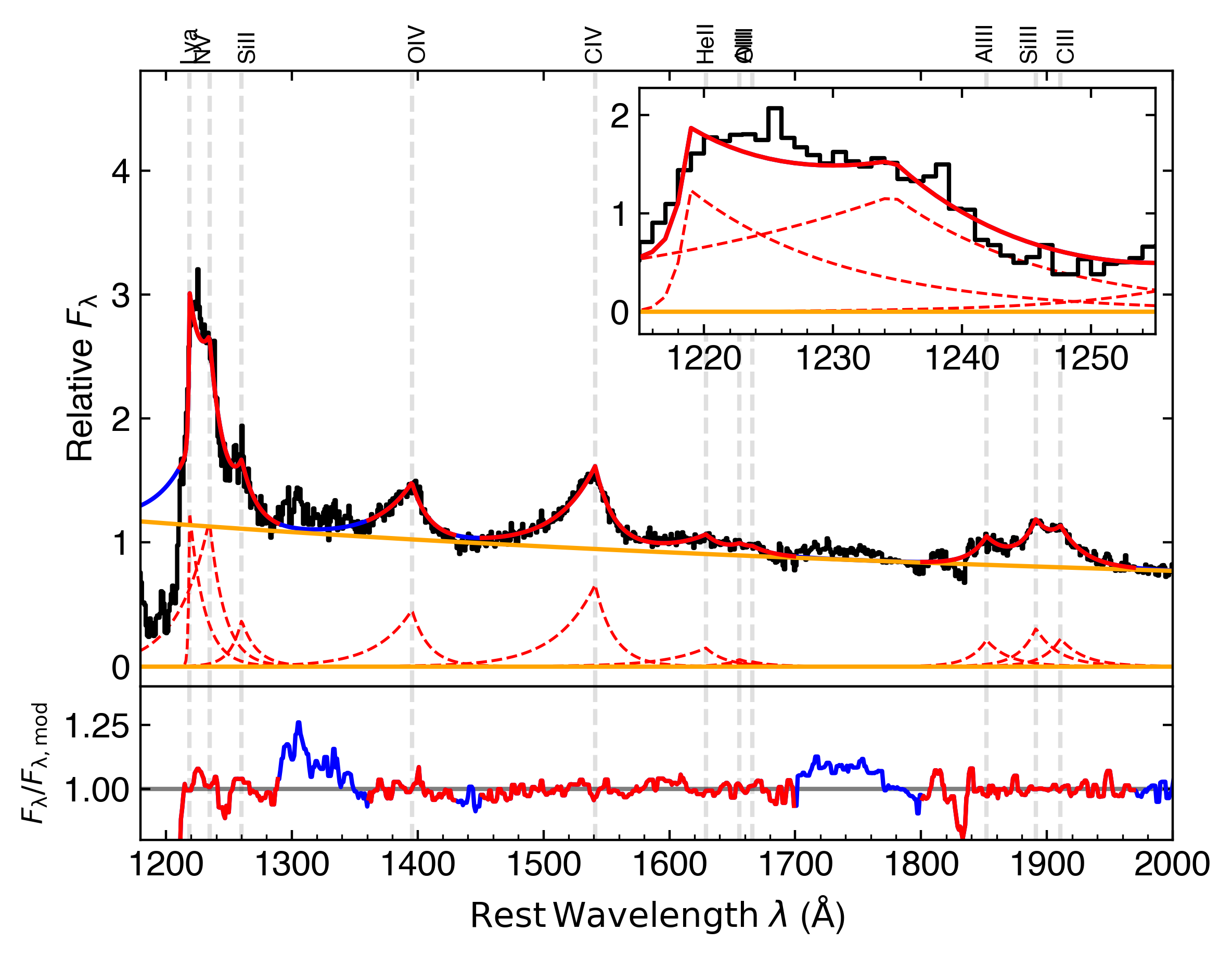} &   \includegraphics[width=80mm]{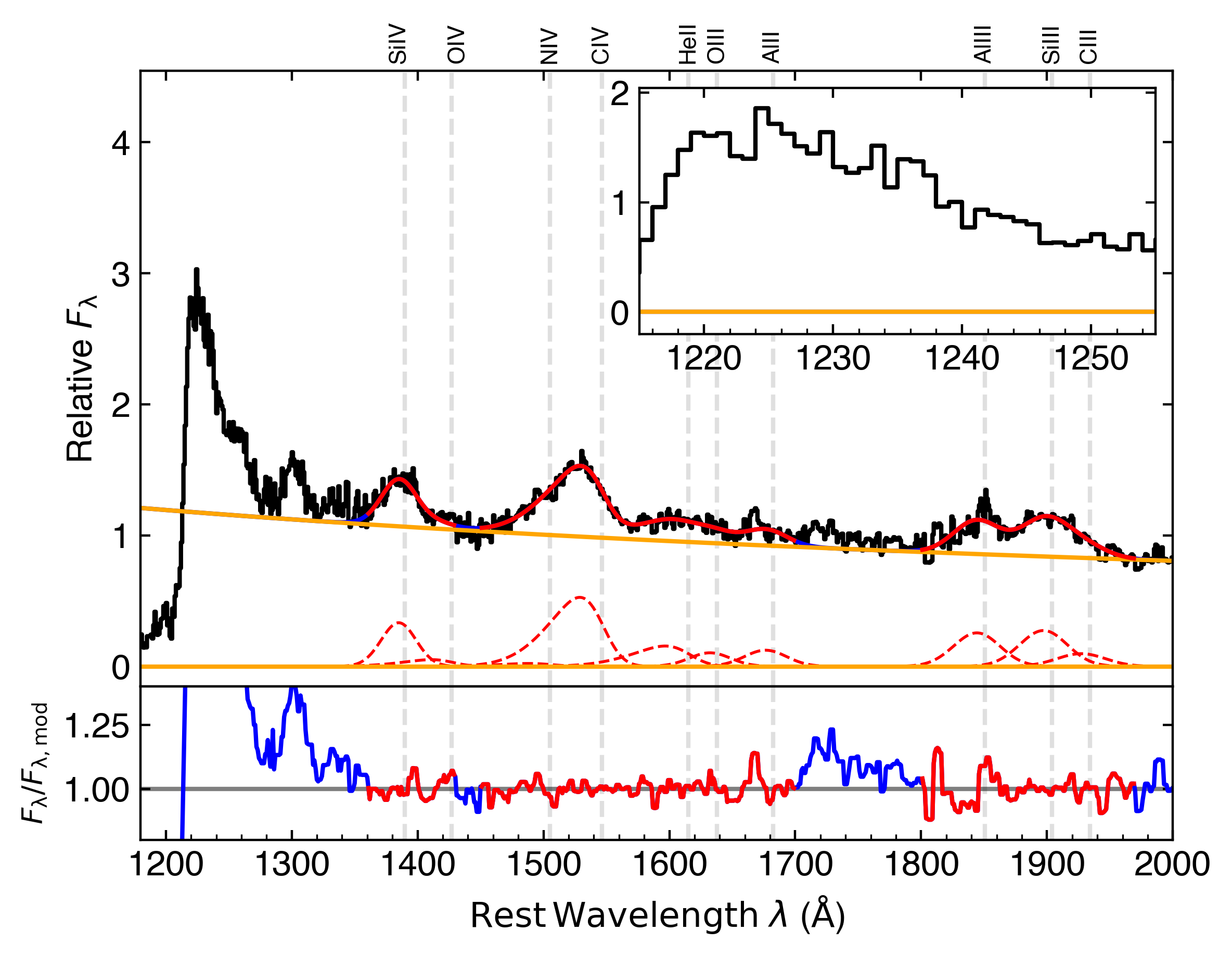} \\ [6pt]
\end{tabular}
\caption{Example fits to composites stacked in bins of \civ\ blueshift ordered by increasing blueshift from panels a-f. Line fitting and continuum fitting windows have been changed slightly between fits. For blueshifts greater than 4000 km $s^{-1}$, a skewed Gaussian function is fit to the emission lines instead of a piece-wise power-law. The red lines indicate the emission line fits as well as the extent of the individual line-fitting windows. All fitted emission lines are labeled and their individual line profiles are shown. In the bottom-right panel, \nv\ was difficult to disentangle from the \lya\ emission, so only the \siivoivciv\ line ratio is fit. A skewed Gaussian was used to fit emission lines because of the highly blueshifted \civ\ spectral feature. } \label{fig:set-blueshift-fits}
\end{figure*}

The following section contains additional tables to supplement the main text. Tables \ref{tab:composite_results}, \ref{tab:composite_mass}, and \ref{tab:composite_blueshift} contain information of the line flux ratios from luminosity composites, black hole mass composites, and \civ\ blueshift composites respectively, each normalised to the flux of the \civ\ emission line. The error for each line in the tables represents the measurement uncertainty of that line, without propagating the error of the \civ\ line from the normalisation. Figure \ref{fig:other_line_ratios} presents the line flux ratios of \oiiialiiciv, \aliiiciv, \siiiiciv, and \ciiiciv\ as functions of the bolometric luminosity, black hole mass, and \civ\ blueshift. Figure \ref{fig:NV_SiIV_Zcompare} is a comparison between the inferred metallicities using the \nvciv\ line ratio and the \siivoivciv\ line ratio. Figure sets \ref{fig:set-individual-fits}, \ref{fig:set-lbol-fits}, \ref{fig:set-mbh-fits}, and \ref{fig:set-blueshift-fits} show sample fits of selected individual spectra, bolometric luminosity composites, black hole mass composites, and \civ\ blueshift composites. 

\begingroup
\begin{table*}
\caption {\label{tab:composite_results} Table of line flux ratios from the bolometric luminosity composites normalised to the flux of the \civ\ emission line. The uncertainty represents the measurement error of the displayed line, independent of the \civ\ uncertainty.}  
\begin{tabular}{lccccccc}
\hline \hline
Line / $\log\left(\rm{L}_{\rm{bol}}/\rm{erg \, s}^{-1}\right)$ & 46.72-46.79 & 46.79-46.95 & 46.95-47.17 & 47.17-47.30 & 47.30-47.40 & 47.40-47.70 \\ \hline
\lya & 0.62 $\pm$ 0.05 & 0.45 $\pm$ 0.08 & 1.49 $\pm$ 0.09 & 1.86 $\pm$ 0.81 & 1.32 $\pm$ 0.07 & 0.59 $\pm$ 0.02 \\ 
\nv & 0.91 $\pm$ 0.05 & 1.06 $\pm$ 0.12 & 0.92 $\pm$ 0.09 & 1.01 $\pm$ 0.03 & 0.63 $\pm$ 0.04 & 0.84 $\pm$ 0.02 \\ 
\siii & 0.23 $\pm$ 0.02 & 0.48 $\pm$ 0.07 & 0.23 $\pm$ 0.02 & 0.14 $\pm$ 0.01 & 0.07 $\pm$ 0.02 & 0.21 $\pm$ 0.01 \\ 
\siiv & 0.07 $\pm$ 0.04 & 0.22 $\pm$ 0.13 & 0.12 $\pm$ 0.03 & 0.28 $\pm$ 0.01 & 0.40 $\pm$ 0.02 & 0.39 $\pm$ 0.01 \\ 
\oiv & 0.28 $\pm$ 0.07 & 0.23 $\pm$ 0.14 & 0.37 $\pm$ 0.04 & --- & --- & --- \\ 
\niv & 0.09 $\pm$ 0.02 & --- & --- & 0.04 $\pm$ 0.01 & --- & --- \\ 
\civ & 1.00 $\pm$ 0.04 & 1.00 $\pm$ 0.08 & 1.00 $\pm$ 0.02 & 1.00 $\pm$ 0.01 & 1.00 $\pm$ 0.02 & 1.00 $\pm$ 0.01 \\ 
\heii & 0.22 $\pm$ 0.02 & 0.24 $\pm$ 0.03 & 0.18 $\pm$ 0.01 & 0.21 $\pm$ 0.01 & 0.18 $\pm$ 0.01 & 0.18 $\pm$ 0.00 \\ 
\oiii & --- & 0.11 $\pm$ 0.03 & 0.00 $\pm$ 0.03 & 0.05 $\pm$ 0.04 & 0.09 $\pm$ 0.04 & 0.06 $\pm$ 0.00 \\ 
\alii & 0.16 $\pm$ 0.02 & 0.09 $\pm$ 0.03 & 0.10 $\pm$ 0.02 & 0.06 $\pm$ 0.04 & 0.02 $\pm$ 0.04 & 0.05 $\pm$ 0.01 \\ 
\aliii & 0.16 $\pm$ 0.01 & 0.12 $\pm$ 0.03 & 0.08 $\pm$ 0.01 & 0.09 $\pm$ 0.00 & 0.09 $\pm$ 0.01 & 0.15 $\pm$ 0.00 \\ 
\siiii & 0.21 $\pm$ 0.01 & 0.22 $\pm$ 0.04 & 0.12 $\pm$ 0.01 & 0.06 $\pm$ 0.02 & --- & 0.18 $\pm$ 0.01 \\ 
\ciii & 0.29 $\pm$ 0.02 & 0.40 $\pm$ 0.03 & 0.47 $\pm$ 0.04 & 0.53 $\pm$ 0.02 & 0.76 $\pm$ 0.03 & 0.47 $\pm$ 0.01 \\ 
\hline \hline
\end{tabular}
\end{table*}
\endgroup

\begingroup
\begin{table*}
\caption {\label{tab:composite_mass} Table of line flux ratios from the \hl{black hole mass} composites normalised to the flux of the \civ\ emission line. The uncertainty represents the measurement error of the displayed line, independent of the \civ\ uncertainty.}  
\begin{tabular}{lccccccc}
\hline \hline
Line / $\log(\rm{M}_{\rm{BH}}/\rm{M}_{\odot}$) & 8.40-8.75 & 8.75-8.87 & 8.87-8.98 & 8.98-9.20 & 9.20-9.40 & 9.40-9.80 \\ \hline
\lya & 1.59 $\pm$ 0.26 & 0.68 $\pm$ 0.21 & 1.58 $\pm$ 0.27 & 1.49 $\pm$ 0.21 & 0.87 $\pm$ 0.05 & 1.10 $\pm$ 0.41 \\ 
\nv & 0.66 $\pm$ 0.17 & 0.73 $\pm$ 0.24 & 0.98 $\pm$ 0.23 & 0.77 $\pm$ 0.10 & 0.91 $\pm$ 0.08 & 0.78 $\pm$ 0.35 \\ 
\siii & 0.36 $\pm$ 0.18 & 0.26 $\pm$ 0.16 & 0.29 $\pm$ 0.10 & 0.15 $\pm$ 0.04 & 0.20 $\pm$ 0.04 & 0.37 $\pm$ 0.24 \\ 
\siiv & 0.13 $\pm$ 0.13 & 0.24 $\pm$ 0.15 & 0.25 $\pm$ 0.15 & 0.30 $\pm$ 0.05 & 0.09 $\pm$ 0.07 & 0.35 $\pm$ 0.33 \\ 
\oiv & 0.12 $\pm$ 0.08 & 0.07 $\pm$ 0.11 & 0.24 $\pm$ 0.22 & --- & 0.33 $\pm$ 0.13 & --- \\ 
\niv & 0.05 $\pm$ 0.05 & --- & --- & --- & --- & --- \\ 
\civ & 1.00 $\pm$ 0.15 & 1.00 $\pm$ 0.23 & 1.00 $\pm$ 0.16 & 1.00 $\pm$ 0.05 & 1.00 $\pm$ 0.07 & 1.00 $\pm$ 0.18 \\ 
\heii & 0.10 $\pm$ 0.05 & 0.25 $\pm$ 0.08 & 0.24 $\pm$ 0.07 & 0.17 $\pm$ 0.02 & 0.23 $\pm$ 0.03 & 0.20 $\pm$ 0.08 \\ 
\oiii & 0.10 $\pm$ 0.10 & 0.03 $\pm$ 0.06 & 0.10 $\pm$ 0.07 & 0.11 $\pm$ 0.03 & 0.09 $\pm$ 0.04 & 0.05 $\pm$ 0.07 \\ 
\alii & 0.09 $\pm$ 0.11 & 0.06 $\pm$ 0.09 & 0.03 $\pm$ 0.05 & 0.00 $\pm$ 0.02 & 0.06 $\pm$ 0.04 & 0.15 $\pm$ 0.09 \\ 
\aliii & 0.12 $\pm$ 0.07 & 0.24 $\pm$ 0.09 & 0.17 $\pm$ 0.05 & 0.11 $\pm$ 0.02 & 0.11 $\pm$ 0.02 & 0.25 $\pm$ 0.15 \\ 
\siiii & --- & 0.17 $\pm$ 0.10 & 0.07 $\pm$ 0.10 & 0.14 $\pm$ 0.08 & 0.21 $\pm$ 0.06 & 0.38 $\pm$ 0.31 \\ 
\ciii & 0.55 $\pm$ 0.20 & 0.46 $\pm$ 0.13 & 0.60 $\pm$ 0.18 & 0.52 $\pm$ 0.12 & 0.38 $\pm$ 0.09 & 0.51 $\pm$ 0.34 \\ 
\hline \hline
\end{tabular}
\end{table*}
\endgroup

\begingroup
\begin{table*}
\caption {\label{tab:composite_blueshift} Table of line flux ratios from the \civ\ blueshift composites normalised to the flux of the \civ\ emission line. The uncertainty represents the measurement error of the displayed line, independent of the \civ\ uncertainty.}  
\begin{tabular}{lcccccc}
\hline \hline
Line / \civ\ Blueshift & $-$200-680 & 680-1500 & 1500-2500 & 2500-3000 & 3000-4000 & 4000-5000  \\ 
& (km s$^{-1}$) & (km s$^{-1}$) & (km s$^{-1}$) & (km s$^{-1}$) & (km s$^{-1}$) &  (km s$^{-1}$)\\ \hline
\lya & 1.08 $\pm$ 0.15 & 0.80 $\pm$ 0.09 & 0.94 $\pm$ 0.12 & 1.12 $\pm$ 0.51 & 0.49 $\pm$ 0.13 & --- \\ 
\nv & 0.83 $\pm$ 0.12 & 0.84 $\pm$ 0.08 & 0.76 $\pm$ 0.08 & 0.91 $\pm$ 0.19 & 1.39 $\pm$ 0.24 & --- \\ 
\siii & 0.22 $\pm$ 0.08 & 0.13 $\pm$ 0.02  & 0.08 $\pm$ 0.03 & 0.18 $\pm$ 0.07 & 0.29 $\pm$ 0.08 & ---\\ 
\siiv & 0.09 $\pm$ 0.09 & 0.22 $\pm$ 0.03 & 0.21 $\pm$ 0.08 & 0.16 $\pm$ 0.16 & 0.09 $\pm$ 0.16 & 0.35 $\pm$ 0.15 \\ 
\oiv & 0.22 $\pm$ 0.09 & --- & 0.15 $\pm$ 0.09 & 0.32 $\pm$ 0.22 & 0.51 $\pm$ 0.25 & 0.17 $\pm$ 0.20 \\ 
\niv & 0.06 $\pm$ 0.05 & 0.01 $\pm$ 0.02 & --- & --- & --- & 0.15 $\pm$ 0.12 \\ 
\civ & 1.00 $\pm$ 0.11 & 1.00 $\pm$ 0.06 & 1.00 $\pm$ 0.07 & 1.00 $\pm$ 0.20 & 1.00 $\pm$ 0.13 & 1.00 $\pm$ 0.16 \\ 
\heii & 0.14 $\pm$ 0.03 & 0.19 $\pm$ 0.03 & 0.20 $\pm$ 0.03 & 0.27 $\pm$ 0.09 & 0.23 $\pm$ 0.06 & 0.34 $\pm$ 0.09 \\ 
\oiii & 0.11 $\pm$ 0.06 & 0.11 $\pm$ 0.03 & 0.07 $\pm$ 0.04 & 0.02 $\pm$ 0.03 & 0.07 $\pm$ 0.05 & 0.14 $\pm$ 0.06 \\ 
\alii & --- & --- & --- & 0.07 $\pm$ 0.05 & --- & 0.16 $\pm$ 0.07 \\
\aliii & 0.04 $\pm$ 0.03 & 0.07 $\pm$ 0.02 & 0.10 $\pm$ 0.03 & 0.15 $\pm$ 0.03 & 0.20 $\pm$ 0.04 & 0.43 $\pm$ 0.09 \\ 
\siiii & 0.05 $\pm$ 0.05 & 0.09 $\pm$ 0.07 & 0.15 $\pm$ 0.06 & 0.27 $\pm$ 0.06 & 0.31 $\pm$ 0.04 & 0.32 $\pm$ 0.16 \\ 
\ciii & 0.43 $\pm$ 0.09 & 0.48 $\pm$ 0.11 & 0.47 $\pm$ 0.08 & 0.29 $\pm$ 0.06 & 0.27 $\pm$ 0.04 & 0.27 $\pm$ 0.13 \\ 
\hline \hline
\end{tabular}
\end{table*}
\endgroup

\bsp	
\label{lastpage}
\end{document}